\documentclass[review,11pt,authoryear,fleqn]{elsarticle}
\usepackage[applemac]{inputenc}
\usepackage[english]{babel}
\usepackage{natbib}
\usepackage{subfig}
\usepackage{morefloats}
\usepackage{latexsym}
\usepackage{graphicx}
\usepackage{amsmath}
\allowdisplaybreaks 
\usepackage{amsfonts}
\usepackage{amssymb,tikz}
\usepackage{array}
\usepackage{dsfont}
\usepackage[counterclockwise]{rotating}
\usepackage{multirow}
\usepackage{bigints}
\usepackage{bbm}

\usepackage{grffile}
\graphicspath{{Figures/}}
\usepackage{epstopdf}       
\usepackage[capposition=top]{floatrow}

\usepackage{geometry}
\usepackage{setspace}
\newcommand{\slfrac}[2]{\scalebox{0.55}{$\left.#1\middle/#2\right.$}}

\usepackage{color,soul}
\usepackage{xcolor}
\usepackage{framed}
\colorlet{shadecolor}{yellow}

\usepackage[pdftex,bookmarks=false,pdfpagelabels=false,hidelinks=true,hyperfootnotes=false,hyperfigures = false,hyperindex=false,pageanchor=false,colorlinks=true,citecolor = blue]{hyperref}

\usepackage{ulem}		
\usepackage{textcomp}	

\usepackage{pdflscape}
\usepackage{afterpage}
\usepackage{capt-of}

\usepackage{rotating}

\usepackage{tabularx}
\usepackage{longtable}
\usepackage{threeparttable}

\newtheorem{assum}{Assumption}[section]

\newtheorem{thm}{Theorem}[section]
\newtheorem{defi}{Definition}

\newcommand{\wtl}{\widetilde}

\newcommand{\EE}{\mathbf{E}}
\newcommand{\PP}{\mathbf{P}}
\newcommand{\btheta}{\boldsymbol{\theta}}

\makeatletter
\def\env@cases{%
  \let\@ifnextchar\new@ifnextchar
  \left.
  \def\arraystretch{1.2}%
  \array{@{}l@{\,\,}l@{}}%
}%
\makeatother

\usepackage{algorithm}
\usepackage[noend]{algpseudocode}
\makeatletter
\def\BState{\State\hskip-\ALG@thistlm}
\makeatother

\usepackage{xpatch}
\makeatletter
\xpatchcmd{\algorithmic}{\itemsep\z@}{\itemsep=1.2ex plus2pt}{}{}
\makeatother

\newcommand{\mysetminusD}{\hbox{\tikz{\draw[line width=0.6pt,line cap=round] (3pt,0) -- (0,6pt);}}}
\newcommand{\mysetminusT}{\mysetminusD}
\newcommand{\mysetminusS}{\hbox{\tikz{\draw[line width=0.45pt,line cap=round] (2pt,0) -- (0,4pt);}}}
\newcommand{\mysetminusSS}{\hbox{\tikz{\draw[line width=0.4pt,line cap=round] (1.5pt,0) -- (0,3pt);}}}

\newcommand{\mysetminus}{\mathbin{\mathchoice{\mysetminusD}{\mysetminusT}{\mysetminusS}{\mysetminusSS}}}

\newcommand{\cmmnt}[1]{\ignorespaces}

\usepackage{verbatim}

\usepackage[draft,colorinlistoftodos,textsize=tiny]{todonotes}
\usepackage{soul}
\makeatletter
\if@todonotes@disabled

\else

\fi
\makeatother

\usepackage{xr}

\begin{document}

\begin{onehalfspacing}

\title{\huge Bayesian MIDAS Penalized Regressions:\\ Estimation, Selection, and Prediction}

\author{Matteo Mogliani\corref{cor1}\fnref{fn1}}
\ead{matteo.mogliani@banque-france.fr}
\address{Banque de France - Conjunctural Analysis and Forecasting Division}
\cortext[cor1]{Banque de France, 46-1383 DGSEI-DCPM-DIACONJ, 31 Rue Croix des Petits Champs, 75049 Paris CEDEX 01 (France). Phone: +33(0)142925939.}
\fntext[fn1]{We wish to thank Florian Pelgrin, Clement Marsilli, Daniele Siena, and participants at the 
5th conference of the International Association of Applied Econometrics (26-29 June 2018, Montreal, Canada), the 4th Vienna Workshop on High Dimensional Time Series in Macroeconomics and Finance (16-17 May 2019, Vienna, Austria), the 62nd ISI World Statistics Congress (18-23 August 2019, Kuala Lumpur, Malaysia), and research seminars at Banque de France and the Institute for Mathematics and Statistics of the Vienna University (18 January 2019, Vienna, Austria). The usual disclaimer applies. The views expressed in this paper are those of the authors and do not necessarily reflect those of the Banque de France.}

\author{Anna Simoni\corref{cor2}}
\ead{simoni.anna@gmail.com}
\address{CREST, CNRS}
\cortext[cor2]{CREST, CNRS, \'{E}cole Polytechnique, ENSAE, 5 Avenue Henry Le Chatelier, 91120 Palaiseau (France). Phone: +33(0)170266837.}


\begin{abstract}
We propose a new approach to mixed-frequency regressions in a high-dimensional environment that resorts to Group Lasso penalization and Bayesian techniques for estimation and inference. In particular, to improve the prediction properties of the model and its sparse recovery ability, we consider a Group Lasso with a spike-and-slab prior. Penalty hyper-parameters governing the model shrinkage are automatically tuned via an adaptive MCMC algorithm. We establish good frequentist asymptotic properties of the posterior of the in-sample and out-of-sample prediction error, we recover the optimal posterior contraction rate, and we show optimality of the posterior predictive density. Simulations show that the proposed models have good selection and forecasting performance in small samples, even when the design matrix presents cross-correlation. When applied to forecasting U.S. GDP, our penalized regressions can outperform many strong competitors. Results suggest that financial variables may have some, although very limited, short-term predictive content.
\vspace{0.3cm}
\end{abstract}

\begin{keyword}
Bayesian MIDAS regressions \sep Penalized regressions \sep Predictive distribution \sep Forecasting \sep Posterior contraction
\JEL C11 \sep C22 \sep C53 \sep E37
\end{keyword}

\maketitle
\end{onehalfspacing}

\newpage
\setcounter{page}{1}
\setcounter{footnote}{0}
\section{Introduction}

Mixed-data sampling (MIDAS) regressions provide a parsimonious and theoretically efficient treatment of the time-aggregation problem when data are sampled at different frequencies \citep{Ghysels2005,Ghysels2007,Andreou2010}. They have been therefore intensively and successfully used to forecast low frequency series, such as GDP, using monthly, weekly or daily predictors \citep{Clements2008,Clements2009,Kuzin2011,Andreou2013}. However, one important issue with MIDAS regressions is the estimation and prediction in high dimension, as the presence of many high-frequency covariates may easily lead to overparameterized models, in-sample overfitting, and poor predictive performance. This happens because the MIDAS approach can efficiently address the dimensionality issue arising from the number of high-frequency lags in the model, but not that arising from the number of high-frequency variables.
In this respect, a number of strategies have been proposed in the literature, such as unrestricted MIDAS regressions (U-MIDAS; \citealp{Foroni2015}) with automatic model selection of relevant predictors and high-frequency lags \citep{Castle2010,Bec2015}, factor-MIDAS regressions \citep{Marcellino2010}, and \textit{targeted} factor-MIDAS regressions (\citealp{Bai2008}; see also \citealp{Bessec2013}, \citealp{Bulligan2015}, and \citealp{Girardi2017}).

Recently, the literature has been increasingly focusing on Machine Learning and penalized regression techniques for macroeconomic applications in a high-dimensional environment (see \citealp{Korobilis2013}; \citealp{NG2013752}; \citealp{Gefang2014}; \citealp{ng_2017}; \citealp{Koop2019}; \citealp{Korobilis2019}; to name only a few). Nevertheless, only a few contributions have paid attention to MIDAS regressions in high-dimension so far \citep{Marsilli2014,Siliverstovs2017,Uematsu2019,Babii2020}.\footnote{\citet{Marsilli2014} proposes a functional MIDAS combined with a Lasso objective function, which is solved in 1-step through a non-linear optimization algorithm. \citet{Siliverstovs2017} proposes a 2-step targeted factor-MIDAS approach, where the soft-thresholding rule is built around U-MIDAS regressions combined with an Elastic-Net objective function. Very recently, \citet{Babii2020} have proposed a sparse-group Lasso estimator for MIDAS regression based on Legendre polynomials, which is closely related to our approach.}
In particular, \citet{Uematsu2019} propose a general theoretical framework for penalized regressions where the number of covariates diverges sub-exponentially from the number of observations. Their framework is then especially fitted for U-MIDAS regressions, as in this case the number of parameters to estimate grows with both the number of high-frequency regressors and the length of the unrestricted lag function. However, this approach might not be generally suited for macroeconomic applications with monthly or daily high-frequency predictors, as penalized U-MIDAS regressions can saturate in presence of more predictors than observations, while unrestricted lags of the high-frequency predictors, by construction highly correlated, may be subject to random selection, leaving most of the remaining lag coefficients shrunk to zero.

In the present paper, we address these issues by proposing a novel penalized Bayesian MIDAS approach, based on Almon lag polynomials (a classic approximating function of the distributed lag model) and a prior that induces a Group Lasso penalty. The group penalty seems particularly attractive for the MIDAS framework, as one can assign the lag polynomial of a given predictor into one single group. Shrinkage is then performed simultaneously over the entire lag polynomial, rather than on individual terms of the MIDAS weighting function, overcoming the problem of extremely high correlation within the distributed lags.
We then introduce two models: the Bayesian MIDAS Adaptive Group Lasso (BMIDAS-AGL) and the Bayesian MIDAS Adaptive Group Lasso with spike-and-slab prior (BMIDAS-AGL-SS). The latter combines the penalized likelihood approach of the Group Lasso prior with a spike-and-slab prior at the group level, which is expected to  improve the sparsity recovery ability of the model \citep{Zhang2014,Zhao2015,Xu2015,Rockova2018}. Finally, penalty hyper-parameters governing the group selection are here tuned through a computationally efficient and data-driven approach based on stochastic approximations \citep{Atchade2011,Atchade2011a} that does not resort to extremely time consuming pilot runs.

\indent We theoretically validate our procedures by establishing good frequentist asymptotic properties of both the posterior predictive error (in- and out-of-sample) and the posterior predictive distribution. For this purpose, we focus on the BMIDAS-AGL-SS model and we adopt a frequentist point of view, \textit{i.e.} we admit the existence of a true parameter that generates the data. The asymptotic theory is developed by allowing the in-sample size $T$, the number of groups $G$, and their maximum size $g_{\max}$, to increase to infinity. First, we establish consistency and derive the posterior contraction rate for both in- and out-of-sample prediction error. When the number of nonzero groups times $\log(G)$ is larger than $\log(T)$, our rate is the same (up to a logarithmic factor) as the minimax rate over a class of group sparse vectors derived in \cite{lounici2011}. Then, we obtain the posterior contraction rate for the parameter of interest and, under stronger assumptions, we establish consistent posterior dimension recovery of the true model. Our asymptotic results for in-sample predictive error and parameter recovery are similar to those reported in \cite{Ning2020} for a multivariate linear regression model with group sparsity, although their assumptions and prior structure differ from ours. In-sample posterior contraction and dimension recovery for a continuous variant of the spike-and-slab prior, which is a mixture of two Laplace densities, have been also recently considered in \cite{Rockova2018} and \cite{Baiinpress}.
Concerning the posterior predictive distribution, we provide two results. The first one holds asymptotically and shows consistency of the predictive distribution. The second one establishes optimality of the predictive distribution, in the sense that if one has a prior then our posterior predictive distribution dominates any other estimator of the distribution of a new observation.

Small-sample estimation, selection, and predictive performance is assessed numerically through Monte Carlo simulations. Results show that the proposed Bayesian MIDAS penalized models present very good in-sample properties. In particular, variable selection is achieved with high probability in a very sparse setting, quite irrespective of the size of the design matrix (up to 50 high-frequency predictors in the Monte Carlo experiments) and the choice of the shape of the weighting scheme in the DGP. However, both estimation and selection performance generally deteriorate with very high cross-correlation among the original high-frequency predictors, as the Group Lasso is not designed to handle strong collinearity between regressors. Simulations also point to good out-of-sample performance, especially in comparison with alternative penalized regressions, such as Lasso, Elastic-Net, SCAD, and MC$+$. Finally, we illustrate our approach in an empirical forecasting application to U.S. GDP growth with 134 real and financial predictors sampled at monthly, weekly, and daily frequencies. We show that our models can provide superior point and density forecasts at short-term horizons (nowcasting and 1-quarter-ahead) compared to simple as well as sophisticated competing models, such as folded-concave penalties, Bayesian Model Averaging, optimally combined univariate Bayesian MIDAS models, and Factor MIDAS models.

The paper is structured as follows. Sections \ref{sec:MIDAS} and \ref{sec:BMIDAS} introduce the MIDAS penalized regressions and the Bayesian MIDAS framework. Section \ref{sec:Asymptotic_Analysis} presents the theoretical analysis. In Section \ref{sec:Gibbs} we describe the Gibbs sampling and we discuss the Empirical Bayes approach used to automatically tune the penalty hyper-parameters. Section \ref{sec:MCsim} presents simulation experiments and Section \ref{sec:emp_app} provides an empirical application to the U.S. GDP growth. Finally, Section \ref{sec:conclusions} concludes. The appendix contains technical proofs on the theorems reported in Section \ref{sec:Asymptotic_Analysis}, while additional results and proofs of technical lemmas are collected in the Supplementary Appendix available on-line.

\section{MIDAS penalized regressions}\label{sec:MIDAS}

\subsection{Basic MIDAS setup}\label{sec:basic_MIDAS}
Consider the variable $y_{t}$, which is observed at discrete times (\textit{i.e.} only once between $t-1$ and $t$), and suppose that we want to use information stemming from a set of $K$ predictors $\mathbf{x}_{t}^{(m)}:=(x_{1,t}^{(m)},\dots,x_{K,t}^{(m)})^{\prime}$, which are observed $m$ times between $t-1$ and $t$, for forecasting purposes. The variables $y_{t}$ and $x_{k,t}^{(m)}$, for $k=1,\dots,K$, are said to be sampled at different frequencies. For instance, quarterly and monthly frequencies, respectively, in which case $m=3$. Let us define the high-frequency lag operator $L^{\slfrac{1}{m}}$, such that $L^{\slfrac{1}{m}}x_{k,t}^{(m)}=x_{k,t-\slfrac{1}{m}}^{(m)}$. Further, let $h=0,1/m,2/m,3/m,\dots$ be an (arbitrary) forecast horizon, where $h=0$ denotes a nowcast with high-frequency information fully matching the low-frequency sample.

The traditional, and most simple, way of dealing with mixed-frequency data is to aggregate the high-frequency predictors by averaging (\textit{i.e.} assigning equal weights), and estimate the regression through least-squares. However, this approach (flat-weighting scheme) may result in an omitted variable bias if the true weighting scheme is not flat. For a set of processes governing the high-frequency predictors, \citet{Andreou2010} show analytically and numerically that the flat least squares estimator is asymptotically inefficient (in terms of asymptotic bias and/or asymptotic variance) compared to an alternative estimator (linear or non-linear), provided for instance within the MIDAS framework, that can account for a curvature in the weighting scheme. More specifically, the MIDAS approach plugs-in the high-frequency lagged structure of predictors $x_{k,t-h}^{(m)}$ in a regression model for the low-frequency response variable $y_{t}$ as follows:
\begin{equation}\label{eq:MIDAS}
y_{t}=\alpha+\sum_{k=1}^{K}\sum_{c=0}^{C-1}B\left(c;\boldsymbol\theta_{k}\right)L^{\slfrac{c}{m}}x_{k,t-h}^{(m)}+\epsilon_{t},
\end{equation}
where $\epsilon_{t}$ is i.i.d. with mean zero and variance $\sigma^{2}<\infty$, and $B\left(c;\boldsymbol\theta_{k}\right)$ is a weighting function which depends on a vector of parameters $\boldsymbol\theta_{k}$ and a lag order $c=0,\dots,C-1$. In this study, we consider a simple polynomial approximation of the underlying true weighting structure provided by the Almon lag polynomial $B\left(c;\boldsymbol\theta_{k}\right)=\sum_{i=0}^{p_{k}}\theta_{k,i}c^{i}$, where $\boldsymbol\theta_{k}:=(\theta_{k,0},\theta_{k,1},\dots,\theta_{k,p_{k}})^{\prime}$. Under the so-called ``direct method'' \citep{Cooper1972}, Equation \eqref{eq:MIDAS} with Almon lag polynomials can be reparameterized as:
\begin{equation}\label{eq:MIDAS_Almon2b}
y_{t}=\alpha+\boldsymbol{\theta}^{\prime}\mathbf{Z}_{t-h}^{(m)}+\epsilon_{t},
\end{equation}
where $\boldsymbol\theta:=\left(\boldsymbol\theta_{1}^{\prime},\dots,\boldsymbol\theta_{K}^{\prime}\right)^{\prime}$ is a vector featuring $\sum_{k=1}^{K}(p_{k}+1)$ parameters and $\mathbf{Z}_{t}^{(m)}:=\left(\mathbf{z}_{1,t}^{(m)\prime},\dots,\mathbf{z}_{K,t}^{(m)\prime}\right)^{\prime}$ a $\left(\sum_{k=1}^{K}(p_{k}+1)\times 1\right)$ vector of linear combinations of the observed high-frequency regressors, where each sub-vector is defined as $\mathbf{z}_{k,t}^{(m)}:=\mathbf{Q}_{k}\mathbf{x}_{k,t}^{(m)}$, with $\mathbf{x}_{k,t}^{(m)}:=\left(x_{k,t}^{(m)},x_{k,t-\slfrac{1}{m}}^{(m)},\dots,x_{k,t-\slfrac{(C-1)}{m}}^{(m)}\right)^{\prime}$ a $(C\times 1)$ vector of high-frequency lags, and
$\mathbf{Q}_{k}$ is a $(p_{k}+1\times C)$ polynomial weighting matrix, with $(i+1)$-th row $[0^{i},1^{i},2^{i},\dots,(C-1)^i]$ for $i=0,\dots,p_k$. The $h$-step-ahead direct forecast $\widehat{y}_{T\vert T-h}^{}$, conditional on sample information available up to $T-h$, can be hence obtained using \eqref{eq:MIDAS_Almon2b}:
\begin{equation}\label{eq:MIDAS_Almon2b_fcst}
\widehat{y}_{T\vert T-h}^{}=\widehat{\alpha}+\widehat{\boldsymbol{\theta}}^{\prime}\mathbf{Z}_{T-h}^{(m)},
\end{equation}
for some point estimators $\widehat\alpha$ and $\widehat{\boldsymbol\theta}$ of $\alpha$ and $\boldsymbol\theta$, respectively. The main advantage of the Almon lag polynomial is that \eqref{eq:MIDAS_Almon2b} is linear in $\boldsymbol{\theta}$ and parsimonious, as it depends only on $\sum_{k=1}^{K}(p_{k}+1)$ parameters, and can be estimated consistently and efficiently via standard methods. However, two additional features make the Almon lag polynomial particularly attractive in the present framework. First, linear restrictions on the value and slope of the lag polynomial $B\left(c;\boldsymbol\theta_{k}\right)$ may be placed for any $c\in(0,C-1)$. Endpoint restrictions, such as $B\left(C-1;\boldsymbol\theta_{k}\right) = 0$ and $\nabla_{c}B\left(c;\boldsymbol\theta_{k}\right)\vert_{c = C-1}=0$, may be desirable and economically meaningful, as they jointly constrain the weighting structure to tail off slowly to zero. This can be obtained by modifying the $\mathbf{Q}_{k}$ matrix consistently with the form and the number of restrictions considered \citep{Smith1976}. As a result, the number of parameters in \eqref{eq:MIDAS_Almon2b} reduces from $\sum_{k=1}^{K}(p_{k}+1)$ to $\sum_{k=1}^{K}(p_{k}-r_{k}+1)$, where $r_{k}\le p_{k}$ is the number of restrictions. Second, a slope coefficient that captures the overall impact of lagged values of $x_{k,t-h}^{(m)}$ on $y_{t}$ can be easily computed as $\widehat{\beta}_{k}=\widehat{\boldsymbol{\theta}}_{k}^{\prime}\mathbf{Q}_{k}\boldsymbol\iota_{C}^{}$, where $\boldsymbol\iota_{C}^{}$ is a $(C\times 1)$ vector of ones.

\subsection{MIDAS penalized regressions}\label{sec:MIDAS-pen}
Although appealing, the MIDAS regression presented above may be easily affected by over-parameterization and multicollinarity in presence of a large number of potentially correlated predictors.\footnote{The direct method used in regression \eqref{eq:MIDAS_Almon2b} may be also hampered by multicollinearity in the artificial variables $\mathbf{Z}_{t}^{(m)}$ \citep{Cooper1972}. However, if $p$ is small, the imprecision arising from multicollinearity may be compensated by the lower number of coefficients to be estimated.}
To achieve variable selection and parameter estimation simultaneously, penalized least squares procedures such as the Lasso \citep{Tibshirani1996} have been recently investigated in a high-dimensional mixed-frequency framework by \citet{Uematsu2019} and \citet{Lima2020}. Cognizant of the limits of the Lasso (estimation bias and lack of selection consistency), the Adaptive Lasso (\citealp{Zou2006}; AL hereafter), which enjoys selection consistency under typically weaker assumptions than the so-called \textit{irrepresentable condition} on the design matrix, represents a tempting alternative to the Lasso that has not yet been explored in this literature.\footnote{The irrepresentable condition states that the predictors not in the model are not representable by predictors in the true model (\textit{i.e.} the irrelevant predictors are roughly orthogonal to the relevant ones). This represents a necessary and sufficient condition for the Lasso for exact recovery of the non-zero coefficients, but it can be easily violated in cases where the design matrix exhibits too strong (empirical) correlations (collinearity between predictors).}

However, in this paper we argue that the AL might still not be suited in a mixed-frequency framework. The rationale is that since the distributed lags in the MIDAS regression \eqref{eq:MIDAS_Almon2b} are by construction highly correlated, the AL estimator would tend to randomly select only one term of each lag polynomial in the active set and shrink the remaining (relevant) coefficients to zero. The theoretical rationale for a failure in the selection ability of the AL in our mixed-frequency setting is hence similar to the one pointed out by \citet{Efron2004} and \citet{Zou2005}, and it is mostly related to the lack of strict convexity in the Lasso penalty. To address this issue, we propose a solution based on the Adaptive Group Lasso estimator (\citealp{Yuan2006, Wang2008}; AGL hereafter). This approach introduces a penalty to a group of regressors, rather than a single regressor, that may lead (if the group structure is carefully set by the researcher) to a finite sample improvement of the AL. In the present framework, it seems reasonable to define a group as each of the $k$ vectors of lag polynomials in the model. This grouping structure is motivated by the fact that if one high-frequency predictor is irrelevant, it is also expected that zero-coefficients occur in all the parameters of its lag polynomial. Hence, this strategy should overcome, at least in part, the limitations of both Lasso and AL in presence of strong correlation in the design matrix arising from the correlation among lags of the transformed high-frequency predictors $\mathbf{Z}_{t}^{(m)}$.

Let us now assume that $y_{t}$ is centered at $0$ and regressors $\mathbf{Z}_{t}^{(m)}$ in \eqref{eq:MIDAS_Almon2b} are standardized. Let us also consider, for ease of exposition, that the order of the Almon lag polynomial and the number of linear restrictions are the same across variables, such that $p_{k}=p$ and $r_{k}=r$, for $k=1,\dots,K$, and the total number of parameters is $K(p-r+1)$. Further, partition the parameter vector $\boldsymbol\theta$ into $G$ disjoint groups, $\boldsymbol\theta_{j}$, for $j=1,\dots,G$, each of size $g_{j}$ and including the lag polynomial of a given predictor. Despite the change in notation (necessary to avoid confusion), it is straightforward that $G=K$, $\boldsymbol\theta_{j}=\boldsymbol\theta_{k}$, $g_{j}=p-r+1$, and $\widetilde{g}:=\sum_{j=1}^{G}g_{j}=K(p-r+1)$. The objective function of the AGL takes the form:
\begin{equation}\label{eq:AGL}
\mathcal{Q}^{}_{\textrm{AGL}}(\boldsymbol\theta):=T^{-1}\mathcal{L}^{}_{T}(\boldsymbol\theta)+\sum_{j=1}^{G}\lambda_{j}\Vert\boldsymbol\theta_{j}\Vert_{2}^{},
\end{equation}
where $\mathcal{L}^{}_{T}(\boldsymbol\theta)$ denotes the negative log-likelihood function, $\Vert\boldsymbol\theta_{j}\Vert_{2}^{}=(\boldsymbol\theta_{j}^{\prime}\boldsymbol\theta_{j})^{1/2}$ the $\ell_{2}$-norm, and $\lambda_{j}$ the group penalty parameter. Estimation and selection consistency of the AGL estimator are established by \citet{Wang2008}. However, as suggested by \citet{Callot2014}, the AGL possesses a variant of the oracle property only if the predictors are partitioned in the correct groups. This happens because selection consistency concerns the inactive groups (\textit{i.e.} including only parameters whose true value is zero), while the asymptotic distribution of those parameters whose true value is zero but are located in active groups is equivalent to the one of least squares including all variables. Hence, the AGL only performs better than least squares including all variables if one is able to identify the true inactive groups. In the present framework, we expect that partitioning each lag polynomial into one group should attenuate this issue.

\section{Bayesian MIDAS penalized regressions}\label{sec:BMIDAS}

Since the work of \citet{Yuan2006}, several algorithms based on convex optimization methods (\textit{e.g.} block-wise and proximal gradient descent) have been proposed to compute the solution paths of the Adaptive Group Lasso \citep{Wang2008}. In this paper, we consider instead a Bayesian hierarchical approach based on MCMC methods \citep{Kyung2010}, which has several advantages compared to the frequentist approach. First, Bayesian methods exploit model inference via posterior distributions of parameters, which usually provide a valid measure of standard errors based on a geometrically ergodic Markov chain \citep{Khare2013}.\footnote{It is nevertheless worth noting that the results in \citet{Khare2013} hold as long as the penalty hyper-parameters are assumed fixed, while convergence properties of the MCMC algorithm for the full Bayesian penalized regression models are still unknown (see also \citealp{Roy2017}).}
Second, they provide a flexible way of estimating the penalty parameters, along with other parameters in the model. Lastly, they provide forecasts via predictive distributions. In what follows, we present in detail the hierarchical structure of the proposed Bayesian MIDAS penalized models.

\subsection{Bayesian MIDAS Adaptive Group Lasso}\label{sec:BMIDAS-AGL}
As noted by \citet{Tibshirani1996}, the Lasso estimator can be interpreted as the Bayes posterior mode using normal likelihood and independent Laplace (double-exponential) prior for the regression coefficients. \citet{Park2008} propose a Bayesian Lasso where the $\ell_{1}$ penalty corresponds to a conditional Laplace prior that can be represented as a scale mixture of Normals with an exponential mixing density. For the Bayesian Adaptive Group Lasso, we consider a multivariate generalization of the double exponential prior as in \citet{Kyung2010} and \citet{Leng2014}, such that the conditional prior of $\boldsymbol\theta$ given $\sigma^2$ can be expressed as a scale mixture of Normals with Gamma hyper-priors. The conditional prior of $\boldsymbol\theta_{j}$, with $j=1,\dots,G$, is then (see Appendix \ref{sec:marg_prior_appendix}):
\begin{align}\label{eq:prior_thetaGL2}
\Pi(\boldsymbol\theta_{j}|\sigma^{2}) & \propto \int_{0}^{\infty}\left(\frac{1}{2\pi\sigma^{2}\tau^{2}_{j}}\right)^{\frac{g_{j}+1}{2}}\exp\left(-\frac{\Vert\boldsymbol\theta_{j}\Vert_{2}^{2}}{2\sigma^{2}\tau^{2}_{j}}\right)\left(\frac{\lambda_j^2}{2}\right)^{\frac{g_j+1}{2}}\frac{\tau_j^{g_j -1}}{\Gamma\left(\frac{g_j+1}{2}\right)}\exp\left(-\frac{\lambda_j^2}{2}\tau_j^2\right) d\tau_j^2 \nonumber \\[1ex]
& \propto \left(\frac{\lambda_{j}^2}{\sigma^2}\right)^{\frac{g_{j}}{2}}\exp\left(-\frac{\lambda_{j}}{\sigma}\Vert\boldsymbol\theta_{j}\Vert_{2}^{}\right).
\end{align}
\noindent This suggests the following hierarchical representation of the Bayesian MIDAS Adaptive Group Lasso model (BMIDAS-AGL):
\begin{align*}
y \vert \mathbf{Z},\boldsymbol\theta,\sigma^2 &\sim \mathcal{N}\left(\mathbf{Z}\boldsymbol{\theta},\sigma^2\mathbf{I}_{T}\right) \\[1ex]
\boldsymbol\theta_{j} \vert \tau_{j}^{2},\sigma^2 &\sim \mathcal{N}(\mathbf{0},\sigma^2\tau_{j}^{2}\mathbf{I}_{g_{j}}) \\[1ex]
\tau_{j}^{2} &\sim \textrm{Gamma}\left(\frac{g_{j}+1}{2},\frac{\lambda^{2}_{j}}{2}\right) \\[1ex]
\sigma^{2} &\sim \textrm{Inv-Gamma}\left(a_{1},b_{1}\right),
\end{align*}
where $y:=(y_{1},\dots,y_{T})^{\prime}$ is centered at $0$ and $\mathbf{Z}:=\left(\mathbf{Z}_{1}^{(m)},\dots,\mathbf{Z}_{T}^{(m)}\right)^{\prime}$ is the $T \times \wtl g$ matrix of standardized regressors, and $\mathbf{I}_{g_{j}}$ is an identity matrix of order $g_{j}$, for $j=1,\dots,G$. By integrating out $\tau_{j}^{2}$ in the hierarchy above, we obtain that the marginal prior for $\boldsymbol\theta_{j}$, given $\sigma^2$, follows a $g_{j}$-dimensional Multi-Laplace distribution with density function $\text{M-Laplace}(\boldsymbol\theta_{j} \vert \mathbf{0},\sigma/\lambda_{j}) \propto (\lambda_{j}/\sigma)^{g_{j}}\exp(-\lambda_{j}\Vert\boldsymbol\theta_{j}\Vert_{2}/\sigma)$ (see Appendix \ref{sec:marg_prior_appendix}). Let $\boldsymbol\tau := (\tau_{1}^{2},\dots,\tau_{G}^{2})$ and $\boldsymbol\lambda := (\lambda_{1}^{2},\dots,\lambda_{G}^{2})$. The full posterior distribution of the unknown parameters conditional on the data and for some penalty hyper-parameters $\boldsymbol\lambda$ is:
\begin{align}\label{eq:full_posterior_AGL}
\Pi(\boldsymbol\theta,\boldsymbol\tau,\sigma^{2} \vert \boldsymbol\lambda,y,\mathbf{Z}) & \propto \left(\sigma^{2}\right)^{-\frac{T+\widetilde{g}-1}{2}-a_{1}-1}\exp\left[-\frac{1}{2\sigma^2}\Vert y-\mathbf{Z}\boldsymbol{\theta}\Vert_{2}^{2}-\frac{b_{1}}{\sigma^{2}}\right] \nonumber \\[1ex]
& \times \prod_{j=1}^{G}\left(\frac{1}{2\pi\sigma^{2}\tau^{2}_{j}}\right)^{\frac{g_{j}}{2}}\exp\left(-\frac{\Vert\boldsymbol\theta_{j}\Vert_{2}^{2}}{2\sigma^{2}\tau^{2}_{j}}\right) \nonumber \\[1ex]
& \times \prod_{j=1}^{G}\left(\lambda_{j}^{2}\right)^\frac{g_{j}+1}{2}\left(\tau_{j}^{2}\right)^{\frac{g_{j}+1}{2}-1}\exp\left(-\frac{\lambda^{2}_{j}}{2}\tau_{j}^{2}\right).
\end{align}

\subsection{Bayesian MIDAS Adaptive Group Lasso with Spike-and-Slab Prior}\label{sec:BMIDAS-AGL-SS}
The Bayesian model outlined above would typically induce sparsity by shrinking the coefficients of the inactive set towards zero, but usually not exactly to zero. To achieve exact sparse solutions, recent literature has increasingly focused on introducing probabilistic sparse recovery by adding a point mass mixture prior to penalized regressions \citep{Zhang2014,Zhao2015,Rockova2018}. In the present study, we follow \citet{Xu2015} and we consider a Bayesian Group Lasso with spike-and-slab prior for group variable selection, which provides two shrinkage effects: a point mass at $\mathbf{0}$ leading to exact zero coefficients (the spike part of the prior) and a Group Lasso prior on the slab part. The combination of these two components together is expected to facilitate variable selection at the group level and to shrink coefficients in the selected groups simultaneously. Similarly to the BMIDAS-AGL, the hierarchical Bayesian MIDAS Adaptive Group Lasso with spike-and-slab prior  (BMIDAS-AGL-SS) is:
\begin{align*}
y \vert \mathbf{Z},\boldsymbol\theta,\sigma^2 &\sim \mathcal{N}\left(\mathbf{Z}\boldsymbol{\theta},\sigma^2\mathbf{I}_{T}\right) \\[1ex]
\boldsymbol\theta_{j} \vert \tau_{j}^{2},\sigma^2,\pi_{0} &\sim (1-\pi_{0})\mathcal{N}(\mathbf{0},\sigma^2\tau_{j}^{2}\mathbf{I}_{g_{j}}) + \pi_{0}\delta_{0}(\boldsymbol\theta_{j}) \\[1ex]
\tau_{j}^{2} &\sim \textrm{Gamma}\left(\frac{g_{j}+1}{2},\frac{\lambda^{2}_{j}}{2}\right) \\[1ex]
\sigma^{2} &\sim \textrm{Inv-Gamma}\left(a_{1},b_{1}\right) \\[1ex]
\pi_{0} &\sim \textrm{Beta}\left(c,d\right),
\end{align*}
where $\delta_{0}(\boldsymbol\theta_{j})$ denotes a point mass at $\mathbf{0}\in \mathbb{R}_{}^{g_{j}}$, for $j=1,\dots,G$. Note that our spike-and-slab prior conditional on $\sigma^2$ is the same as in \citet{Ning2020}, except that \textit{i)} they specify an independent prior for $\btheta$ and the model variance parameter, and they have the same penalty hyper-parameter across the groups, \textit{ii)} their prior for the dimension is not necessarily a Binomial distribution as in our case, and \textit{iii)} we place a conjugate Beta prior on $\pi_{0}$ while they consider $\pi_0$ as fixed. Adding a Beta prior to the hierarchy allows for mixing over the sparsity level. We use typical non-informative priors for the error variance $\sigma^2$ and we set the hyper-parameters $c$ and $d$ according to the theoretical results presented is Section \ref{sec:Asymptotic_Analysis} (see also Section \ref{sec:cond_post} for the exact values used in the present paper). By integrating out $\tau_{j}^{2}$, we now obtain that the marginal prior for $\boldsymbol\theta_{j}$ is a mixture of point mass at $\mathbf{0}\in \mathbb{R}_{}^{g_{j}}$ and a $g_{j}$-dimensional Multi-Laplace distribution (see Appendix \ref{sec:marg_prior_appendix}). The full posterior distribution of the unknown parameters conditional on the data and for some penalty hyper-parameters $\boldsymbol\lambda$ is:
\begin{align}\label{eq:full_posterior_AGLSS}
\Pi(\boldsymbol\theta,\boldsymbol\tau,\sigma^{2},\pi_{0} \vert \boldsymbol\lambda,y,\mathbf{Z}) & \propto \left(\sigma^{2}\right)^{-\frac{T+\widetilde{g}-1}{2}-a_{1}-1}\exp\left[-\frac{1}{2\sigma^2}\Vert y-\mathbf{Z}\boldsymbol{\theta}\Vert_{2}^{2}-\frac{b_{1}}{\sigma^{2}}\right]\pi_{0}^{c-1}(1-\pi_{0})^{d-1} \nonumber \\[1ex]
& \times \prod_{j=1}^{G}\left[(1-\pi_{0})\left(\frac{1}{2\pi\sigma^{2}\tau^{2}_{j}}\right)^{\frac{g_{j}}{2}}\exp\left(-\frac{\Vert\boldsymbol\theta_{j}\Vert_{2}^{2}}{2\sigma^{2}\tau^{2}_{j}}\right)\mathbf{I}_{\boldsymbol\theta_{j}\neq 0}+\pi_{0}\delta_{0}(\boldsymbol\theta_{j})\right] \nonumber \\[1ex]
& \times \prod_{j=1}^{G}\left(\lambda_{j}^{2}\right)^\frac{g_{j}+1}{2}\left(\tau_{j}^{2}\right)^{\frac{g_{j}+1}{2}-1}\exp\left(-\frac{\lambda^{2}_{j}}{2}\tau_{j}^{2}\right).
\end{align}

\section{Theoretical properties}\label{sec:Asymptotic_Analysis}
This section provides the theoretical validation of our Bayesian MIDAS procedure. In Section \ref{ss:asymptotic_analysis} we investigate the asymptotic properties of our procedure by adopting a frequentist point of view. In Section \ref{ss:Optimality} we address the question of optimality of the corresponding posterior predictive density. Further results on dimension recovery of the model and distributional approximation of the posterior of $\btheta$ are provided in the Supplementary Appendix B.2. As we assume sparsity in the data generating process, in the asymptotic analysis we focus on the BMIDAS-AGL-SS model, which can provide exact sparse recovery of the true model. We use the notation $\Pi(\cdot|y,\mathbf{Z})$ (resp. $\Pi(\cdot)$) to denote both the posterior (resp. prior) distribution and its Lebesgue density function.
\subsection{Asymptotic analysis}\label{ss:asymptotic_analysis}
In this section we establish the posterior contraction rate for both the in-sample and the out-of-sample prediction error, as well as the consistency of both the marginal posterior of $\btheta$ and the posterior predictive density of a new observation. We adopt a frequentist point of view, in the sense that we admit the existence of a true value of $(\btheta,\sigma^2)$, denoted by $(\btheta_0,\sigma_0^2)$, that generates the data. We denote by $\EE_0[\cdot]$ the expectation taken with respect to the true data distribution $\mathcal{N}(\mathbf{Z}\boldsymbol{\theta}_0,\sigma_0^2 \mathbf{I}_T)$ conditional on $(\mathbf{Z},\btheta_0,\sigma_0^2)$.\\
\indent Recall that $g_j$ is the size of the $j$-th group and $\wtl g:= \sum_{j =1}^G g_j$ is the total number of slope parameters in \eqref{eq:MIDAS_Almon2b}. Let $g_{\max} := \max_{1 \leq j\leq G} g_j$ be the maximum group size and let $S_0 \subseteq \{1,\ldots, G\}$ denote the set containing the indices of the true nonzero groups with $s_0 := \vert S_0\vert$ its cardinality. For a generic vector
$\mathbf{v}$, denote by $s_{\mathbf{v}}$ the number of nonzero groups of $\mathbf{v}$. Moreover, let $\|\btheta\|_{\infty} := \max_{1\leq j\leq \wtl g}|\theta_{j}|$, $\|\mathbf{Z}\|_{op}$ be the spectral norm and $\|\mathbf{Z}\|_o := \max\{\|\mathbf{Z}_{j}\|_{op}; 1 \leq j \leq G\}$ where $\mathbf{Z}_{j}$ is the $(T \times g_j)$-submatrix of $\mathbf{Z}$ made of all the rows and the columns corresponding to the indices in the $j$-th block.

We introduce the following assumptions, which are quite mild in that they put weak restrictions on the prior hyper-parameters and allow $g_{\max}$ and the largest element of $\btheta_0$ to increase with $T\rightarrow \infty$.

\begin{assum}\label{Ass:1}
  For positive and bounded constants $\underline{\sigma}^2$, $\overline{\sigma}^2$, $c_g$ and $c_{\btheta}$, suppose that: (i) $0< \underline{\sigma}^2 \leq \sigma_0^2 \leq \overline{\sigma}^2<\infty$; (ii) $g_{\max} \leq c_g \log(G)/\log(T)$; (iii) $\|\btheta_{0}\|_{\infty} \leq c_{\btheta} \sqrt{\log(G)}$. 
\end{assum}
\noindent Define the region $\overline\Theta_0 := \{\btheta\in\mathbb{R}^{\wtl g};\, \|\btheta\|_{\infty} \leq c_{\btheta} \sqrt{\log(G)}\}$, where $c_{\btheta}$ is as defined in Assumption \ref{Ass:1}. By Assumption \ref{Ass:1} \textit{(iii)}, $\btheta_0\in\overline\Theta_0$.
\begin{assum}\label{Ass:2}
  Let $\lambda_{\min} := \min\{\lambda_j;\, j\leq G\}$ and $\lambda_{\max} := \max\{\lambda_j;\, j\leq G\}$ and assume that $\|\mathbf{Z}\|_0/\sqrt{T}< \sqrt{\log(T)}$ with probability $1$. The scale parameters $\lambda_j$ are allowed to change with $T$ and belong to the range:
  $$\|\mathbf{Z}\|_0 \underline{c}/\sqrt{T} \leq \lambda_{\min} \leq \lambda_{\max} \leq \overline{C} \min\left\{\sqrt{\log(T)},\sqrt{\log(G)}\|\mathbf{Z}\|_o\right\}$$
  for two positive constants $1 <\underline{c}< \overline{C}<\infty$.
\end{assum}

\begin{assum}\label{Ass:3}
  The hyper-parameters $c,d$ of the Beta prior for $\pi_0$ satisfy: $(d+j-1)/(c + G - j) \leq j/[G^u(G - j + 1)]$ for every $0<u<s_0$ and $\forall j\in\{1,\ldots,G\}\subseteq\mathbb{N}$, $d \leq s_0$ and $\log(G + c) \leq c_{\pi_0} \log(G)$ for some fixed constant $0<c_{\pi_0}<\infty$. Moreover, the hyperparameter $a_1$ of the prior for $\sigma^2$ satisfies $a_1>1$.
\end{assum}

Assumption \ref{Ass:1} \textit{(ii)} restricts the rate at which the maximum group size increases to infinity. Assumption \ref{Ass:1} \textit{(iii)} restricts the growth rate of the maximum component of the true $\btheta_0$. Assumption \ref{Ass:1} \textit{(i)} excludes degenerate cases by restricting the model variance. In the following we establish results that hold uniformly over $(\btheta_0,\sigma_0^2)\in\overline{\Theta}_0\times [\underline{\sigma}^2, \overline{\sigma}^2]$. Assumption \ref{Ass:2} restricts the limiting behaviour of the parameters $\boldsymbol\lambda$ of the M-Laplace distribution. The upper bound prevents to shrink the non-zero groups too much towards zero while the lower bound prevents many false signals. Assumption \ref{Ass:3} is a mild assumption on the prior on $\pi_0$, it allows $c$ of the form $c= \bar{\kappa} G^v$, for constants $\bar{\kappa}>1$ and $v>1$ as in \citet[Example 5]{Castillo2015} but it excludes a $v$ that increases with $G$.

\subsubsection{In-sample asymptotic properties}\label{sec:Asymptotic_Analysis_insample}
\indent We now investigate the in-sample asymptotic properties of our model $y|\mathbf{Z},\mathbf{\btheta},\sigma^2 \sim \mathcal{N}_T(\mathbf{Z}\mathbf{\btheta},\sigma^2 \mathbf{I}_T)$, where the parameter $(\btheta',\sigma^2)'$ is endowed with the BMIDAS-AGL-SS prior \eqref{eq:marginal:prior}. The first theorem establishes posterior consistency of the in-sample prediction error.

\begin{thm}\label{thm:1}
  Suppose Assumptions \ref{Ass:1} - \ref{Ass:3} hold, and let $\epsilon := \max\{\sqrt{s_0 \log(G)/T},\sqrt{\log(T)/T}\} \allowbreak\rightarrow 0$. Then, for a sufficiently large $M >0$:
  \begin{equation}
    \sup_{\btheta_0\in \overline\Theta_0,\sigma_0^2 \in[\underline{\sigma}^2, \overline{\sigma}^2]}\EE_0 \left[\Pi\left(\left.\btheta\in \mathbb{R}^{\wtl g}; \|\mathbf{Z}(\btheta - \btheta_0)\|_2^2 \geq M T\epsilon^2\right|y,\mathbf{Z}\right)\right] \rightarrow 0.
  \end{equation}
\end{thm}

The assumption $\epsilon \rightarrow 0$ in Theorem \ref{thm:1} implies that $\log(G) = o(T)$. The latter becomes $\log(dimension(\btheta)) = o(T)$ when the number of groups is equal to the number of parameters, which has been shown in the literature to be a necessary condition for sparse recovery (see \textit{e.g.} \citealp{lounici2011}). The contraction rate $\epsilon$ is the same as the one in \cite{Ning2020} under our assumptions. The first term of the rate coincides (up to a logarithmic factor) with the minimax rate over a class of group sparse vectors derived in \cite{lounici2011}. As discussed in that paper, the Group Lasso estimator has some advantages over the Lasso estimator in some important cases. One of these cases is the one we consider where $g_{\max} < \log(G)$ under our Assumption \ref{Ass:1} \textit{(ii)}. In this case the upper bound $\epsilon$ of the posterior contraction rate for the in-sample prediction error is faster than the lower bound on the prediction of the Lasso estimator (see \citealp[Section 7]{lounici2011}). Finally, when the number of groups is equal to the number of parameters, then the first term of the rate corresponds to the optimal one for this problem (see \textit{e.g.} \citealp{buhlmann2011}, and \citealp{Castillo2015}).\\
\indent The proof of Theorem \ref{thm:1} is provided in Appendix \ref{App:proofs} and follows the strategy of \cite{Ning2020} adapted to our slightly different prior and model. Instead of relying on the general posterior contraction theory based on the Hellinger distance, which is not appropriate in this setting, this strategy of proof relies on the average R\'{e}nyi divergence, which can be easily used to obtain the rate in terms of Euclidean distance. To obtain the posterior contraction rate with respect to the R\'{e}nyi divergence, we use the general posterior contraction procedure for independent observations, as in \cite{ghosal2007}. That is, we first show that our prior \eqref{eq:marginal:prior} puts enough mass in a shrinking neighborhood of the true density function. Then, we show that the prior puts a small mass, decreasing to zero, on the complement of a finite dimensional space with increasing dimension that approximates well the infinite dimensional model. Finally, we prove the existence of exponential tests of the true density against the complement of balls around the truth. These results are proved in Lemmas B.1.2 and B.1.5 in the Supplementary Appendix.\\
\indent Given the result in Theorem \ref{thm:1}, we are now ready to establish a result on parameter recovery of our procedure, that is, consistency of the marginal posterior of $\btheta$. As discussed in \cite{bickel2009}, the parameter $\btheta$ is not estimable without a condition on $\mathbf{Z}$ because of its large dimension. In particular, it is well known from the literature (see \textit{e.g.} \citealp{Castillo2015}, and \citealp{Ning2020}) that if $\btheta$ is sparse, then local invertibility of the Gram matrix $\mathbf{Z}'\mathbf{Z}$ is sufficient for estimability of $\btheta$. We then introduce the following quantity.
\begin{defi}
  For every $s>0$, the smallest scaled sparse singular value of dimension $s$ is defined as
  \begin{equation}\label{def:1}
    \wtl\phi(s) : = \inf\left\{\frac{\|\mathbf{Z}\btheta\|_2^2}{\|\mathbf{Z}\|_o^2 \|\btheta\|_2^2},\, 0\leq s_{\btheta}\leq s\right\}.
  \end{equation}
\end{defi}

The restricted eigenvalue condition requires that $\wtl\phi(s) > 0$. This means that the smallest eigenvalue for the sub-matrix of $\mathbf{Z}$ made of columns corresponding to the non-zero groups is strictly larger than zero. We call it ``scaled'' eigenvalue because we divide by the maximum operator norm of submatrices of $\mathbf{Z}$. We point out that since $\wtl g$ is large, possibly larger than $T$, then it would be unrealistic to require that the minimum eigenvalue of $\mathbf{Z}'\mathbf{Z}$ is strictly positive. Our restricted sparse eigenvalue condition is instead much weaker than this condition, and it is hence more realistic.\\
\indent Let $\wtl  s_0 := \max\{s_0, \log(T)/\log(G)\}$. Lemma B.1.4 in the Supplementary Appendix shows that $\EE_0[\Pi(\btheta; s_{\btheta} \geq M_2 \wtl  s_0|y,\mathbf{Z})] \rightarrow 0$ uniformly over $(\btheta_0,\sigma_0^2)\in\overline{\Theta}_0\times [\underline{\sigma}^2, \overline{\sigma}^2]$. This means that with probability approaching $1$, the posterior puts mass one on the $\btheta$'s with dimension $\wtl s_0$. This implies that when $\log(T)/\log(G) > s_0$, the posterior can go beyond the true dimension $s_0$. In spite of this feature, the next theorem shows that the posterior is still able to recover the true parameter $\btheta_0$, similarly to \citet[Corollary 3.3]{Ning2020}.
\begin{thm}\label{thm2:contraction:rate:theta}
Suppose Assumptions {\ref{Ass:1}} - {\ref{Ass:3}} hold, and let $\epsilon := \max\{\sqrt{s_0 \log(G)/T},\sqrt{\log(T)/T}\} \allowbreak \rightarrow 0$. Then, for a sufficiently large $M >0$ and for $\wtl\phi(s)$ as defined in \eqref{def:1} we have:
  \begin{equation}
     \sup_{\btheta_0\in \overline\Theta_0,\sigma_0^2 \in[\underline{\sigma}^2, \overline{\sigma}^2]}\EE_0 \left[\Pi\left(\left.\btheta\in \mathbb{R}^{\wtl g}; \|\btheta - \btheta_0\|_2^2 \geq \frac{M T\epsilon^2}{\wtl\phi(s_0 + M_2\wtl s_0)\|\mathbf{Z}\|_o^2}\right|y,\mathbf{Z}\right)\right] \rightarrow 0.
  \end{equation}
\end{thm}
The proof of Theorem \ref{thm2:contraction:rate:theta} is provided in Appendix \ref{App:proofs}. The first term of the rate coincides (up to a logarithmic factor) with the minimax rate over a class of group sparse vectors derived in \cite{lounici2011}. The posterior contraction rate provided in Theorem \ref{thm2:contraction:rate:theta} deteriorates when $\wtl\phi(s_0 + M_2\wtl s_0)$ is small.

\subsubsection{Out-of-sample asymptotic properties}
\indent We now analyse the behaviour of the out-of-sample prediction error, as well as the behaviour of the predictive density associated with our BMIDAS-AGL-SS model. For this, consider the posterior predictive density
$$f_{\mathbf{Z}_{\tau-h},y,\mathbf{Z}}(y_{\tau}):=f(y_{\tau}|\mathbf{Z}_{\tau-h},y,\mathbf{Z}) := \int f_0(y_{\tau}|\btheta,\sigma^2,\mathbf{Z}_{\tau-h})\Pi(\btheta,\sigma^2|y,\mathbf{Z})d\btheta d\sigma^2,\qquad \forall \tau > T,$$
where $\mathbf{Z}_{\tau-h}$ is a $(\wtl{g}\times 1)$ vector, $f_0(\cdot|\mathbf{Z}_{\tau-h},\btheta,\sigma^2)$ is the Lebesgue density of the one-dimensional $\mathcal{N}(\mathbf{Z}_{\tau - h}'\btheta,\sigma^2)$ distribution, and denote by $\|\cdot\|_{TV}$ the Total Variation distance. In addition, we denote by $f_{\mathbf{Z}_{\tau-h},\btheta_0,\sigma_0^2}(\cdot):=f_0(\cdot|\mathbf{Z}_{\tau-h},\btheta_0,\sigma_0^2)$ the Lebesgue density of the one-dimensional $\mathcal{N}(\mathbf{Z}_{\tau - h}'\btheta_0,\sigma_0^2)$ distribution. As before, the expectation $\EE_0[\cdot]$ denotes the expectation taken with respect to the true data distribution of $y|\mathbf{Z},\btheta_0,\sigma_0^2 \sim \mathcal{N}(\mathbf{Z}\boldsymbol{\theta}_0,\sigma_0^2 \mathbf{I}_T)$ conditional on $(\mathbf{Z},\btheta_0,\sigma_0^2)$. In addition, $\EE_{\mathbf{Z}_{\tau-h}}$ has to be understood as the expectation taken with respect to the distribution of $\mathbf{Z}_{\tau - h}$. Finally, we denote by $\mathbf{Z}_{\tau - h,j}$ the $(g_j \times 1)$-subvector of $\mathbf{Z}_{\tau - h}$ made of all the elements corresponding to the indices in the $j$-th block of $\mathbf{Z}_{\tau - h}$.\\
\indent As before, we assume that $(\btheta',\sigma^2)'$ is endowed with the BMIDAS-AGL-SS prior \eqref{eq:marginal:prior}. We start by establishing the contraction rate of the posterior of the out-of-sample prediction error.
\begin{thm}\label{thm4:out-of-sample1}
  Suppose the assumptions of Theorem \ref{thm2:contraction:rate:theta} be satisfied. For every $\tau > T$ and every fixed $\mathbf{Z}_{\tau - h}$ define: $$\eta_0 := \sup_{\{\btheta; 0 \leq s_{\btheta} \leq M_2 \wtl s_0 + s_0\}}\frac{|\mathbf{Z}_{\tau - h}'\btheta|^2}{\sum_{j=1}^G \|\mathbf{Z}_j' \btheta_j/\sqrt{T}\|_2^2}$$
  \noindent for a constant $M_2 > \frac{2 + C_1}{u} + 1$ that does not depend on $(\btheta_0,\sigma_0^2)$ and $C_1>0$ the constant in Lemma B.1.1. Moreover, let $\epsilon := \max\{\sqrt{s_0 \log(G)/T},\sqrt{\log(T)/T}\} \rightarrow 0$. Then, $\forall \tau > T$:
  \begin{equation}\label{th3:eq1}
    \sup_{\btheta_0\in \overline\Theta_0,\sigma_0^2 \in[\underline{\sigma}^2, \overline{\sigma}^2]}\EE_0 \left[\Pi\left(\left.\btheta\in \mathbb{R}^{\wtl g}; |\mathbf{Z}_{\tau - h}'(\btheta - \btheta_0)|^2 \geq \eta_0\frac{M\epsilon^2}{\wtl\phi(s_0 + M_2\wtl s_0)}\right|y,\mathbf{Z}\right)\right]\rightarrow 0.
  \end{equation}
\end{thm}

The proof of Theorem \ref{thm4:out-of-sample1} is provided in Appendix \ref{App:proofs}. Result \eqref{th3:eq1} in the previous theorem provides a contraction rate for the out-of-sample prediction error for a given $\mathbf{Z}_{\tau - h}$. This rate might be slower than the rate for the in-sample prediction error if $\eta_0/\wtl\phi(s_0 + M_2\wtl s_0)$ is not bounded in probability. To get this rate we have exploited the sparsity structure and the property of the posterior distribution to concentrate on vectors $\btheta$ with dimension $\wtl s_0$. On the other hand, if we assumed that $\EE_{\mathbf{Z}_{\tau - h}}[\mathbf{Z}_{\tau - h}^4] < \kappa$ for some constant $\kappa$, as in \cite{CarrascoRossi2016}, which in turn implies $\|\mathbf{Z}_{\tau - h}\|_2 = O_p(1)$, then we would get the same rate as in Theorem \ref{thm2:contraction:rate:theta}. Here we do not impose this restriction. It is worth noting that the numerator of $\eta_0$ can be upper bounded by $\max_{1\leq j\leq G}|\mathbf{Z}_{\tau-h,j}'\btheta_j|^2(M_2\wtl s_0 + s_0)^2$, which diverges. However, $\eta_0$ does not necessarily diverge if its denominator diverges at least at the same rate.\\
\indent We now state the consistency of the posterior predictive density function.
\begin{thm}\label{thm3:out-of-sample1}
  Suppose the assumptions of Theorem \ref{thm2:contraction:rate:theta} be satisfied. Let $\epsilon := \max\{\sqrt{s_0 \log(G)/T}, \allowbreak \sqrt{\log(T)/T}\} \rightarrow 0$. Then, $\forall \tau > T$ such that  $\EE_{\mathbf{Z}_{\tau - h}}[\mathbf{Z}_{\tau - h,j}\mathbf{Z}_{\tau - h,k}'] = 0$ for every $j \neq k$ and $\epsilon^2\frac{\left\|\EE_{\mathbf{Z}_{\tau - h}}[\mathbf{Z}_{\tau - h}\mathbf{Z}_{\tau - h}']\right\|_{o}}{\|\mathbf{Z}/\sqrt{T}\|_o^2\wtl \phi(s_0 + M_2 \wtl s_0)} = o_p(1)$ the following holds:
  \begin{equation}\label{th3:eq2}
    \sup_{\btheta_0\in \overline\Theta_0,\sigma_0^2 \in[\underline{\sigma}^2, \overline{\sigma}^2]}\EE_{\mathbf{Z}_{\tau-h}}\left[\EE_0\left[\|f_{\mathbf{Z}_{\tau-h},y,\mathbf{Z}} - f_{\mathbf{Z}_{\tau-h},\btheta_0,\sigma_0^2}\|_{TV}\right]\right] \rightarrow 0.
  \end{equation}
\end{thm}

The theorem establishes consistency of the posterior predictive distribution under the condition that the out-of-sample predictors are uncorrelated across groups. This assumption seems reasonable given the group structure. However, if this condition is not verified then the consistency result remains true either under the condition that $\epsilon^2\frac{\left\|\EE_{\mathbf{Z}_{\tau - h}}[\mathbf{Z}_{\tau - h}\mathbf{Z}_{\tau - h}']\right\|_{op}}{\|\mathbf{Z}/\sqrt{T}\|_o^2\wtl \phi(s_0 + M_2 \wtl s_0)} = o_p(1)$, which is a little bit stronger, or under the condition $\epsilon^2 \sup_{\{\btheta; 0 \leq s_{\btheta} \leq M_2 \wtl s_0 + s_0\}}\frac{\EE_{\mathbf{Z}_{\tau-h}}|\mathbf{Z}_{\tau - h}'\btheta|^2}{\sum_{j=1}^G \|\mathbf{Z}_j' \btheta_j/\sqrt{T}\|_2^2 \wtl \phi(s_0 + M_2 \wtl s_0)} = o_p(1)$, which involves a quantity similar to the rate in Theorem \ref{thm4:out-of-sample1}.\\

\subsection{Optimality of the predictive density}\label{ss:Optimality}
\indent While Theorem \ref{thm3:out-of-sample1} states good asymptotic properties of the posterior predictive density function, it is also possible to argue that the posterior predictive density function has optimal properties in finite sample. Indeed, if one has two contenders for estimating $f_{\mathbf{Z}_{\tau-h},\btheta_0,\sigma_0^2}(y_{\tau})$, say our predictive density $f_{\mathbf{Z}_{\tau-h},y,\mathbf{Z}}(y_{\tau})$ and another density $\wtl f_{\mathbf{Z}_{\tau-h},y,\mathbf{Z}}(y_{\tau})$, hence $f_{\mathbf{Z}_{\tau-h},y,\mathbf{Z}}$ is better than $\wtl f_{\mathbf{Z}_{\tau-h},y,\mathbf{Z}}$ if the following expected difference of the Kullback-Leibler divergences with respect to the true $f_{\mathbf{Z}_{\tau-h},\btheta_0,\sigma_0^2}$ is positive:
\begin{multline}\label{eq_out_of_sample_optimality}
  \int \Pi(\btheta_0,\sigma_0^2) d\btheta_0 d\sigma_0^2\EE_0\left[K\left(\left.f_{\mathbf{Z}_{\tau-h},\btheta_0,\sigma_0^2}\right\Vert \wtl f_{\mathbf{Z}_{\tau-h},y,\mathbf{Z}}(y_{\tau})\right) - K\left(\left.f_{\mathbf{Z}_{\tau-h},\btheta_0,\sigma_0^2}\right\Vert f_{\mathbf{Z}_{\tau-h},y,\mathbf{Z}}(y_{\tau})\right)\right]\\
  = \int \Pi(\btheta_0,\sigma_0^2) d\btheta_0 d\sigma_0^2\int f_{\mathbf{Z},\btheta_0,\sigma_0^2}(y) dy\int f_{\mathbf{Z}_{\tau-h},\btheta_0,\sigma_0^2}(y_{\tau})\log\frac{f_{\mathbf{Z}_{\tau-h},y,\mathbf{Z}}(y_{\tau})}{\wtl f_{\mathbf{Z}_{\tau-h},y,\mathbf{Z}}(y_{\tau})} d y_{\tau}\\
  = \int f_{\mathbf{Z}}(y) dy\int f_{\mathbf{Z}_{\tau-h},y,\mathbf{Z}}(y_{\tau})\log\frac{f_{\mathbf{Z}_{\tau-h},y,\mathbf{Z}}(y_{\tau})}{\wtl f_{\mathbf{Z}_{\tau-h},y,\mathbf{Z}}(y_{\tau})} d y_{\tau},
\end{multline}
\noindent where $K(f_1\Vert f_2):= \int\log(f_1/f_2)f_1$ is the Kullback-Leibler divergence between two probability measures with Lebesgue densities $f_1$ and $f_2$, $f_{\mathbf{Z},\btheta_0,\sigma_0^2}(y)$ is the Lebesgue density of the $\mathcal{N}(\mathbf{Z}\boldsymbol{\theta}_0,\sigma_0^2 \mathbf{I}_T)$ distribution, and $f_{\mathbf{Z}}(y) = \int f_{\mathbf{Z},\btheta_0,\sigma_0^2}(y) \Pi(\btheta_0,\sigma_0^2|y,\mathbf{Z})d\btheta_{0}d\sigma_0^{2}$. Since the right hand side of \eqref{eq_out_of_sample_optimality} is equal to $\int f_{\mathbf{Z}}(y) dy \,K(f_{\mathbf{Z}_{\tau-h},y,\mathbf{Z}} \Vert \wtl f_{\mathbf{Z}_{\tau-h},y,\mathbf{Z}})$, then it is strictly positive. This shows that if one has a prior, and exploits it, then our posterior predictive density dominates every other density estimator. In particular, it dominates any density estimator that replaces $(\btheta',\sigma^2)'$ by a consistent estimator. Of course, this comparison does not hold when no prior information is available. However, for some distributions, it is known that the Bayesian predictive density outperforms other density estimators even without a prior (see \textit{e.g.} \citealp{Aitchison1975}).

\section{Gibbs sampler}\label{sec:Gibbs}

\subsection{Conditional posteriors}\label{sec:cond_post}
We now provide details on the Gibbs sampler we use to simulate from the posterior distributions \eqref{eq:full_posterior_AGL} and \eqref{eq:full_posterior_AGLSS}. For this purpose, we consider an efficient block Gibbs sampler \citep{Hobert1998}. Let us denote $\boldsymbol\theta_{\mysetminus j}:=(\boldsymbol\theta_{1}^{\prime},\dots,\boldsymbol\theta_{j-1}^{\prime},\boldsymbol\theta_{j+1}^{\prime},\dots,\boldsymbol\theta_{G}^{\prime})^{\prime}$ the $\boldsymbol\theta$ vector without the $j$th high-frequency lag polynomial, and $\mathbf{Z}_{j}$ and denote $\mathbf{Z}_{\mysetminus j}$ partitions of the design matrix corresponding to $\boldsymbol\theta_{j}$ and $\boldsymbol\theta_{\mysetminus j}$, respectively. With a conjugate Gamma prior placed on the penalty hyper-parameters, $\lambda^{2}_{j}\sim\textrm{Gamma}\left(a_{2},b_{2}\right)$, the full conditional posteriors for the BMIDAS-AGL model are:
\begin{align*}
\boldsymbol\theta_{j} \vert \boldsymbol\theta_{\mysetminus j},\sigma^{2},\boldsymbol\tau,\boldsymbol\lambda,y,\mathbf{Z} &\sim  \mathcal{N}\left(\mathbf{A}_{j}^{-1}\mathbf{C}_{j}^{},\sigma^{2}\mathbf{A}_{j}^{-1}\right) \\[1ex]
\tau_{j}^{-2} \vert \boldsymbol\theta,\sigma^{2},\boldsymbol\lambda,y,\mathbf{Z} &\sim \textrm{Inv-Gaussian}\left(\frac{\lambda_{j}\sigma}{\Vert\boldsymbol\theta_{j}\Vert_{2}^{}},\lambda^{2}_{j}\right) \\[1ex]
\sigma^{2} \vert \boldsymbol\theta,\boldsymbol\tau,\boldsymbol\lambda,y,\mathbf{Z} &\sim \textrm{Inv-Gamma}\left(\frac{T+\widetilde{g}-1}{2}+a_{1},\frac{1}{2}\Vert y-\mathbf{Z}\boldsymbol{\theta}\Vert_{2}^{2} + \frac{1}{2}\sum_{j=1}^{G}\frac{\Vert\boldsymbol\theta_{j}\Vert_{2}^{2}}{\tau_{j}^{2}}+b_{1}\right) \\[1ex]
\lambda^{2}_{j} \vert \boldsymbol\theta,\sigma^{2},\boldsymbol\tau,y,\mathbf{Z} &\sim \textrm{Gamma}\left(\frac{g_{j}+1}{2}+a_{2},\frac{\tau_{j}^{2}}{2}+b_{2}\right)
\end{align*}
where $\mathbf{A}_{j}:=\mathbf{Z}_{j}^{\prime}\mathbf{Z}_{j}^{}+\tau_{j}^{-2}\mathbf{I}_{g_{j}}$ and $\mathbf{C}_{j} := \mathbf{Z}_{j}^{\prime}\left(y-\mathbf{Z}_{\mysetminus j}\boldsymbol\theta_{\mysetminus j}\right)$, for $j=1,\dots,G$.

\bigskip

For the BMIDAS-AGL-SS model, we place again a conjugate gamma prior on the penalty hyper-parameters. Then, for $j=1,\dots,G$, we have the following full conditional posteriors :
\begin{align*}
\boldsymbol\theta_{j} \vert \boldsymbol\theta_{\mysetminus j},\sigma^{2},\boldsymbol\tau,\boldsymbol\lambda,\boldsymbol\pi_{1},\pi_{0},y,\mathbf{Z} &\sim  (1-\pi_{1,j})\mathcal{N}\left(\mathbf{A}_{j}^{-1}\mathbf{C}_{j}^{},\sigma^{2}\mathbf{A}_{j}^{-1}\right)  + \pi_{1,j}\delta_{0}(\boldsymbol\theta_{j}) \\[1ex]
\tau_{j}^{-2} \vert \boldsymbol\theta,\sigma^{2},\boldsymbol\lambda,\boldsymbol\pi_{1},\pi_{0},y,\mathbf{Z} &\sim \left\{
\begin{array}{cc}
\textrm{Inv-Gaussian}\left(\frac{\lambda_{j}\sigma}{\Vert\boldsymbol\theta_{j}\Vert_{2}^{}},\lambda^{2}_{j}\right) & ~~~~\text{if}~~ \boldsymbol\theta_{j}\neq\mathbf{0}\\
\textrm{Gamma}\left(\frac{g_{j}+1}{2}, \frac{\lambda^{2}_{j}}{2}\right)  & ~~~~\text{if}~~ \boldsymbol\theta_{j}=\mathbf{0}
\end{array}
\right.\\[1ex]
\sigma^{2} \vert \boldsymbol\theta,\boldsymbol\tau,\boldsymbol\lambda,\boldsymbol\pi_{1},\pi_{0},y,\mathbf{Z} &\sim \textrm{Inv-Gamma}\left(\frac{T+\widetilde{G}-1}{2}+a_{1},\frac{1}{2}\Vert y-\mathbf{Z}\boldsymbol{\theta}\Vert_{2}^{2} + \frac{1}{2}\sum_{j=1}^{G}\frac{\Vert\boldsymbol\theta_{j}\Vert_{2}^{2}}{\tau_{j}^{2}}+b_{1}\right) \\[1ex]
\lambda^{2}_{j} \vert \boldsymbol\theta,\sigma^{2},\boldsymbol\tau,\boldsymbol\pi_{1},\pi_{0},y,\mathbf{Z} &\sim \textrm{Gamma}\left(\frac{g_{j}+1}{2}+a_{2},\frac{\tau_{j}^{2}}{2}+b_{2}\right) \\[1ex]
\pi_{0} \vert \boldsymbol\theta,\sigma^{2},\boldsymbol\tau,\boldsymbol\lambda,\boldsymbol\pi_{1},y,\mathbf{Z} &\sim \textrm{Beta}\left(\sum_{j=1}^{G}(1-\gamma_{j})+c,\sum_{j=1}^{G}\gamma_{j}+d\right)
\end{align*}
where $\mathbf{A}_{j} := \mathbf{Z}_{j}^{\prime}\mathbf{Z}_{j}+\tau_{j}^{-2}\mathbf{I}_{g_{j}}$, $\mathbf{C}_{j} := \mathbf{Z}_{j}^{\prime}\left(y-\mathbf{Z}_{\mysetminus j}\boldsymbol\theta_{\mysetminus j}\right)$, $\boldsymbol\pi_{1} := (\pi_{1,1},\dots,\pi_{1,G})$, $\widetilde{G} := \sum_{j=1}^{G}g_{j}\gamma_{j}$, $d=1$ and $c=\bar{\kappa}G^v$ with $\bar{\kappa}=v=(1+G^{-1})$ (see Section \ref{ss:asymptotic_analysis}, Assumption \ref{Ass:3}), and
\begin{align*}
\gamma_{j} &=
\left\{
\begin{array}{cc}
1 & ~~~~\text{if}~~ \boldsymbol\theta_{j}\neq\mathbf{0}\\
0 & ~~~~\text{if}~~ \boldsymbol\theta_{j}=\mathbf{0}
\end{array}
\right.\\[1ex]
\pi_{1,j} &= \Pi(\boldsymbol\theta_{j}=\mathbf{0} \vert \boldsymbol\theta_{\mysetminus j},\sigma^{2},\boldsymbol\tau,\boldsymbol\lambda,\pi_{0},y,\mathbf{Z}) = \frac{\pi_{0}}{\pi_{0}+(1-\pi_{0})\left[(\tau_{j}^{2})^{-\frac{g_{j}}{2}}\vert\mathbf{A}_{j}\vert^{-\frac{1}{2}}\exp\left(\frac{1}{2\sigma^2}\mathbf{C}_{j}^{\prime}\mathbf{A}_{j}^{-1}\mathbf{C}_{j}^{}\right)\right]}.
\end{align*}

\subsection{Tuning the penalty hyper-parameters}\label{sec:EB}
The hierarchical models presented above treat the penalty parameters as hyper-parameters, \textit{i.e.} as random variables with Gamma prior distributions $\Pi(\boldsymbol\lambda)$ and Gamma conditional posterior distributions $\Pi(\boldsymbol\lambda\vert \boldsymbol\phi,y,\mathbf{Z})$, given a vector of parameters $\boldsymbol\phi$. However, the main drawback of this approach is that the resulting posterior distributions can be sensitive to the choice of the prior. \citet{Park2008} and \citet{Kyung2010} suggest to address this issue by implementing the Monte Carlo EM algorithm (MCEM) proposed by \citet{Casella2001}, which complements the Gibbs sampler and provides marginal maximum likelihood estimates of the hyper-parameters  when the marginal distribution $f(y\vert\boldsymbol\lambda,\mathbf{Z})=\int f(y\vert\boldsymbol\phi,\boldsymbol\lambda,\mathbf{Z})\Pi(\boldsymbol\phi\vert\boldsymbol\lambda)d\boldsymbol\phi$ is not available in closed form. This can be easily obtained by repeatedly maximizing $N$ times the expectation function
\begin{displaymath}
Q(\boldsymbol\lambda\vert\boldsymbol\lambda^{(n)})=\int \log f(y,\boldsymbol\phi\vert\boldsymbol\lambda,\mathbf{Z})\Pi(\boldsymbol\phi\vert \boldsymbol\lambda^{(n)},y,\mathbf{Z})d\boldsymbol\phi,
\end{displaymath}
and updating $\boldsymbol\lambda$ at each $n$-th iteration. A run of the Gibbs sampler is nevertheless required to simulate from the intractable distribution $\Pi(\boldsymbol\phi\vert \boldsymbol\lambda^{(n)},y,\mathbf{Z})$.

Although very attractive, the MCEM algorithm may be extremely costly from a computational point of view, as a large number of Monte Carlo iterations, each requiring a fully converged Gibbs sampling, is usually needed to maximize the marginal log-likelihood $\log f(y\vert\boldsymbol\lambda,\mathbf{Z})$. Hence, a serious trade-off between computational efficiency ($N$ Monte Carlo iterations) and accuracy of the results ($S$ Gibbs iterations) may arise. In the present framework, careful attention must be paid to this feature, because the computational burden implied by the Group Lasso increases dramatically as the number of predictors increases \citep{Yuan2006}. To deal with this issue, we adopt an alternative Empirical Bayes approach that relies on the so-called \textit{internal} adaptive MCMC algorithms (see \citealp{Atchade2011a}). Within this family of algorithms, the specific class of \textit{controlled} MCMC resorts to stochastic approximation algorithms to solve maximization problems when the likelihood function is intractable, by mimicking standard iterative methods such as the gradient algorithm. This approach is therefore computationally efficient, because it requires only a single Monte Carlo run $(N=1)$. Following \citet{Atchade2011}, using a stochastic approximation to solve the maximization problem and updating $\boldsymbol\lambda$ in each Gibbs iteration $s=1,\dots,S$, the solution to the EM algorithm takes the form:
\begin{displaymath}
\boldsymbol\lambda^{(s+1)}=\boldsymbol\lambda^{(s)}+a^{(s)}\nabla_{\boldsymbol\lambda}Q(\boldsymbol\lambda^{(s)}\vert\boldsymbol\lambda^{(s)})=\boldsymbol\lambda^{(s)}+a^{(s)}\int H(\boldsymbol\lambda^{(s)},\boldsymbol\phi)\Pi(\boldsymbol\phi\vert \boldsymbol\lambda^{(s)},y,\mathbf{Z})d\boldsymbol\phi,
\end{displaymath}
where $H(\boldsymbol\lambda,\boldsymbol\phi):=\nabla_{\boldsymbol\lambda}\log\left[f(y\vert\boldsymbol\phi,\boldsymbol\lambda,\mathbf{Z})\Pi(\boldsymbol\phi\vert\boldsymbol\lambda)\right]=\nabla_{\boldsymbol\lambda}\log\Pi(\boldsymbol\phi\vert\boldsymbol\lambda)$, as the likelihood does not usually depend on the hyper-parameters $\boldsymbol\lambda$, and $a^{(s)}$ is a step-size taking a Robbins-Monro form $a^{(s)}=1/s^{q}$, with $q\in(0.5,1)$ \citep{Lange1995}. If the integral $\int H(\boldsymbol\lambda^{(s)},\boldsymbol\phi)\Pi(\boldsymbol\phi\vert \boldsymbol\lambda^{(s)},y,\mathbf{Z})d\boldsymbol\phi$ is approximated by $H(\boldsymbol\lambda^{(s)},\boldsymbol\phi^{(s+1)})$, we get an approximate EM algorithm, where both E- and M-steps are approximately implemented. Hence, marginal maximum likelihood estimates of the hyper-parameters, $\widehat{\boldsymbol\lambda}$, and draws from the posterior distribution of the parameters, $\Pi(\boldsymbol\phi\vert \widehat{\boldsymbol\lambda},y,\mathbf{Z})$, are both obtained using a single run of the Gibbs sampler.
In the present framework, making the transformation $\boldsymbol\omega=\frac{1}{2}\log(\boldsymbol\lambda)$, from Section \ref{sec:BMIDAS} the function $H(\boldsymbol\omega,\boldsymbol\phi)=\nabla_{\boldsymbol\omega}\log\Pi(\boldsymbol\phi\vert\boldsymbol\omega)$ takes the form:
\begin{displaymath}
H(\boldsymbol\omega,\boldsymbol\phi) = (\mathbf{g}+1)-\exp(2\boldsymbol\omega)\odot\boldsymbol\tau,
\end{displaymath}
where $\mathbf{g}=(g_{1}^{},\dots,g_{G}^{})^{\prime}$ and $\odot$ is the element-wise product. Hence, the updating rule for $\boldsymbol\omega$ is:
\begin{displaymath}
\omega_{j}^{(s+1)} = \omega_{j}^{(s)}+a^{(s)}\left[(g_{j}+1)-\exp\left(2\omega_{j}^{(s)}\right)\tau_{j}^{2,(s+1)}\right]
\end{displaymath}
from which we get $\boldsymbol\lambda^{(s+1)}=\exp(2\boldsymbol\omega^{(s+1)})$.
The algorithm can be completed by allowing for a stabilization procedure (\textit{e.g.} truncation on random boundaries; \citealp{Andrieu2005}; \citealp{Atchade2011}) ensuring the convergence of $\boldsymbol\lambda$ and the posterior distribution of $\boldsymbol\phi$ towards $\widehat{\boldsymbol\lambda}$ and $\Pi(\boldsymbol\phi\vert \widehat{\boldsymbol\lambda},y,\mathbf{Z})$, respectively. Details on the stabilization algorithm are reported in Appendix \ref{sec:stab_alg_appendix}.

\subsection{Numerical illustration}\label{sec:EB_sim}
We illustrate the main features and the computational advantage of the proposed methodology using simulated data generated from the MIDAS model \eqref{eq:MIDAS} with $K=4$, $m=3$, and $C=12$. We calibrate $B\left(c;\boldsymbol\theta\right)$ such that the MIDAS weights decay monotonically to almost zero after four periods. The regressors and the error term are i.i.d. draws from a standard normal distribution of length $T=500$. Only the second regressor $(k=2)$ enters the active set, with slope coefficient $\beta_{2}=1$, while $\beta_{1}=\beta_{3}=\beta_{4}=0$. We estimate the models presented in Section \ref{sec:BMIDAS} using $p=3$ and we tune the penalty hyper-parameters $\boldsymbol\lambda$ using the stochastic approximation approach. We update $\boldsymbol\lambda$ in a single run of the Gibbs sampler by drawing $S=400,000$ samples. The analysis is carried out using MATLAB R2017a on a workstation with a 2.50GHz Intel Core i7-6500U CPU.

\begin{figure}[!t]
\caption{Tuning the penalty hyper-parameters}
\begin{center}
\includegraphics[width=1\textwidth]{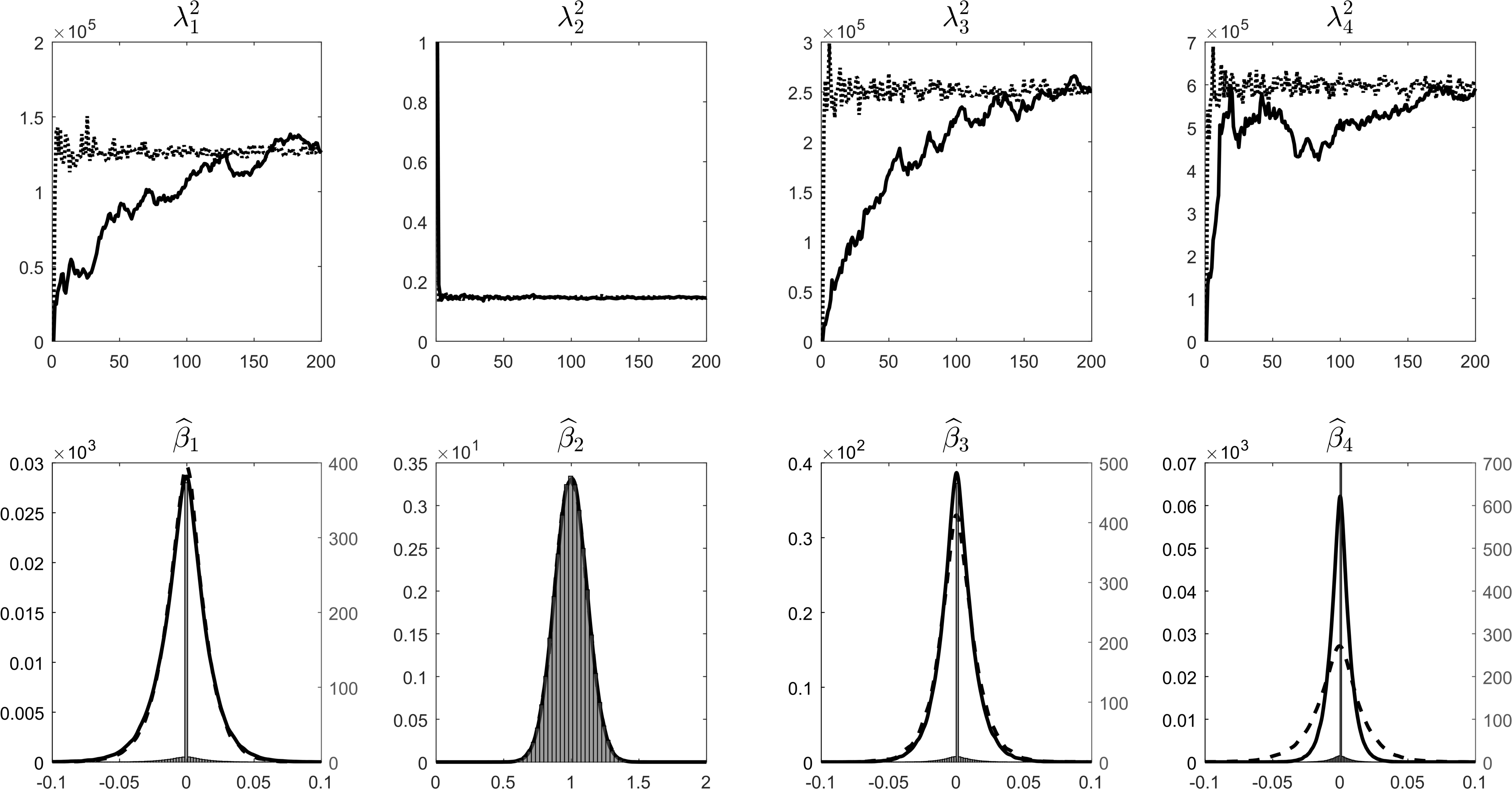}
\end{center}
\floatfoot{Note: The first panel illustrates the evolution of the penalty hyper-parameters $\boldsymbol\lambda$ across iterations of the stochastic approximation approach for BMIDAS-AGL (solid lines) and BMIDAS-AGL-SS model (dotted lines). The second panel illustrates the posterior distributions of parameters $\boldsymbol\beta$ for BMIDAS-AGL (solid lines) and BMIDAS-AGL-SS model (histogram) using the stochastic approximation approach,  and the BIMDAS-AGL model using the MCEM algorithm (dashed lines).}
\label{fig:tuning_hyperparam}
\end{figure}

The evolution of $\boldsymbol\lambda$ across iterations (starting with $\lambda_{j}^{(0)}=1$) is reported in the first panel of Figure \ref{fig:tuning_hyperparam}. Each point in the plots represents the $2000$th update of $\boldsymbol\lambda$ provided by the stochastic approximation approach for the BMIDAS-AGL model (solid lines) and the BMIDAS-AGL-SS model (dotted lines). For both models, the hyper-parameters converge to fairly similar values. However, while the convergence is steady and extremely fast for the active variable, the BMIDAS-AGL model displays slower convergence for the penalty terms of the inactive set compared to the BMIDAS-AGL-SS. Hence, it turns out that the spike-and-slab prior may not only improve the sparse recovery ability of the model but also enhance the convergence of the penalty hyper-parameters, reducing the variance of the posterior distribution for the parameters of the inactive set around the zero-point mass when draws are assigned (even with some low probability) to the slab part of the model. This is indeed confirmed from the inspection of the posterior densities of $\boldsymbol\beta$ in the second panel of Figure \ref{fig:tuning_hyperparam}. The two models feature correct variable selection and consistent estimates of the regression coefficients, with virtually identical densities for $\beta_{2}$ and largest mass at zero for $\beta_{1}$, $\beta_{3}$, and $\beta_{4}$. However, compared to the BMIDAS-AGL, the BMIDAS-AGL-SS model displays the lowest variation around the point mass at exactly zero for the parameters of the inactive set. Finally, these outcomes are compared to those obtained by tuning the penalty hyper-parameters of the BMIDAS-AGL model using the MCEM algorithm with a fairly reasonable amount of Monte Carlo runs ($N=200$) and Gibbs draws ($S=50,000$). Looking at the posterior densities, the results for the MCEM algorithm (dashed lines) appear almost indistinguishable from those obtained using the stochastic approximation approach, with the only exception of $\beta_{4}$. However, the computational burden differs substantially across algorithms: for this simple simulation experiment and the settings described above, the analysis is performed in less than 2 minutes with stochastic approximations, against 30 minutes required by the MCEM algorithm.

\section{Monte Carlo experiments}\label{sec:MCsim}

\subsection{Design of the experiments}\label{sec:MCsim_design}
\begin{figure}[!t]
\caption{Normalized MIDAS weights in the Monte Carlo simultations}
\begin{center}
\includegraphics[scale=0.77]{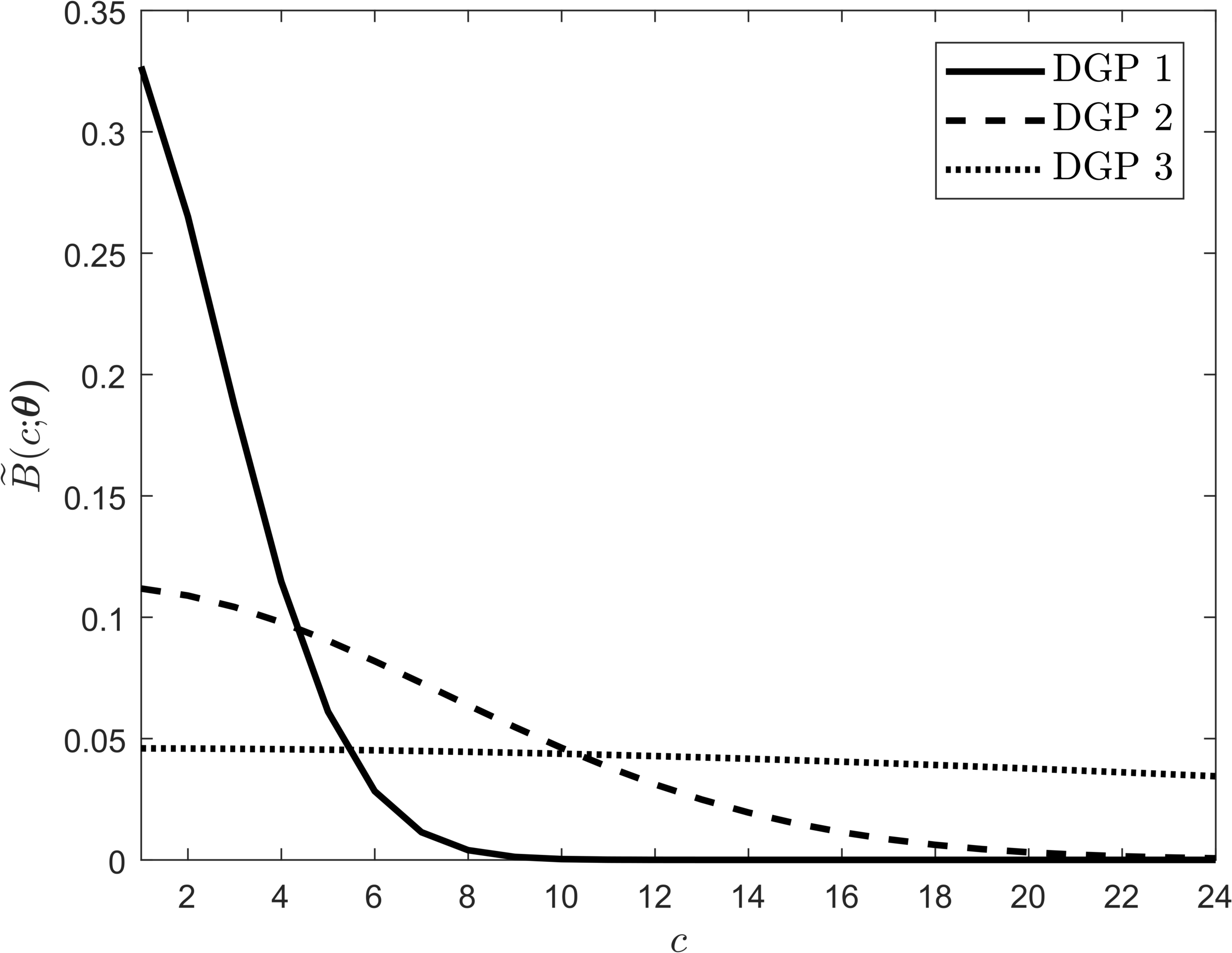}
\end{center}
\label{fig:MIDAS_weights}
\end{figure}

We evaluate the small-sample performance of the proposed models through Monte Carlo experiments. For this purpose, we consider the following DGP, involving $K=\{30,50\}$ predictors sampled at frequency $m=3$ and $T=200$ in-sample observations:
\begin{align*}
y_{t} &= \alpha+\sum_{k=1}^{K}\beta_{k}\sum_{c=0}^{C-1}\widetilde{B}\left(c;\boldsymbol\theta\right)L^{\slfrac{c}{m}}x_{k,t-h}^{(m)}+\epsilon_{t} \\[1.5ex]
x_{k,t}^{(m)} &= \mu + \rho x_{k,t-\slfrac{1}{m}}^{(m)} + \varepsilon_{k,t},
\end{align*}
where $\widetilde{B}\left(c;\boldsymbol\theta\right)$ denotes the normalized weigths (\textit{i.e.} summing up to 1). Following \citet{Andreou2010}, we investigate three alternative weighting schemes that correspond to fast-decaying weights (DGP 1), slow-decaying weights (DGP 2), and near-flat weights (DGP 3). These three weighting schemes are represented in Figure \ref{fig:MIDAS_weights}. In all simulations we set the lag length $C=24$. Note that the same weighting structure applies to all the predictors entering the DGPs. Further, for ease of analysis we assume $h=0$, \textit{i.e.} a nowcasting model with high-frequency information fully matching the low frequency. In this specification, $\epsilon_{t}$ and $\boldsymbol\varepsilon_{t} := (\varepsilon_{1,t},\dots, \varepsilon_{K,t})^{\prime}$ are i.i.d. with distribution:
\begin{displaymath}
\left( \begin{array}{c}
\epsilon_{t}\\
\boldsymbol\varepsilon_{t}
\end{array} \right) \thicksim \textrm{i.i.d.}~\mathcal{N}
\left[ \left( \begin{array}{c}
0 \\
\mathbf{0}
\end{array} \right),
\left( \begin{array}{c c}
\sigma^{2} & \mathbf{0} \\
\mathbf{0} & \boldsymbol\Sigma_{\varepsilon} \\
\end{array} \right)
\right],
\end{displaymath}
where $\boldsymbol\Sigma_{\varepsilon}$ has elements $\sigma_{\varepsilon}^{\vert k-k^{\prime} \vert}$, such that the diagonal elements are equal to one and the off-diagonal elements control for the correlation between $x_{k,t}^{(m)}$ and $x_{k^{\prime},t}^{(m)}$, with $k\neq k^{\prime}$. We set $\sigma_{\varepsilon}=\{0.50,0.95\}$, \textit{i.e.} from moderate to extremely high correlation structure in the design matrix $\mathbf{x}_{t}^{(m)}$. As for the parameters in the DGP, we choose $\alpha=0.5$, $\mu=0.1$, $\rho=0.9$, and $\boldsymbol\beta=(0,0.3,0.5,0,0.3,0.5,0,0,0.8,\mathbf{0})^{\prime}$. The latter implies that only five out of $K$ predictors are relevant. Conditional on these parameters, we set $\sigma$ such that the noise-to-signal ratio of the mixed-frequency regression is 0.20.

We estimate our BMIDAS-AGL and BMIDAS-AGL-SS models with Almon lag polynomial order $p=3$ and $r=2$ endpoint restrictions (both tail and derivative; see Section \ref{sec:basic_MIDAS}). The hyper-parameters $\boldsymbol\lambda$ are tuned using the stochastic approximation approach described in Section \ref{sec:EB}, with step-size $a^{(s)}=1/s^{0.8}$ (preliminary results suggest that this sequence is sufficient to achieve convergence). We set the number of Monte Carlo replications at $R=500$. The Gibbs sampler is run for $S=300,000$ iterations, with the first $100,000$ used as a burn-in period, and every 10th draw is saved.

\subsection{Estimation and selection performance}\label{sec:selection}
We evaluate the estimation performance of our models by computing the average mean squared error (MSE), the average variance (VAR), and the average squared bias (BIAS$^2$) over $R$ Monte Carlo replications and the full set of $K$ estimated overall slope parameters $\widehat{\boldsymbol\beta}$.\footnote{For $R$ Monte Carlo replications, $K$ variables, and $S$ Gibbs draws, we have that:
\begin{displaymath}
\text{MSE}=\text{VAR}+\text{BIAS$^2$}=\frac{1}{RKS}\sum_{i=1}^{R}\sum_{k=1}^{K}\sum_{s=1}^{S}\left[\widehat{\beta}_{k,i}^{(s)}-\mathbb{E}\left(\widehat{\beta}_{k,i}\right)\right]^2 + \frac{1}{RK}\sum_{i=1}^{R}\sum_{k=1}^{K}\left[\mathbb{E}\left(\widehat{\beta}_{k,i}\right)-\beta_{k}\right]^2,
\end{displaymath}
where $\mathbb{E}\left(\widehat{\beta}_{k}\right)=\frac{1}{S}\sum_{s=1}^{S}\widehat{\beta}_{k}^{(s)}$ and $\widehat{\beta}_{k}=\widehat{\boldsymbol{\theta}}_{k}^{\prime}\mathbf{Q}\boldsymbol\iota_{C}^{}$, for $k=1,\dots,K$. Note that for the BMIDAS-AGL-SS model we use the median estimator.
}
To evaluate the selection ability of the models, we consider the True Positive Rate (TPR), the False Positive Rate (FPR), and the Matthews correlation coefficient (MCC), the latter measuring the overall quality of the classification. We follow different approaches to implement variable selection. For the BMIDAS-AGL model, we rely on the credible interval criterion suggested by \citet{Kyung2010}, which excludes a predictor $k$, for $k=1,\dots,K$, from the estimated active set if the credible interval, at say 95\% level, of the posterior distribution of its slope coefficient $\widehat{\beta}_{k}$ includes zero. For the BMIDAS-AGL-SS model, we resort to the posterior median estimator \citep{Barbieri2004}, that is, under some conditions, a soft thresholding estimator presenting model selection consistency and optimal asymptotic estimation rate \citep{Xu2015}.

\subsection{Forecast evaluation}\label{sec:fcst}
Forecasts are obtained from the following posterior predictive density for $y_{T\vert T-h}^{}$:
\begin{equation}\label{eq:pred_dens}
f(y_{T\vert T-h}^{}\vert\mathcal{D})=\int_{\boldsymbol\phi,\boldsymbol\lambda} f(y_{T\vert T-h}^{}\vert\boldsymbol\phi,\boldsymbol\lambda,\mathcal{D})\Pi(\boldsymbol\phi,\boldsymbol\lambda\vert\mathcal{D}) d\boldsymbol\phi d\boldsymbol\lambda
\end{equation}
where $\Pi(\boldsymbol\phi,\boldsymbol\lambda\vert\mathcal{D})$ denotes the joint posterior distribution of the BMIDAS parameters conditional on past available information, $\mathcal{D}$. According to the framework described in Sections \ref{sec:BMIDAS} and \ref{sec:EB}, draws $y_{T\vert T-h}^{(s)}$, $s=1,\dots,S$, from the predictive distribution \eqref{eq:pred_dens} can be obtained directly from the Gibbs sampler.\footnote{It is worth noting that we do not condition on a fixed value $\widehat{\boldsymbol\lambda}$, such as the maximum likelihood estimate that can be obtained, for instance, by averaging over the Gibbs samples of $\boldsymbol\lambda$, because this would ignore the uncertainty  around the estimate of the penalty parameters.}
This leads to a distribution of predictions that can be used for out-of-sample evaluation of the model. Point forecasts are computed by averaging over these draws, \textit{i.e.} $\widehat{y}_{T\vert T-h}^{}=(S-\bar{s}+1)^{-1}\sum_{s=\bar{s}+1}^{S}y_{T\vert T-h}^{(s)}$, where $\bar{s}$ is the last burn-in iteration, and evaluated through the average root mean squared forecast error (RMSFE) over the $R$ Monte Carlo replications. Further, since draws from the predictive density are available, an evaluation of the entire predictive distribution is performed through the average log-score (LS), \textit{i.e.} the average of the log of the predictive likelihood evaluated at the out-turn of the forecast \citep{Mitchell2011}, and the average continuously ranked probability score (CRPS), which measures the average distance between the empirical CDF of the out-of-sample observations and the empirical CDF associated with the predictive density of each model \citep{Gneiting2007}.

\subsection{Monte Carlo results}\label{sec:MCsim_results}
\begin{figure}[!t]
\caption{Breakdown of MSE by active $(\mathcal{A})$ and inactive set $(\mathcal{A}^{c})$}
\begin{center}
\includegraphics[width=1\textwidth]{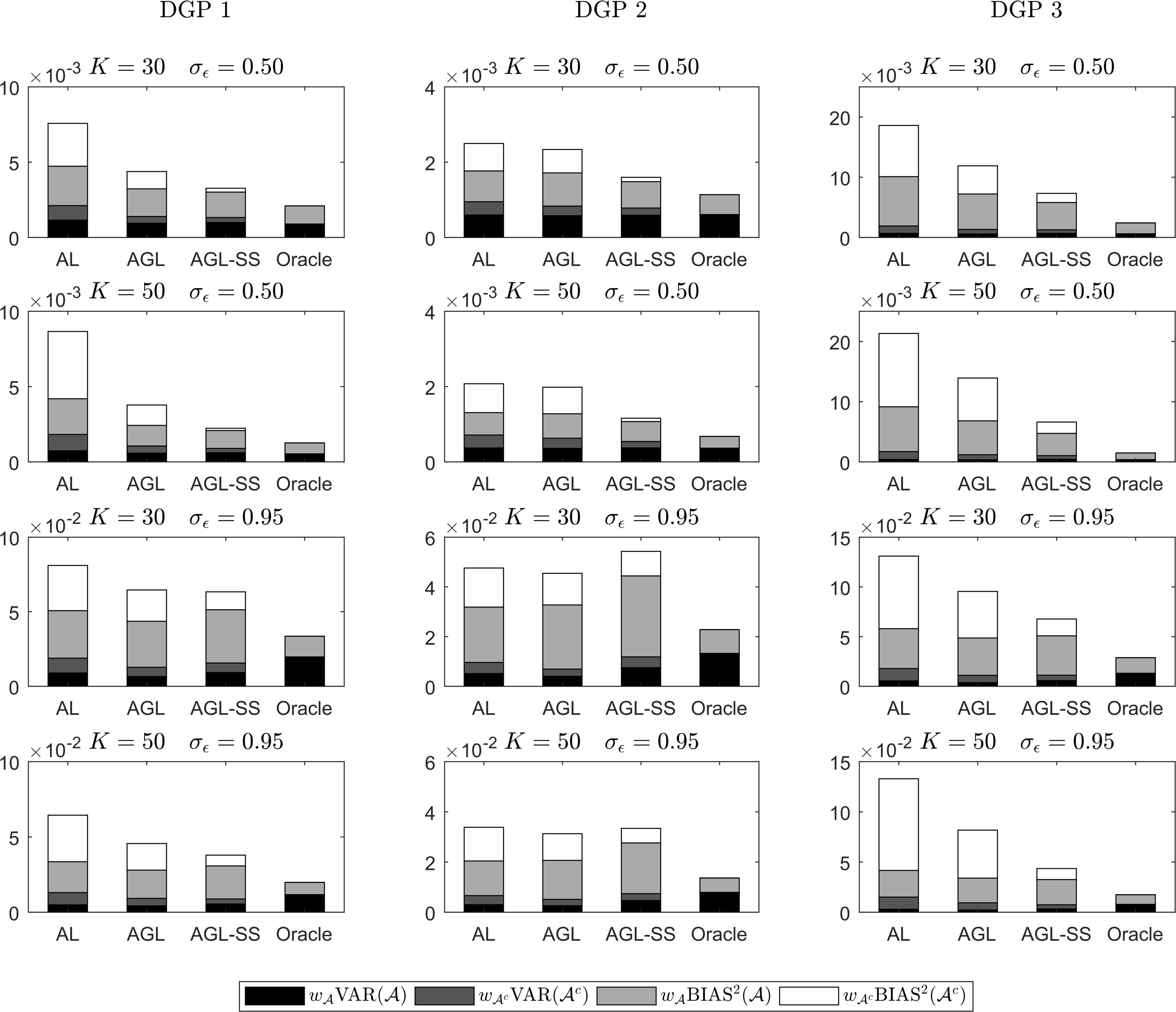}
\end{center}
\label{fig:MSE_breakdown}
\end{figure}

Simulation results are reported in Table \ref{t:mc_results1} and point to a number of interesting features. First, the models perform overall quite similarly in terms of MSE, although the BMIDAS-AGL-SS seems to perform somewhat better across DGPs by mainly providing the smallest bias. This leads to highest TPR and lowest FPR for this model, entailing better classification of active and inactive sets across simulations. Second, the MSE increases substantially with the degree of correlation in the design matrix (governed by the value of $\sigma_{\varepsilon}$). However, the MSE tends to decrease with more irrelevant predictors in the DGP. To understand the latter result, it is useful to look at the bias/variance breakdown of the MSE stemming from both the active $(\mathcal{A})$ and inactive set $(\mathcal{A}^{c})$ reported in Figure \ref{fig:MSE_breakdown}, where we refer to our models as AGL and AGL-SS. Results suggest that while the contribution of the inactive set to the total MSE is broadly stable when $K$ increases, the contribution of the active set decreases for both bias and variance. These findings can nevertheless be attributed, at least in part, to the lower relative weight of the active set $(w_{\mathcal{A}})$ in the total bias and variance, as the number of relevant predictors $(K_{\mathcal{A}})$ is fixed while the number of irrelevant predictors $(K_{\mathcal{A}^{c}})$ is allowed to increase.\footnote{For $w_{\mathcal{A}}=K_{\mathcal{A}}/K$ and $w_{\mathcal{A}^{c}}=1-w_{\mathcal{A}}$, we have that:
\begin{displaymath}
\text{MSE}=w_{\mathcal{A}}\text{MSE}(\mathcal{A})+w_{\mathcal{A}^{c}}\text{MSE}(\mathcal{A}^{c})=w_{\mathcal{A}}\left[\text{VAR}(\mathcal{A})+\text{BIAS$^2$}(\mathcal{A})\right]+w_{\mathcal{A}^{c}}\left[\text{VAR}(\mathcal{A}^{c})+\text{BIAS$^2$}(\mathcal{A}^{c})\right].
\end{displaymath}
}
Looking at the MSE by active/inactive set reported in Table \ref{t:mc_results1} (columns 5 and 6), we note that the $\text{MSE}(\mathcal{A})$ appears broadly stable or it increases only moderately with more irrelevant variables in DGPs 1 and 2.\footnote{For DGP 3, the $\text{MSE}(\mathcal{A})$ tends to increase more substantially with more irrelevant predictors. One explanation is that the linear restrictions imposed to the lag polynomials are incorrect under this DGP.}
Hence, our interpretation is that the performance of our BMIDAS models in selecting and estimating the coefficients in the active set seems only marginally affected by the increase in the degree of sparsity.

This result is confirmed by the TPR, which is relatively high and hovers around 90\% for moderate correlation, and it's overall stable across different values of $K$, suggesting that the models can select the correct sparsity pattern with a high probability even in finite samples. It is worth noting that the TPR drops to 30-50\% with very high correlation in the design matrix, while the FPR remains overall very low. Note that this outcome is nevertheless not unexpected, as the Group Lasso can address the issue of strong collinearity within the lag polynomials but is not designed to handle strong collinearity between the high-frequency regressors. For comparison purposes, we also report in Figure \ref{fig:MSE_breakdown} the breakdown of the MSE for the Bayesian MIDAS Adaptive Lasso (BMIDAS-AL) and the Oracle BMIDAS, the former estimated following the same approach as in Sections \ref{sec:BMIDAS} and \ref{sec:Gibbs} and the latter using the algorithm described in \citet{Pettenuzzo2016} on the set of relevant variables only.\footnote{We consider the same restricted Almon lag polynomial as for our models. Further, for the Oracle BMIDAS, we follow \citet{Pettenuzzo2016} and we use relatively diffuse priors on both the coefficient covariance matrix and the regression variance. As for the prior mean coefficients, we set all the coefficients but the intercept to zero.}
A visual inspection of Figure \ref{fig:MSE_breakdown} seems to confirm the intuition discussed in Section \ref{sec:MIDAS-pen}, that the AL may not be suited in the present framework. Results suggest that higher MSE provided by the AL can be mainly attributed to higher bias and variance in the inactive set. It is worth noting that while these findings appear quite conclusive for DGPs 1 and 3, evidence is less clear-cut for DGP 2, especially when high correlation in the design matrix is considered. When compared to the Oracle, our models perform fairly well overall. Not surprisingly, in most cases the main difference lies in the bias of the active set, as the Group Lasso would typically trade off more bias for less variance.

\begin{table}[!t]
\footnotesize
\caption{Monte Carlo simulations: estimation and selection accuracy}
\centering
\setlength{\tabcolsep}{0.25cm}
\begin{tabular}{ccc|ccccc|ccc}
\hline\hline
 &  & & \multicolumn{5}{c|}{} & \multicolumn{3}{c}{} \\[-2.0ex]
$K$ & $\sigma_{\varepsilon}$ & $\overline\sigma$ & MSE & MSE$(\mathcal{A})$ & MSE$(\mathcal{A}^{c})$ & VAR & \multicolumn{1}{c|}{BIAS$^2$} & TPR & FPR & MCC \\[0.5ex]
\hline
 &  &  & \\[-1.0ex]
 &  &  & \multicolumn{8}{c}{DGP 1} \\[1.5ex]
\hline
 &  &  & \\[-1.5ex]
 &  &  & \multicolumn{8}{c}{BMIDAS-AGL} \\[1.0ex]
\hline
30 & 0.50 & 1.3 & 4.4E-03 & 1.7E-02 & 1.9E-03 & 1.4E-03 & 3.0E-03 & 0.96 & 0.03 & 0.90 \\
   & 0.95 & 2.1 & 6.5E-02 & 2.2E-01 & 3.3E-02 & 1.3E-02 & 5.2E-02 & 0.36 & 0.04 & 0.41 \\[1.0ex]
50 & 0.50 & 1.3 & 3.8E-03 & 1.9E-02 & 2.0E-03 & 1.1E-03 & 2.7E-03 & 0.94 & 0.03 & 0.85 \\
   & 0.95 & 2.1 & 4.6E-02 & 2.3E-01 & 2.5E-02 & 9.2E-03 & 3.6E-02 & 0.35 & 0.03 & 0.40 \\
\hline
 &  &  & \\[-1.5ex]
 &  &  & \multicolumn{8}{c}{BMIDAS-AGL-SS} \\[1.0ex]
\hline
30 & 0.50 & 1.3 & 3.3E-03 & 1.6E-02 & 7.1E-04 & 1.3E-03 & 1.9E-03 & 0.96 & 0.01 & 0.95 \\
   & 0.95 & 2.1 & 6.3E-02 & 2.7E-01 & 2.2E-02 & 1.6E-02 & 4.8E-02 & 0.36 & 0.02 & 0.49 \\[1.0ex]
50 & 0.50 & 1.3 & 2.2E-03 & 1.8E-02 & 4.7E-04 & 8.9E-04 & 1.3E-03 & 0.94 & 0.00 & 0.95 \\
   & 0.95 & 2.1 & 3.8E-02 & 2.7E-01 & 1.2E-02 & 8.8E-03 & 2.9E-02 & 0.35 & 0.01 & 0.50 \\
\hline
 &  &  & \\[-1.0ex]
 &  &  & \multicolumn{8}{c}{DGP 2} \\[1.5ex]
\hline
 &  &  & \\[-1.5ex]
 &  &  & \multicolumn{8}{c}{BMIDAS-AGL} \\[1.0ex]
\hline
30 & 0.50 & 1.1 & 2.3E-03 & 8.7E-03 & 1.1E-03 & 8.3E-04 & 1.5E-03 & 0.99 & 0.02 & 0.93 \\
   & 0.95 & 1.8 & 4.5E-02 & 1.8E-01 & 1.9E-02 & 6.9E-03 & 3.9E-02 & 0.48 & 0.04 & 0.53 \\[1.0ex]
50 & 0.50 & 1.1 & 2.0E-03 & 1.0E-02 & 1.1E-03 & 6.3E-04 & 1.4E-03 & 0.98 & 0.02 & 0.90 \\
   & 0.95 & 1.8 & 3.1E-02 & 1.8E-01 & 1.5E-02 & 5.1E-03 & 2.6E-02 & 0.46 & 0.03 & 0.51 \\
\hline
 &  &  & \\[-1.5ex]
 &  &  & \multicolumn{8}{c}{BMIDAS-AGL-SS} \\[1.0ex]
\hline
30 & 0.50 & 1.1 & 1.6E-03 & 7.7E-03 & 3.7E-04 & 7.8E-04 & 8.2E-04 & 0.99 & 0.01 & 0.98 \\
   & 0.95 & 1.8 & 5.4E-02 & 2.4E-01 & 1.7E-02 & 1.2E-02 & 4.2E-02 & 0.47 & 0.02 & 0.58 \\[1.0ex]
50 & 0.50 & 1.1 & 1.2E-03 & 9.0E-03 & 2.9E-04 & 5.4E-04 & 6.2E-04 & 0.98 & 0.00 & 0.97 \\
   & 0.95 & 1.8 & 3.3E-02 & 2.5E-01 & 9.5E-03 & 7.4E-03 & 2.6E-02 & 0.45 & 0.01 & 0.60 \\
\hline
 &  &  & \\[-1.0ex]
 &  &  & \multicolumn{8}{c}{DGP 3} \\[1.5ex]
\hline
 &  &  & \\[-1.5ex]
 &  &  & \multicolumn{8}{c}{BMIDAS-AGL} \\[1.0ex]
\hline
30 & 0.50 & 1.0 & 1.2E-02 & 3.9E-02 & 6.5E-03 & 1.3E-03 & 1.1E-02 & 0.85 & 0.15 & 0.61 \\
   & 0.95 & 1.6 & 9.5E-02 & 2.5E-01 & 6.5E-02 & 1.1E-02 & 8.4E-02 & 0.35 & 0.10 & 0.27 \\[1.0ex]
50 & 0.50 & 1.0 & 1.4E-02 & 6.0E-02 & 8.8E-03 & 1.2E-03 & 1.3E-02 & 0.77 & 0.16 & 0.46 \\
   & 0.95 & 1.6 & 8.2E-02 & 2.7E-01 & 6.1E-02 & 9.4E-03 & 7.2E-02 & 0.32 & 0.10 & 0.22 \\
\hline
 &  &  & \\[-1.5ex]
 &  &  & \multicolumn{8}{c}{BMIDAS-AGL-SS} \\[1.0ex]
\hline
30 & 0.50 & 1.0 & 7.3E-03 & 3.1E-02 & 2.5E-03 & 1.3E-03 & 6.0E-03 & 0.87 & 0.06 & 0.78 \\
   & 0.95 & 1.6 & 6.8E-02 & 2.7E-01 & 2.7E-02 & 1.1E-02 & 5.7E-02 & 0.33 & 0.03 & 0.41 \\[1.0ex]
50 & 0.50 & 1.0 & 6.6E-03 & 4.1E-02 & 2.8E-03 & 1.0E-03 & 5.6E-03 & 0.81 & 0.05 & 0.71 \\
   & 0.95 & 1.6 & 4.3E-02 & 2.8E-01 & 1.7E-02 & 7.4E-03 & 3.6E-02 & 0.30 & 0.02 & 0.41 \\
\hline\hline
\multicolumn{11}{l}{\scriptsize \parbox[t]{14.5cm}{Notes: BMIDAS-AGL and BMIDAS-AGL-SS refer to the models described in Section \ref{sec:BMIDAS}. $\overline\sigma$ is the average variance of the error process across MC simulations. MSE, VAR, and BIAS$^2$ denote the Mean Squared Error, the Variance, and the Squared Bias, respectively. $\text{MSE}(\mathcal{A})$ and $\text{MSE}(\mathcal{A}^{c})$ denote the MSE for the active and inactive set, respectively. TPR, FPR, and MCC denote the True Positive Rate, the False Positive Rate, and the Matthews Correlation Coefficient, respectively.}}
\end{tabular}
\vspace{7mm}
\label{t:mc_results1}
\end{table}

\begin{table}[!t]
\footnotesize
\caption{Monte Carlo simulations: restricted vs unrestricted weighting schemes}
\centering
\setlength{\tabcolsep}{0.21cm}
\begin{tabular}{cc|ccc|ccc}
\hline\hline
 &  &  \multicolumn{3}{c|}{} & \multicolumn{3}{c}{} \\[-2.0ex]
 &  &  \multicolumn{3}{c|}{BMIDAS-AGL}  &  \multicolumn{3}{c}{BMIDAS-AGL-SS} \\[0.5ex]
\hline
 &  &  \multicolumn{3}{c|}{} & \multicolumn{3}{c}{} \\[-2.0ex]
$K$ & $\sigma_{\varepsilon}$ & TPR & FPR & \multicolumn{1}{c|}{MCC} & TPR & FPR & MCC \\[0.5ex]
\hline
 &  &  \\[-1.0ex]
 &  &  \multicolumn{6}{c}{DGP 1} \\[1.5ex]
\hline
30 & 0.50 & 0.15 & -0.02 & 0.15 & 0.14 & -0.01 & 0.11 \\
   & 0.95 & 0.06 & 0.01 & 0.03 & 0.06 & -0.01 & 0.08 \\[1.0ex]
50 & 0.50 & 0.25 & -0.04 & 0.27 & 0.15 & -0.01 & 0.12 \\
   & 0.95 & 0.08 & 0.00 & 0.09 & 0.06 & 0.00 & 0.09 \\
\hline
 &  &  \\[-1.0ex]
 &  &  \multicolumn{6}{c}{DGP 2} \\[1.5ex]
\hline
30 & 0.50 & 0.09 & -0.02 & 0.09 & 0.08 & -0.01 & 0.07 \\
   & 0.95 & 0.10 & -0.01 & 0.12 & 0.11 & -0.01 & 0.13 \\[1.0ex]
50 & 0.50 & 0.18 & -0.03 & 0.21 & 0.11 & -0.01 & 0.11 \\
   & 0.95 & 0.14 & -0.01 & 0.15 & 0.09 & 0.00 & 0.11 \\
\hline
 &  &  \\[-1.0ex]
 &  &  \multicolumn{6}{c}{DGP 3} \\[1.5ex]
\hline
30 & 0.50 & -0.10 & 0.11 & -0.27 & -0.08 & 0.04 & -0.15 \\
   & 0.95 & -0.09 & 0.06 & -0.20 & -0.10 & 0.01 & -0.12 \\[1.0ex]
50 & 0.50 & -0.12 & 0.12 & -0.33 & -0.11 & 0.04 & -0.19 \\
   & 0.95 & -0.08 & 0.07 & -0.23 & -0.10 & 0.01 & -0.13 \\
\hline\hline
\multicolumn{8}{l}{\scriptsize \parbox[t]{8.1cm}{Notes: See Table \ref{t:mc_results1}. The reported values denote the difference in TPR, FPR, and MCC between models estimated with restricted $(r=2)$ and unrestricted $(r=0)$ Almon lag polynomials.}}
\end{tabular}
\vspace{7mm}
\label{t:mc_results2}
\end{table}

Third, the in-sample results shown in Table \ref{t:mc_results1} deteriorate when the DGP with near-flat weights is considered, and mostly when $\sigma_{\varepsilon}=0.95$. This happens because the linear restrictions imposed on the lag polynomials force the weighting structure to tail off to zero, while DGP 3 assumes that the weighting scheme under the null is almost uniform over the lag window. It follows that relaxing the restrictions on the Almon lag polynomial should lead to an improvement of the results under this DGP. However, it is not clear how much of the selection results under DGPs 1 and 2 can be attributed to the imposed linear restrictions. Table \ref{t:mc_results2} provides an answer to these questions by reporting the difference in TPR, FPR, and MCC obtained with restricted $(r=2)$ and unrestricted $(r=0)$ Almon lag polynomials. For DGPs with fast- or slow-decaying weights, the results suggest that imposing correct linear restrictions that are valid under the null seems to improve the selection ability of the models. The gain in terms of TPR ranges 10-25 percentage points for moderate correlation and 5-15 percentage points for very high correlation in the design matrix, while the gain in terms of FPR ranges 1-4 percentage points. Interestingly enough, the results show that the BMIDAS-AGL-SS model is relatively less affected than the BMIDAS-AGL model when correct linear restrictions are relaxed. For the DGP with near-flat weights we observe an opposite outcome, as expected. However, the magnitude of these results must be considered with care, as the number of relevant and irrelevant predictors in the simulated true model is here strongly asymmetric.

\clearpage
\begin{table}[!t]
\footnotesize
\caption{Monte Carlo simulations: forecasting performance}
\centering
\setlength{\tabcolsep}{0.09cm}
\begin{tabular}{l|ccc|ccc|ccc|ccc}
\hline\hline
 &  \multicolumn{3}{c|}{} & \multicolumn{3}{c|}{} & \multicolumn{3}{c|}{} & \multicolumn{3}{c}{} \\[-2.0ex]
 & RMSFE & LS & CRPS & RMSFE & LS & CRPS & RMSFE & LS & CRPS & RMSFE & LS & CRPS \\[0.5ex]
\hline
 &  \multicolumn{6}{c|}{} & \multicolumn{6}{c}{} \\[-2.0ex]
\multicolumn{1}{c|}{$K$} & \multicolumn{6}{c|}{$30$} & \multicolumn{6}{c}{$50$} \\[0.5ex]
\hline
 &  \multicolumn{3}{c|}{} & \multicolumn{3}{c|}{} & \multicolumn{3}{c|}{} & \multicolumn{3}{c}{} \\[-2.0ex]
\multicolumn{1}{c|}{$\sigma_{\varepsilon}$} & \multicolumn{3}{c|}{0.50} & \multicolumn{3}{c|}{0.95} & \multicolumn{3}{c|}{0.50} & \multicolumn{3}{c}{0.95}  \\[0.5ex]
\hline
 &  \multicolumn{12}{c}{} \\[-1.0ex]
 &  \multicolumn{12}{c}{DGP 1} \\[1.5ex]
\hline
Oracle & $\mathit{1.17}$ & -$\mathit{1.61}$ & $\mathit{0.68}$ & $\mathit{1.87}$ & -$\mathit{2.13}$ & $\mathit{1.10}$ & $\mathit{1.12}$ & -$\mathit{1.59}$ & $\mathit{0.65}$ & $\mathit{1.88}$ & -$\mathit{2.12}$ & $\mathit{1.10}$ \\[1.0ex]
\hline
BMIDAS-AGL & 1.15 & -0.11 & 1.14 & 1.11 & -0.04 & 1.09 & 1.22 & -0.15 & 1.20 & 1.18 & -0.09 & 1.15 \\[1.0ex]
BMIDAS-AGL-SS & 1.13 & -0.09 & 1.12 & 1.10 & -0.04 & 1.08 & 1.14 & -0.10 & 1.13 & 1.15 & -0.07 & 1.12 \\[1.0ex]
\hline
BMIDAS-AL & 1.29 & -0.22 & 1.28 & 1.19 & -0.11 & 1.16 & 1.53 & -0.38 & 1.50 & 1.31 & -0.20 & 1.27 \\[1.0ex]
MIDAS-L & 1.29 & -0.22 & 1.28 & 1.18 & -0.09 & 1.14 & 1.59 & -0.42 & 1.58 & 1.35 & -0.23 & 1.31 \\[1.0ex]
MIDAS-EN & 1.32 & -0.24 & 1.31 & 1.17 & -0.09 & 1.14 & 1.64 & -0.45 & 1.63 & 1.37 & -0.25 & 1.33 \\[1.0ex]
MIDAS-SCAD & 1.31 & -0.25 & 1.31 & 1.20 & -0.11 & 1.17 & 1.75 & -0.61 & 1.77 & 1.39 & -0.30 & 1.36 \\[1.0ex]
MIDAS-MC$+$ & 1.39 & -0.31 & 1.40 & 1.18 & -0.09 & 1.15 & 1.69 & -0.53 & 1.69 & 1.41 & -0.32 & 1.39 \\
\hline
 &  \multicolumn{12}{c}{} \\[-1.0ex]
 &  \multicolumn{12}{c}{DGP 2} \\[1.5ex]
\hline
Oracle & $\mathit{0.97}$ & -$\mathit{1.42}$ & $\mathit{0.56}$ & $\mathit{1.55}$ & -$\mathit{1.91}$ & $\mathit{0.90}$ & $\mathit{0.93}$ & -$\mathit{1.40}$ & $\mathit{0.54}$ & $\mathit{1.51}$ & -$\mathit{1.88}$ & $\mathit{0.87}$ \\[1.0ex]
\hline
BMIDAS-AGL & 1.08 & -0.06 & 1.07 & 1.09 & -0.05 & 1.08 & 1.14 & -0.10 & 1.13 & 1.14 & -0.07 & 1.11 \\[1.0ex]
BMIDAS-AGL-SS & 1.07 & -0.05 & 1.06 & 1.09 & -0.06 & 1.08 & 1.11 & -0.07 & 1.10 & 1.12 & -0.07 & 1.10 \\[1.0ex]
\hline
BMIDAS-AL & 1.13 & -0.08 & 1.11 & 1.19 & -0.12 & 1.16 & 1.23 & -0.16 & 1.20 & 1.27 & -0.17 & 1.23 \\[1.0ex]
MIDAS-L & 1.10 & -0.06 & 1.09 & 1.08 & -0.04 & 1.07 & 1.14 & -0.09 & 1.12 & 1.13 & -0.07 & 1.10 \\[1.0ex]
MIDAS-EN & 1.11 & -0.07 & 1.10 & 1.08 & -0.04 & 1.07 & 1.16 & -0.11 & 1.14 & 1.13 & -0.07 & 1.11 \\[1.0ex]
MIDAS-SCAD & 1.08 & -0.05 & 1.07 & 1.11 & -0.06 & 1.10 & 1.11 & -0.06 & 1.09 & 1.22 & -0.14 & 1.20 \\[1.0ex]
MIDAS-MC$+$ & 1.08 & -0.04 & 1.06 & 1.09 & -0.04 & 1.08 & 1.11 & -0.06 & 1.09 & 1.18 & -0.11 & 1.16 \\
\hline
 &  \multicolumn{12}{c}{} \\[-1.0ex]
 &  \multicolumn{12}{c}{DGP 3} \\[1.5ex]
\hline
Oracle & $\mathit{0.95}$ & -$\mathit{1.41}$ & $\mathit{0.55}$ & $\mathit{1.44}$ & -$\mathit{1.86}$ & $\mathit{0.85}$ & $\mathit{0.91}$ & -$\mathit{1.39}$ & $\mathit{0.53}$ & $\mathit{1.52}$ & -$\mathit{1.88}$ & $\mathit{0.88}$ \\[1.0ex]
\hline
BMIDAS-AGL & 1.20 & -0.13 & 1.18 & 1.17 & -0.09 & 1.14 & 1.26 & -0.18 & 1.24 & 1.21 & -0.16 & 1.21 \\[1.0ex]
BMIDAS-AGL-SS & 1.16 & -0.12 & 1.15 & 1.16 & -0.09 & 1.14 & 1.20 & -0.13 & 1.17 & 1.16 & -0.11 & 1.15 \\[1.0ex]
\hline
BMIDAS-AL & 1.37 & -0.26 & 1.34 & 1.26 & -0.16 & 1.22 & 1.54 & -0.43 & 1.51 & 1.34 & -0.26 & 1.33 \\[1.0ex]
MIDAS-L & 1.37 & -0.28 & 1.36 & 1.33 & -0.21 & 1.30 & 1.36 & -0.26 & 1.34 & 1.30 & -0.24 & 1.29 \\[1.0ex]
MIDAS-EN & 1.38 & -0.28 & 1.36 & 1.31 & -0.19 & 1.27 & 1.40 & -0.29 & 1.37 & 1.30 & -0.22 & 1.28 \\[1.0ex]
MIDAS-SCAD & 1.38 & -0.30 & 1.36 & 1.30 & -0.20 & 1.27 & 1.50 & -0.43 & 1.50 & 1.45 & -0.45 & 1.46 \\[1.0ex]
MIDAS-MC$+$ & 1.42 & -0.32 & 1.41 & 1.27 & -0.16 & 1.23 & 1.47 & -0.39 & 1.46 & 1.37 & -0.32 & 1.36 \\
\hline\hline
\multicolumn{13}{l}{\scriptsize \parbox[t]{15.9cm}{Notes: RMSFE, LS, and CRPS denote respectively the root mean squared forecast error, the log-score, and the continuously ranked probability score, in relative terms with respect to the Oracle benchmark. Values for the Oracle (in italic) denote absolute outcomes. MIDAS regressions with Lasso, Elastic-Net, SCAD, and MC$+$ penalties are estimated using the {\tt penalized} MATLAB toolbox. Regularization parameters are tuned by optimizing the BIC over a fine grid of values.}}
\end{tabular}
\vspace{7mm}
\label{t:mc_results3}
\end{table}

Finally, we report in Table \ref{t:mc_results3} the forecasting performance of our models, in relative terms with respect to the Oracle for ease of comparison.\footnote{We compute RMSFE and CRPS ratios, such that values greater than one indicate that the Oracle performs best. For the LS, we compute the log-score differentials, such that negative values indicate that the Oracle performs best.}
The results are broadly in line with those stemming from the in-sample analysis, and suggest that our BMIDAS models perform quite similarly in terms of point and density forecasts, although the BMIDAS-AGL-SS model seems to perform best overall. The performance of our models deteriorates substantially with higher correlation in the design matrix, although a higher average variance of the error process (see $\overline\sigma$ in Table \ref{t:mc_results1}) might explain, at least in part, the large differences spotted across DGPs. Nevertheless, the performance appears only mildly sensitive to the number of irrelevant predictors ($K_{\mathcal{A}^{c}}$ increasing). We compare these results to those obtained from a set of alternative penalized regressions. In addition to the BMIDAS-AL, we consider the following penalized MIDAS models, in the spirit of \citet{Uematsu2019}: the Lasso (L), the Elastic-Net (EN; \citealp{Zou2005}), and two folded-concave penalizations, such as the smoothly clipped absolute deviation (SCAD; \citealp{Fan2001}) and the minimax concave penalty (MC$+$; \citealp{Zhang2010}). In particular, the SCAD and MC$+$ penalties do not require the irrepresentable condition to achieve selection consistency and can attenuate the estimation bias problem of convex penalty functions \citep{Fan2011}. Compared to these alternative penalized regressions, our models seem to provide a substantially higher predictive performance. Results look quite significant for DGPs 1 and 3, irrespective of the degree of sparsity and the correlation in the design matrix, but appear less clear-cut for DGP 2. Similarly to \citet{Uematsu2019}, we point out that the Lasso often performs best among the set of competing penalties. Taken jointly, these findings are supportive of the proposed Adaptive Group Lasso prior in the present penalized mixed-frequency framework, and suggest that the chosen modeling approach, although computationally demanding, would likely pay off in terms of predictive accuracy.

\section{Empirical application: forecasting U.S. GDP}\label{sec:emp_app}
We apply the proposed Bayesian MIDAS penalized regression to U.S. GDP data. Following the literature, we consider the annualized quarterly growth rate of GDP, $y_{t}=4\log(Y_{t}/Y_{t-1})$. As for the predictors, we consider 125 macroeconomic series sampled at monthly frequency and extracted from the FRED-MD database \citep{McCracken2016}.\footnote{We consider the entire FRED-MD database (at the time of writing), except for the series of non-borrowed reserves of depository institutions, because of the extreme changes observed since 2008 (see \citealp{Uematsu2019}), as well as the effective Federal Funds rate and the spread with the 10-year government bond rate, as we already use these series at a daily frequency.}
Further, we also consider a set of daily and weekly financial data, which have proven to improve short- to medium-term macro forecasts \citep{Andreou2013,Pettenuzzo2016,Adrian2019}: the effective Federal Funds rate; the interest rate spread between the 10-year government bond rate and the Federal Funds rate; returns on the portfolio of small minus big stocks considered by \citet{Fama1993}; returns on the portfolio of high minus low book-to-market ratio stocks studied by \citet{Fama1993}; returns on a winner minus loser momentum spread portfolio; the Chicago Fed National Financial Conditions Index (NFCI), and in particular its three sub-indexes (risk, credit, and leverage). Finally, we consider the Aruoba-Diebold-Scotti (ADS) daily business conditions index \citep{Aruoba2009} to track the real business cycle at high frequency. Overall, the total number of predictors entering the models is $K=134$. The data sample starts in 1980Q1, and we set $\underline{T}=\text{2000Q1}$ and $\overline{T}=\text{2017Q4}$ the first and last out-of-sample observations, respectively. Estimates are carried-out recursively using an expanding window, and $h$-step-ahead posterior predictive densities are generated from \eqref{eq:pred_dens} through a direct forecast approach. We hence dispose of $(\overline{T}-\underline{T}+1)=72$ out-of-sample observations. We consider several forecast horizons, leading to a sequence of 3 nowcasting ($h=0,1/3,2/3$) and 5 short- and medium-term forecasting ($h=1,4/3,5/3,2,4$) exercises. We keep the empirical application as much realistic as possible, but for ease of analysis we do not take into account real-time issues (ragged/jagged-edge data and revisions). The dataset is hence compiled using the latest vintages available at the time of writing.

Forecasts are compared to those from a benchmark model represented by a simple random-walk (RW). Point and density forecasts are evaluated again by the means of relative RMSFE ratios, average LS differentials, and average CRPS ratios. For both the RMSFE and the CRPS, values less than one shall indicate that our penalized mixed-frequency models outperform (in either point or density forecast sense) the RW. For the LS, positive values shall indicate that our models produce more accurate density forecasts than the RW.
To account for sample uncertainty underlying the observed forecast differences, we report results for the \citet{Diebold1995} and \citet{West1996} test (DMW hereafter), which posits the null hypothesis of unconditional equal predictive accuracy. The resulting test statistic is computed using HAC standard errors (for $h>1$) and a small-sample adjustment to the consistent estimate of the variance, and compared with critical values from the Student's $t$ distribution with $(\overline{T}-\underline{T})$ degrees of freedom \citep{Harvey1997}.
As a robustness check, we further consider forecasts from the following competing models:
\begin{itemize}
\item AR(1) model.
\item BMIDAS-AL model.
\item Penalized MIDAS regressions with Lasso, Elastic-Net, SCAD, and MC$+$ penalties. Regularization parameters are tuned by optimizing the BIC over a fine grid of values.
\item Combination of $K$ single-indicator Bayesian MIDAS models (BMIDAS-comb) as in \citet{Pettenuzzo2016}. The combination weights are computed using a discounted version of the optimal prediction pool proposed by \citet{Geweke2011}, with discount factor $\delta=0.9$ \citep{Stock2004,Andreou2013}.
\item Bayesian Model Averaging and Model Selection with MIDAS (BMA-MIDAS and BMS-MIDAS, respectively). We consider a standard $g$-BRIC prior and a reversible-jump MC$^{3}$ algorithm, the latter modified to account for groups of lag polynomials in the addition/deletion/swaption moves. This ensures that model proposals are based on selection of individual predictors rather than isolated terms of lag polynomials.
\item Dynamic Factor and \textit{targeted}-Factor Bayesian MIDAS regressions (Factor-BMIDAS and $t$-Factor-BMIDAS, respectively). We set the number of dynamic factors as in \citet{Bai2002,Bai2007}. Targeted predictors are selected using the Elastic-Net soft-thresholding rule \citep{Bai2008}. Factors are estimated using either principal components (PCA) or the quasi-maximum likelihood approach (QML) of \citet{Doz2012}.
\end{itemize}

\begin{sidewaystable}
\footnotesize
\caption{Out-of-sample results: relative accuracy of point and density forecasts}
\centering
\setlength{\tabcolsep}{0.13cm}
\begin{tabular}{l|ccc|ccc|ccc|ccc}
\hline\hline
 & \multicolumn{3}{c|}{}  & \multicolumn{3}{c|}{}  & \multicolumn{3}{c|}{}  & \multicolumn{3}{c}{} \\[-1.0ex]
 & RMSFE & LS & CRPS & RMSFE & LS & CRPS & RMSFE & LS & CRPS & RMSFE & LS & CRPS \\[1.5ex]
\hline
 & \multicolumn{3}{c|}{$h=0$} & \multicolumn{3}{c|}{$h=1/3$} & \multicolumn{3}{c|}{$h=2/3$} & \multicolumn{3}{c}{$h=1$} \\
\hline
RW & $\mathit{0.028}$ & $\mathit{2.138}$ & $\mathit{0.016}$ & $\mathit{0.028}$ & $\mathit{2.138}$ & $\mathit{0.016}$ & $\mathit{0.028}$ & $\mathit{2.138}$ & $\mathit{0.016}$ & $\mathit{0.029}$ & $\mathit{2.069}$ & $\mathit{0.016}$ \\
\hline
BMIDAS-AGL & 0.60 & 0.49 & 0.60 & 0.53 & 0.59 & 0.53 & 0.65 & 0.45 & 0.64 & 0.78 & 0.31 & 0.76 \\
BMIDAS-AGL-SS & \textbf{0.59} & \textbf{0.55} & \textbf{0.58} & \textbf{0.51} & \textbf{0.70} & \textbf{0.50} & \textbf{0.55} & \textbf{0.62} & \textbf{0.54} & \textbf{0.72} & \textbf{0.40} & \textbf{0.71} \\
\hline
AR(1) & 0.85 & 0.17 & 0.81 & 0.85 & 0.17 & 0.81 & 0.85 & 0.17 & 0.81 & 0.84 & 0.19 & 0.80 \\
BMIDAS-AL & 0.59 & 0.50 & 0.59 & 0.56 & 0.55 & 0.57 & 0.70 & 0.40 & 0.68 & 0.82 & 0.28 & 0.79 \\
MIDAS-L & 0.80 & 0.23 & 0.77 & 0.75 & 0.29 & 0.74 & 0.83 & 0.19 & 0.80 & 0.80 & 0.24 & 0.77 \\
MIDAS-EN & 0.80 & 0.23 & 0.77 & 0.75 & 0.29 & 0.74 & 0.83 & 0.19 & 0.80 & 0.80 & 0.25 & 0.77 \\
MIDAS-SCAD & 0.63 & 0.48 & 0.62 & 0.76 & 0.34 & 0.73 & 0.74 & 0.30 & 0.73 & 0.80 & 0.24 & 0.77 \\
MIDAS-MC$+$ & 0.68 & 0.36 & 0.69 & 0.67 & 0.44 & 0.66 & 0.68 & 0.40 & 0.67 & 0.81 & 0.30 & 0.77 \\
BMIDAS-comb & 0.64 & 0.45 & 0.64 & 0.66 & 0.45 & 0.65 & 0.68 & 0.44 & 0.65 & 0.78 & 0.33 & 0.75 \\
BMA-MIDAS & 0.61 & 0.52 & 0.60 & 0.59 & 0.56 & 0.58 & 0.55 & 0.61 & 0.54 & 0.81 & 0.25 & 0.81 \\
BMS-MIDAS & 0.62 & 0.51 & 0.60 & 0.59 & 0.54 & 0.59 & 0.59 & 0.57 & 0.58 & 0.82 & 0.23 & 0.83 \\
Factor-BMIDAS PCA & 0.67 & 0.43 & 0.67 & 0.67 & 0.46 & 0.66 & 0.66 & 0.48 & 0.65 & 0.77 & 0.35 & 0.75 \\
Factor-BMIDAS QML & 0.71 & 0.39 & 0.70 & 0.73 & 0.39 & 0.71 & 0.71 & 0.42 & 0.69 & 0.88 & 0.29 & 0.81 \\
$t$-Factor-BMIDAS PCA & 0.66 & 0.45 & 0.66 & 0.62 & 0.50 & 0.62 & 0.65 & 0.46 & 0.66 & 0.74 & 0.36 & 0.74 \\
$t$-Factor-BMIDAS QML & 0.63 & 0.49 & 0.62 & 0.61 & 0.54 & 0.60 & 0.64 & 0.50 & 0.63 & 0.74 & 0.33 & 0.76 \\
\hline
\multicolumn{13}{c}{} \\[3ex]
\hline
& \multicolumn{3}{c|}{$h=4/3$} & \multicolumn{3}{c|}{$h=5/3$} & \multicolumn{3}{c|}{$h=2$} & \multicolumn{3}{c}{$h=4$} \\
\hline
RW & $\mathit{0.029}$ & $\mathit{2.069}$ & $\mathit{0.016}$ & $\mathit{0.029}$ & $\mathit{2.069}$ & $\mathit{0.016}$ & $\mathit{0.033}$ & $\mathit{1.967}$ & $\mathit{0.018}$ & $\mathit{0.036}$ & $\mathit{1.872}$ & $\mathit{0.020}$ \\
\hline
BMIDAS-AGL & 0.91 & 0.20 & 0.89 & 1.16 & 0.08 & 1.05 & 0.87 & 0.18 & 0.86 & 1.05 & -0.03 & 1.05 \\
BMIDAS-AGL-SS & 0.83 & 0.24 & 0.82 & 0.82 & 0.24 & 0.80 & 0.76 & 0.28 & 0.74 & 0.77 & 0.29 & 0.76 \\
\hline
AR(1) & 0.84 & 0.19 & 0.80 & 0.84 & 0.19 & 0.80 & 0.79 & 0.24 & 0.76 & 0.73 & 0.31 & 0.70 \\
BMIDAS-AL & 1.01 & 0.23 & 0.90 & 1.32 & 0.10 & 1.06 & 0.84 & 0.23 & 0.82 & 1.01 & -0.64 & 1.00 \\
MIDAS-L & 0.83 & 0.19 & 0.80 & 0.82 & 0.22 & 0.78 & 0.79 & 0.24 & 0.76 & \textbf{0.71} & 0.33 & \textbf{0.68} \\
MIDAS-EN & 0.83 & 0.19 & 0.80 & 0.82 & 0.22 & 0.78 & 0.79 & 0.25 & 0.76 & 0.71 & 0.32 & 0.69 \\
MIDAS-SCAD & 0.83 & 0.19 & 0.80 & 0.82 & 0.22 & 0.78 & 0.79 & 0.24 & 0.76 & 0.71 & 0.33 & 0.68 \\
MIDAS-MC$+$ & 0.76 & 0.29 & 0.75 & 0.85 & 0.21 & 0.80 & 0.79 & 0.24 & 0.76 & 0.72 & 0.33 & 0.69 \\
BMIDAS-comb & 0.81 & 0.24 & 0.80 & 0.88 & 0.17 & 0.86 & 0.82 & 0.21 & 0.81 & 0.72 & \textbf{0.35} & 0.71 \\
BMA-MIDAS & 0.84 & 0.22 & 0.83 & 0.82 & 0.13 & 0.78 & 0.81 & 0.24 & 0.79 & 0.75 & 0.31 & 0.75 \\
BMS-MIDAS & 0.85 & 0.19 & 0.84 & 0.89 & -0.02 & 0.86 & 0.81 & 0.22 & 0.79 & 0.75 & 0.30 & 0.74 \\
Factor-BMIDAS PCA & 0.86 & 0.27 & 0.81 & 0.96 & 0.21 & 0.87 & 0.92 & 0.18 & 0.85 & 0.83 & 0.20 & 0.81 \\
Factor-BMIDAS QML & 0.91 & 0.22 & 0.87 & 0.95 & 0.18 & 0.91 & 0.97 & 0.13 & 0.93 & 0.91 & 0.14 & 0.88 \\
$t$-Factor-BMIDAS PCA & 0.76 & \textbf{0.32} & \textbf{0.75} & 0.81 & 0.25 & 0.79 & \textbf{0.74} & 0.30 & \textbf{0.72} & 0.72 & 0.32 & 0.70 \\
$t$-Factor-BMIDAS QML & \textbf{0.75} & 0.32 & 0.76 & \textbf{0.78} & \textbf{0.29} & \textbf{0.78} & 0.74 & \textbf{0.31} & 0.74 & 0.73 & 0.32 & 0.70 \\
\hline\hline
\multicolumn{13}{l}{\scriptsize \parbox[t]{18.0cm}{Notes: RMSFE, LS, and CRPS denote respectively the root mean squared forecast error, the log-score, and the continuously ranked probability score, in relative terms with respect to the random-walk (RW) benchmark. Values for the RW (in italic) denote absolute outcomes. Bold values denote the best outcomes.}}
\end{tabular}
\label{t:ea_results1}
\end{sidewaystable}

\begin{sidewaystable}
\footnotesize
\caption{Out-of-sample results: DMW test for equal predictive accuracy with respect to the BMIDAS-AGL-SS model}
\centering
\setlength{\tabcolsep}{0.15cm}
\begin{tabular}{l|ccc|ccc|ccc|ccc}
\hline\hline
 & \multicolumn{3}{c|}{}  & \multicolumn{3}{c|}{}  & \multicolumn{3}{c|}{}  & \multicolumn{3}{c}{} \\[-1.0ex]
 & RMSFE & LS & CRPS & RMSFE & LS & CRPS & RMSFE & LS & CRPS & RMSFE & LS & CRPS \\[1.5ex]
\hline
 & \multicolumn{3}{c|}{$h=0$} & \multicolumn{3}{c|}{$h=1/3$} & \multicolumn{3}{c|}{$h=2/3$} & \multicolumn{3}{c}{$h=1$} \\
\hline
RW & \textbf{0.00} & \textbf{0.00} & \textbf{0.00} & \textbf{0.00} & \textbf{0.00} & \textbf{0.00} & \textbf{0.00} & \textbf{0.00} & \textbf{0.00} & \textbf{0.00} & \textbf{0.00} & \textbf{0.00} \\
AR(1) & \textbf{0.03} & \textbf{0.00} & \textbf{0.00} & \textbf{0.01} & \textbf{0.00} & \textbf{0.00} & \textbf{0.01} & \textbf{0.00} & \textbf{0.00} & 0.18 & \textbf{0.02} & 0.14 \\
BMIDAS-AL & 0.49 & 0.12 & 0.27 & \textbf{0.02} & \textbf{0.00} & \textbf{0.00} & \textbf{0.00} & \textbf{0.00} & \textbf{0.00} & \textbf{0.08} & \textbf{0.02} & \textbf{0.06} \\
MIDAS-L & \textbf{0.03} & \textbf{0.00} & \textbf{0.01} & \textbf{0.00} & \textbf{0.00} & \textbf{0.00} & \textbf{0.02} & \textbf{0.00} & \textbf{0.00} & 0.22 & \textbf{0.05} & 0.21 \\
MIDAS-EN & \textbf{0.03} & \textbf{0.00} & \textbf{0.01} & \textbf{0.00} & \textbf{0.00} & \textbf{0.00} & \textbf{0.02} & \textbf{0.00} & \textbf{0.00} & 0.23 & \textbf{0.05} & \textbf{0.21} \\
MIDAS-SCAD & \textbf{0.03} & \textbf{0.03} & \textbf{0.03} & \textbf{0.00} & \textbf{0.00} & \textbf{0.00} & \textbf{0.00} & \textbf{0.00} & \textbf{0.00} & 0.22 & \textbf{0.05} & 0.21 \\
MIDAS-MC$+$ & \textbf{0.01} & \textbf{0.00} & \textbf{0.00} & \textbf{0.00} & \textbf{0.00} & \textbf{0.00} & \textbf{0.00} & \textbf{0.00} & \textbf{0.00} & 0.19 & 0.12 & 0.18 \\
BMIDAS-comb & \textbf{0.04} & \textbf{0.00} & \textbf{0.01} & \textbf{0.00} & \textbf{0.00} & \textbf{0.00} & \textbf{0.02} & \textbf{0.00} & \textbf{0.00} & 0.23 & 0.14 & 0.28 \\
BMA-MIDAS & 0.12 & 0.16 & 0.13 & \textbf{0.00} & \textbf{0.00} & \textbf{0.00} & 0.33 & 0.31 & 0.43 & \textbf{0.05} & \textbf{0.03} & \textbf{0.04} \\
BMS-MIDAS & 0.11 & 0.11 & 0.11 & \textbf{0.00} & \textbf{0.00} & \textbf{0.00} & \textbf{0.06} & \textbf{0.08} & \textbf{0.06} & \textbf{0.03} & \textbf{0.02} & \textbf{0.02} \\
Factor-BMIDAS PCA & \textbf{0.03} & \textbf{0.01} & \textbf{0.01} & \textbf{0.00} & \textbf{0.00} & \textbf{0.00} & \textbf{0.01} & \textbf{0.00} & \textbf{0.00} & 0.19 & 0.15 & 0.21 \\
Factor-BMIDAS QML & \textbf{0.00} & \textbf{0.00} & \textbf{0.00} & \textbf{0.01} & \textbf{0.00} & \textbf{0.00} & \textbf{0.00} & \textbf{0.00} & \textbf{0.00} & \textbf{0.09} & 0.11 & \textbf{0.10} \\
$t$-Factor-BMIDAS PCA & \textbf{0.08} & \textbf{0.06} & \textbf{0.03} & \textbf{0.01} & \textbf{0.00} & \textbf{0.00} & \textbf{0.01} & \textbf{0.01} & \textbf{0.00} & 0.36 & 0.27 & 0.31 \\
$t$-Factor-BMIDAS QML & \textbf{0.09} & \textbf{0.08} & \textbf{0.08} & \textbf{0.00} & \textbf{0.00} & \textbf{0.00} & \textbf{0.01} & \textbf{0.00} & \textbf{0.00} & 0.34 & 0.13 & 0.19 \\
\hline
\multicolumn{13}{c}{} \\[3ex]
\hline
& \multicolumn{3}{c|}{$h=4/3$} & \multicolumn{3}{c|}{$h=5/3$} & \multicolumn{3}{c|}{$h=2$} & \multicolumn{3}{c}{$h=4$} \\
\hline
RW & \textbf{0.02} & \textbf{0.00} & \textbf{0.00} & \textbf{0.02} & \textbf{0.00} & \textbf{0.00} & \textbf{0.06} & \textbf{0.05} & \textbf{0.01} & \textbf{0.06} & \textbf{0.02} & \textbf{0.03} \\
AR(1) & 0.47 & 0.25 & 0.59 & 0.39 & 0.17 & 0.45 & 0.23 & \textbf{0.09} & 0.24 & 0.66 & 0.61 & 0.78 \\
BMIDAS-AL & 0.15 & 0.45 & 0.18 & \textbf{0.09} & \textbf{0.07} & \textbf{0.05} & \textbf{0.00} & 0.14 & \textbf{0.00} & \textbf{0.00} & \textbf{0.10} & \textbf{0.00} \\
MIDAS-L & 0.49 & 0.26 & 0.61 & 0.52 & 0.32 & 0.60 & 0.24 & 0.12 & 0.27 & 0.81 & 0.72 & 0.88 \\
MIDAS-EN & 0.49 & 0.26 & 0.61 & 0.53 & 0.33 & 0.61 & 0.25 & 0.14 & 0.29 & 0.76 & 0.68 & 0.85 \\
MIDAS-SCAD & 0.49 & 0.26 & 0.61 & 0.52 & 0.32 & 0.60 & 0.24 & 0.12 & 0.27 & 0.81 & 0.72 & 0.88 \\
MIDAS-MC$+$ & 0.88 & 0.82 & 0.87 & 0.34 & 0.31 & 0.47 & 0.22 & \textbf{0.08} & 0.22 & 0.70 & 0.68 & 0.81 \\
BMIDAS-comb & 0.65 & 0.49 & 0.63 & 0.16 & \textbf{0.08} & 0.13 & 0.13 & 0.13 & \textbf{0.08} & 0.82 & 0.83 & 0.82 \\
BMA-MIDAS & 0.37 & 0.26 & 0.34 & 0.52 & 0.22 & 0.60 & 0.27 & 0.36 & 0.22 & 0.81 & 0.89 & 0.84 \\
BMS-MIDAS & 0.19 & 0.11 & 0.20 & 0.13 & 0.12 & 0.15 & 0.26 & 0.24 & 0.23 & 0.82 & 0.72 & 0.83 \\
Factor-BMIDAS PCA & 0.34 & 0.68 & 0.55 & 0.17 & 0.36 & 0.24 & 0.19 & 0.21 & 0.17 & 0.22 & 0.16 & 0.29 \\
Factor-BMIDAS QML & 0.16 & 0.40 & 0.19 & 0.12 & 0.22 & \textbf{0.09} & 0.14 & 0.19 & \textbf{0.09} & \textbf{0.09} & \textbf{0.07} & 0.12 \\
$t$-Factor-BMIDAS PCA & 0.90 & 0.93 & 0.92 & 0.61 & 0.55 & 0.53 & 0.87 & 0.66 & 0.76 & 0.76 & 0.68 & 0.80 \\
$t$-Factor-BMIDAS QML & 0.92 & 0.87 & 0.89 & 0.73 & 0.74 & 0.62 & 0.65 & 0.66 & 0.46 & 0.68 & 0.72 & 0.79 \\
\hline\hline
\multicolumn{13}{l}{\scriptsize \parbox[t]{17.7cm}{Notes: $p$-values of the DMW test for unconditional equal predictive accuracy of model $i$ compared to the BMIDAS-AGL-SS model, according to the one-sided $t$-statistic version of the test. Bold values denote $p$-values $\le 10\%$.}}
\end{tabular}
\label{t:ea_results2}
\end{sidewaystable}

All the models considered in the application include one lag of the growth rate of GDP, which is hence excluded from the selection procedures. To match the sample frequencies, we consider again a restricted Almon lag polynomial, with $p=3$ and $r=2$ endpoint restrictions, and twelve months of past high-frequency observations. As for the MCMC, the Gibbs sampler is run for $S=600,000$ iterations, with the first $200,000$ used as a burn-in period, and every 10th draw is saved. For the BMA/BMS model, we increase the number of iterations to $4,000,000$, in order to let the algorithm sufficiently explore the model space, which is fairly vast in the current application.

Results are reported in Tables \ref{t:ea_results1} and \ref{t:ea_results2}. In the first row of Table \ref{t:ea_results1}, we report RMSFE, LS and CRPS for the benchmark RW, while in the other rows we report the relative scores for all models compared to the RW. For the short horizons between $h=0$ and $h=1$ (\textit{i.e.} nowcast and 1-step-ahead forecast), the results suggest that the BMIDAS-AGL-SS model systematically outperforms all the competitors, with point and density predictive gains often statistically significant (at 10\% level; see Table \ref{t:ea_results2}). The best competing results are given by the BMA/BMS-MIDAS models for $h=0,1/3,2/3$ and the $t$-Factor-BMIDAS models for $h=1$. The predictive gains provided by the BMIDAS-AGL-SS are admittedly small compared to these best competing models, but often fairly large compared to the other models considered. For $h>1$, the BMIDAS-AGL-SS model cannot provide the best predictive outcomes, but it is often ranked among the best models and can often outperform a large number of competitors, although the number of rejections in the DMW test edges down to only a few. Conversely, results for the BMIDAS-AGL model are overall more disappointing, as the model seems to perform reasonably well only up to $h=1$.

A number of additional interesting features arise from the out-of-sample analysis. First, results for the competing penalized MIDAS regressions (Lasso, EN, SCAD, and MC$+$) are very similar, and sometimes virtually identical, to each other but also fairly volatile, as their predictive performance can deteriorate substantially from one forecast horizon to another. Second, for $h>1$ best outcomes are provided by targeted-Factor models, except for $h=4$ where the MIDAS-L regression performs slightly better. The AR is never ranked first but, for relatively long horizons, this model becomes hard to beat, which is broadly in line with the empirical findings reported in previous studies. Finally, Factor MIDAS regressions and MIDAS penalized regressions seem to perform quite similarly, overall. These findings are broadly in line with those reported by \citet{Uematsu2019}, but they additionally reveal that some prior sparse selection (\textit{i.e.} targeting) seems necessary to let Factor models systematically outperform penalized regressions. Nevertheless, to our opinion these findings lack of generality on this point, leaving the debate over sparse $vs$ dense modelling in presence of possibly highly correlated predictors still open \citep{Giannone2017}.

\begin{sidewaysfigure}
\caption{Inclusion probabilities for the BMIDAS-AGL-SS model}
\begin{center}
\includegraphics[scale=0.60]{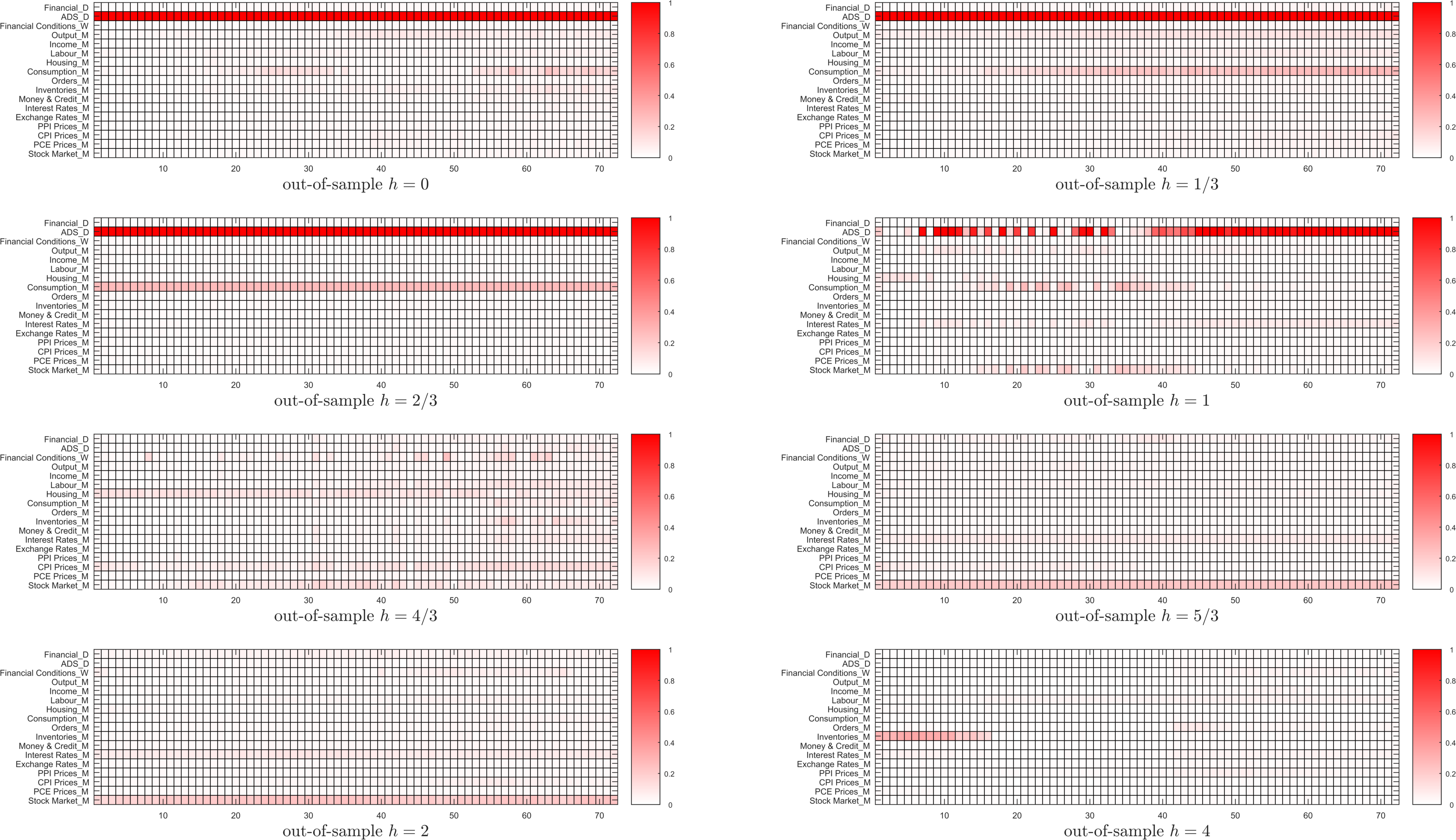}
\end{center}
\label{fig:prob_selection}
\end{sidewaysfigure}

Figure \ref{fig:prob_selection} reports the variables inclusion probabilities obtained from the BMIDAS-AGL-SS model. Given the large number of variables considered in the application, for ease of exposition we aggregate these probabilities according to the nature of the regressors and/or their frequency, as well as to the classification used by \citet{McCracken2016}. The patterns reported in the figure show a systematic inclusion with very high probability of the ADS index for all forecast horizons between $h=0$ and $h=1$. The model tends to select also a bunch of high-frequency predictors related to consumption, output, and inventories, but with much lower probability. This is not unsurprising, as the ADS index already contains signals stemming from the real economy (initial jobless claims, real GDP, payroll employment, industrial production, real personal income less transfers, and real manufacturing and trade sales). The housing market seems to play a very limited role in the model ($h=1$), while virtually no financial indicators are selected for nowcasting purposes. However, this feature tends to progressively attenuate for $h>1$, and financial variables (financial condition indexes and stock market) are selected with somewhat higher probability up to $h=2$. This result seems broadly in line with recent literature \citep{Andreou2013} and suggests that financial variables may convey some, although here very limited, short-term leading information which goes beyond the predictive content of real indicators.

\section{Concluding remarks}\label{sec:conclusions}
We propose a new approach to modeling and forecasting mixed-frequency regressions (MIDAS) that addresses the issue of simultaneously estimating and selecting relevant high-frequency predictors in a high-dimensional environment. Our approach is based on MIDAS regressions resorting to Almon lag polynomials and an adaptive penalized regression approach, namely the Group Lasso objective function. The proposed models rely on Bayesian techniques for estimation and inference. In particular, the penalty hyper-parameters driving the model shrinkage are automatically tuned via an Empirical Bayes algorithm based on stochastic approximations. We establish the posterior contraction rate for the in-sample and the out-of-sample prediction error, as well as the consistency of the marginal posterior of parameters. Simulations show that the proposed models present very good in-sample and out-of-sample performance. When applied to a forecasting model of the U.S. GDP growth with high-frequency real and financial predictors, the results suggest that our models produce significant short-term predictive gains compared to several alternative models. Our findings point to a very limited short-term predictive content of high-frequency financial variables, which is broadly in line with the existing literature.

The models presented in the present paper could be extended in several ways, such as time-varying lag polynomials with stochastic volatility error dynamics \citep{Carriero2015, Pettenuzzo2016}, as well as quantile mixed-frequency regressions \citep{Lima2020}. Further, recent research has been focusing on factor-adjusted sparse regressions to deal with high-dimensional regressions and highly correlated predictors \citep{Kneip2011,Fan2020}. We believe that these extensions to our models represent interesting paths for future research.

\clearpage
\bibliographystyle{elsarticle-harv}
\begin{singlespace}
\bibliography{BMIDAS_AGL_ArXiv_V3}
\end{singlespace}

\clearpage
\renewcommand{\theequation}{A\arabic{equation}}
\renewcommand{\thetable}{A\arabic{table}}
\renewcommand{\thefigure}{A\arabic{figure}}
\renewcommand{\thesubsection}{A.\arabic{subsection}}
\setcounter{equation}{0}  
\setcounter{table}{0}
\setcounter{figure}{0}
\setcounter{subsection}{0}

\section*{Appendix}

\subsection{Marginal prior for the Bayesian Adaptive Group Lasso}\label{sec:marg_prior_appendix}
In this section we detail the computations to get the marginal prior for $\btheta$ conditional on $\sigma^2$ for the prior given in Section \ref{sec:BMIDAS-AGL} or conditional on $(\sigma^2,\pi_0)$ for the prior given in Section \ref{sec:BMIDAS-AGL-SS}. Given the specification of these priors, we have to marginalise the $\mathcal{N}(\mathbf{0},\sigma^2\tau_{j}^{2}\mathbf{I}_{g_{j}})$ distribution with respect to the prior of $\tau_j^2$:
\begin{multline*}
  \bigintss_{0}^{\infty} \left(\frac{1}{2\pi\sigma^2 \tau_j^2}\right)^{g_j/2}\exp\{-\|\btheta_j\|_2^2/(2\sigma^2 \tau_j^2)\}\left(\frac{\lambda_j^2}{2}\right)^{(g_j+1)/2}\frac{\tau_j^{g_j + 1 - 2}}{\Gamma((g_j+1)/2)}\exp\{-\lambda_j^2 \tau_j^2/2\} d\tau_j^2\\
  = \left(\frac{1}{2\pi\sigma^2}\right)^{g_j/2}\left(\frac{\lambda_j^2}{2}\right)^{(g_j+1)/2}\frac{1}{\Gamma((g_j+1)/2)}\\ \hfill
  \times\bigintss_{0}^{\infty} \underbrace{\tau_j^{-1}\exp\left\{-\frac{1}{2}\left(\frac{\|\btheta_j\|_2^2}{\sigma^2 \tau_j^2} + \lambda_j^2 \tau_j^2\right)\right\}}_{= GIG(\tau_j^2;\lambda_j^2,\|\btheta_j\|_2^2/\sigma^2,1/2 )(\lambda_j^2\sigma^2/\|\btheta_j\|_2^2)^{-1/4} 2 K_{1/2}(\lambda_j\|\btheta_j\|_2/\sigma)}d\tau_j^2\\
  = \left(\frac{1}{2\pi\sigma^2}\right)^{g_j/2}\left(\frac{\lambda_j^2}{2}\right)^{(g_j+1)/2}\frac{1}{\Gamma((g_j+1)/2)} (\lambda_j^2\sigma^2/\|\btheta_j\|_2^2)^{-1/4} 2 K_{1/2}(\lambda_j\|\btheta_j\|_2/\sigma)\\
  = \left(\frac{1}{2\pi\sigma^2}\right)^{g_j/2}\left(\frac{\lambda_j^2}{2}\right)^{(g_j+1)/2}\frac{1}{\Gamma((g_j+1)/2)} \left(\frac{\lambda_j^2\sigma^2}{\|\btheta_j\|_2^2}\right)^{-1/4} 2 \sqrt{\frac{\pi\sigma}{2\lambda_j\|\btheta_j\|_2}} \exp\{-\lambda_j\|\btheta_j\|_2/\sigma\}\\
  = \underbrace{\frac{1}{\Gamma((g_j+1)/2)}\pi^{(1-g_j)/2} 2^{-g_j}}_{:=C_j}\left(\frac{1}{\sigma^2}\right)^{g_j/2}\lambda_{j}^{g_j}\exp\{-\lambda_j\|\btheta_j\|_2/\sigma\}\\
  \propto \left(\frac{\lambda_j^2}{\sigma^2}\right)^{g_j/2}\exp\{-\lambda_j\|\btheta_j\|_2/\sigma\}
\end{multline*}
where $GIG(\tau_j^2;a,b,p)$ denotes the pdf of a Generalized Inverse Gaussian distribution with parameters $a$, $b$ and $p$, and $K_{1/2}(\lambda_j\|\btheta_j\|_2/\sigma) = \sqrt{\frac{\pi\sigma}{2\lambda_j\|\btheta_j\|_2}} \exp\{-\lambda_j\|\btheta_j\|_2/\sigma\}$ is a modified Bessel function of the second kind. Remark that the constant $C_j$ can also be written as $C_j = \pi^{-g_j/2}\Gamma(g_j/2)/(2\Gamma(g_j))$ by using the property of the Gamma function $\Gamma(g_j/2 + 1/2) = 2^{1 - g_j}\sqrt{\pi}\Gamma(g_j)/\Gamma(g_j/2)$. Hence, by integrating out $\tau_j^2$ we get:
\begin{eqnarray}\label{eq:marginal:prior1}
  \btheta_j|\sigma^2 & \sim & \textrm{M-Laplace}\left(\btheta_j;\mathbf{0},\frac{\sigma}{\lambda_j}\right),\qquad j=1,\ldots,G
\end{eqnarray}
for the prior of the BMIDAS-AGL model in Section \ref{sec:BMIDAS-AGL}, or
\begin{eqnarray}\label{eq:marginal:prior}
  \btheta_j|\sigma^2,\pi_0 & \sim & (1 - \pi_0)\textrm{M-Laplace}\left(\btheta_j;\mathbf{0},\frac{\sigma}{\lambda_j}\right) + \pi_0\delta_0(\btheta_j),\qquad j=1,\ldots,G
\end{eqnarray}
for the prior of the BMIDAS-AGL-SS model in Section \ref{sec:BMIDAS-AGL-SS}. In both cases, $\textrm{M-Laplace}(\btheta_j;\mathbf{0},\sigma/\lambda_j)$ denotes a $g_j$-dimensional Multi-Laplace distribution.

\subsection{Proofs}\label{App:proofs}
In this section we prove the theorems in Section \ref{sec:Asymptotic_Analysis}. Technical lemmas and further results on dimension recovery and distributional approximation of the posterior of $\btheta$ are reported in the Supplementary Appendix. Recall the notation $y:=(y_1,\ldots,y_T)'$ for the $T$-vector of observations on $y_t$ and $\mathbf{Z}$ for the $T \times \wtl g$ matrix of observations on $\mathbf{Z}_t^{(m)}$. Moreover, hereafter we use the following notation. Let $f$ (resp. $f_0$) denote the joint conditional probability density with respect to the Lebesgue measure of $y$ given $(\mathbf{Z},\btheta,\sigma^2)$ for a generic value of the parameters $(\btheta',\sigma^2)'$ (resp. of $y$ given $(\mathbf{Z},\btheta_0,\sigma_0^2)$ for the true value of the parameters). We denote by $\EE_f[\cdot]$ the expectation taken with respect to the probability distribution with Lebesgue density $f$ and by $\EE_0[\cdot]$ (resp. $\PP_0$) the expectation (resp. the probability) taken with respect to the true conditional distribution of $y$ given $(\mathbf{Z},\btheta_0,\sigma_0^2)$. For two probability densities $f_1$ and $f_2$, the Kullback-Leibler divergence is denoted by $K(f_1,f_2) := \int f_1\log(f_1/f_2)$ while $V(f_1,f_2):= \int f_1(\log(f_1/f_2) - K(f_1,f_2))^2$ denotes the Kullback-Leibler variation. The average R\'{e}nyi divergence of order $1/2$ between two $T$-dimensional densities $f_1$ and $f_2$ is denoted by $d(f_1, f_2) := -\frac{1}{T}\log\int\sqrt{f_1 f_2}$.\\
\indent Let $\wtl  s_0 := \max\{s_0, \log(T)/\log(G)\}$ and, for a vector $\mathbf{v}$, we denote by $s_{\mathbf{v}}$ the number of groups in $\mathbf{v}$ with nonzero components and by $S_{\mathbf{v}}$ the indices of these groups. For two constants $M_2>0$ and $C_2>0$ define the set $\Theta_{\wtl s_0} := \{\btheta\in\Theta; \, s_{\btheta} < M_2 \wtl s_0\}$ and the sieve
\begin{equation}\label{eq:16}
  \mathcal{F}_T(C_2) := \left\{(\btheta,\sigma^2)\in\Theta_{\wtl s_0} \times \mathbb{R}_+; \,\max_{1\leq j \leq G}\|\btheta_j\|_2 \leq \frac{C_2+1}{\underline{c}}\xi,\, T^{-1} \leq \sigma^2 \leq e^{C_2T\epsilon^2}\right\},
\end{equation}
where $\xi := T \wtl s_0\log(G)/\|\mathbf{Z}\|_o$. Let $f = \mathcal{N}_T(\mathbf{Z}\btheta,\sigma^2 \mathbf{I}_T)$ and $f_0 = \mathcal{N}_T(\mathbf{Z}\btheta_0,\sigma_0^2 \mathbf{I}_T)$. The notation $f\in \mathcal{F}_T(C_2)$ must be understood as: $f = \mathcal{N}_T(\mathbf{Z}\btheta,\sigma^2 \mathbf{I}_T)$ with $(\btheta,\sigma^2) \in \mathcal{F}_T(C_2)$. We denote by $N(\eta,\mathcal{F}, \rho)$ the $\eta$-covering number of a set $\mathcal{F}$ with respect to a metric $\rho$, which is the minimal number of $\eta$-balls in the $\rho$-metric needed to cover the set $\mathcal{F}$. 
\subsubsection{Proof of Theorem \ref{thm:1}}
The proof of this theorem follows \cite{Ning2020}. As in the latter, it is made of two parts. In the first part we obtain the posterior contraction rate with respect to the divergence $d(f_0, f)$. In the second part we use this result to show the result of the theorem. However, while the general idea of the proof is the same as in \cite{Ning2020}, many technical intermediate steps are different due to the fact that we have a different prior distribution and different assumptions. We  provide all the details for completeness.\\
\indent Let us consider the set $\Theta_{\wtl s_0} := \{\btheta\in\Theta; s_{\btheta}< M_2 \wtl s_0\}$ with $M_2 > \frac{2 + C_1}{u} + 1$, where $C_1>0$ is a constant that depends only on $a_1,d,c_{\pi_0},\overline{C},\underline{\sigma}^2, c_{\btheta}, c_g$ and that satisfies (B3) in Lemma B.1.2. For every $\varepsilon >0$ let $\mathcal{A}_{\varepsilon}:=\{(\btheta,\sigma^2)\in\Theta_{\wtl s_0}\times \mathbb{R}_+; \, d(f_0,f)> \varepsilon\}$. Then, for every $\varepsilon >0$:
\begin{equation}
  \EE_0 \Pi\left(\left.(\btheta,\sigma^2)\in \mathbb{R}^{\wtl g}\times \mathbb{R}_+; \, d(f_0,f)> \varepsilon\right|y,\mathbf{Z}\right) \leq \EE_0 \Pi\left(\left.\mathcal{A}_{\varepsilon}\right|y,\mathbf{Z}\right) + \EE_0 \Pi\left(\left.\btheta \in\Theta_{\wtl s_0}^c\right|y,\mathbf{Z}\right).\label{eq:12}
\end{equation}
The second term on the right hand side goes to zero by Lemma B.1.4 in the Supplementary Appendix. We then focus on the first term. By Lemma B.1.5 in the Supplementary Appendix there exists a test $\phi_T$ such that for some constant $M_1 > 2M_2(1 + 8c_g) + 4 + 2C_2$ that does not depend on $(\btheta_0,\sigma_0^2)$,
  \begin{equation}\label{proof:test:1}
    \EE_0 \phi_T \leq e^{- M_1 T\epsilon^2/2},\qquad \textrm{and }\qquad \sup_{f\in\mathcal{F}_T(C_2); d(f_0, f) > M_1 \epsilon^2} \EE_f (1 - \phi_T) \leq e^{- M_1 T\epsilon^2}.
  \end{equation}
\noindent Hence, by using the first result in \eqref{proof:test:1} we get the upper bound:
\begin{multline}
  \EE_0 \Pi\left(\left.\mathcal{A}_{\varepsilon}\right|y,\mathbf{Z}\right) = \EE_0 [\Pi\left(\left.\mathcal{A}_{\varepsilon}\right|y,\mathbf{Z}\right) \phi_T] + \EE_0[\Pi\left(\left.\mathcal{A}_{\varepsilon}\right|y,\mathbf{Z}\right)(1 - \phi_T)]\\
  \leq e^{- M_1 T\epsilon^2/2} + \EE_0[\Pi\left(\left.\mathcal{A}_{\varepsilon}\right|y,\mathbf{Z}\right)(1 - \phi_T)]\label{eq:9}
\end{multline}
\noindent and we just need to upper bound the second term in \eqref{eq:9}. To do this and for the constant $C_1$ introduced above, define the event $\mathcal{A}_1 := \{\int (f/f_0) \overline{\Pi}(\btheta,\sigma^2)d\btheta d\sigma^2 \geq \exp\{- (1 + C_1)T\epsilon^2\}\}$, where $\overline{\Pi}(\btheta,\sigma^2)$ denotes the prior in \eqref{eq:marginal:prior} marginalized with respect to the prior of $\pi_0$ and restricted to have support on $\mathcal{B}_0(\epsilon) := \{(\btheta,\sigma^2); K(f_0,f) \leq T\epsilon^2, V(f_0,f)\leq T\epsilon^2\}$. By using the results of Lemmas B.1.1 and B.1.2 in the Supplementary Appendix we obtain that for large $T$:
\begin{multline}
  \EE_0[\Pi\left(\left.\mathcal{A}_{\varepsilon}\right|y,\mathbf{Z}\right)(1 - \phi_T)] = \EE_0\left[\frac{\int_{\mathcal{A}_{\varepsilon}}f/f_0 \Pi(\btheta,\sigma^2)d\btheta d\sigma^2}{\int_{\Theta}\int_{0}^{\infty} f/f_0 \Pi(\btheta,\sigma^2)d\btheta d\sigma^2}(1 - \phi_T)\right]\\
  \leq \EE_0\left[\left.\frac{\int_{\mathcal{A}_{\varepsilon}}f/f_0 \Pi(\btheta,\sigma^2)d\btheta d\sigma^2}{\int_{\mathcal{B}_0(\epsilon)} f/f_0 \overline{\Pi}(\btheta,\sigma^2)d\btheta d\sigma^2\Pi(\mathcal{B}_0(\epsilon))}(1 - \phi_T)\right|\mathcal{A}_1\right]\PP_0(\mathcal{A}_1) + \PP_0(\mathcal{A}_1^c)\\
  \leq \EE_0\left[\int_{\mathcal{A}_{\varepsilon}}f/f_0 \Pi(\btheta,\sigma^2)d\btheta d\sigma^2 e^{(1+2C_1)T\epsilon^2}(1 - \phi_T)\right] + (C_1^2 T \epsilon^2)^{-1}.\label{eq:10}
\end{multline}
\noindent Moreover, let $\mathcal{A}_{\varepsilon,T} := \{(\btheta,\sigma^2)\in \mathcal{F}_T(C_2); \, d(f_0,f)> \varepsilon\}$ where $\mathcal{F}_T(C_2)$ is as defined in \eqref{eq:16} with $C_2 > 2 C_1 + 1$ and remark that $\mathcal{A}_{\varepsilon} \subseteq \mathcal{A}_{\varepsilon,T}\cup \{(\btheta,\sigma^2)\in (\Theta_{\wtl s_0}\times \mathbb{R}_+)\setminus \mathcal{F}_T(C_2)\}$. Then, by letting $\varepsilon = M_1 \epsilon^2$, we get:
\begin{multline}
  \EE_0\left[\int_{\mathcal{A}_{\varepsilon}}f/f_0 \Pi(\btheta,\sigma^2)d\btheta d\sigma^2 e^{(1+2C_1)T\epsilon^2}(1 - \phi_T)\right] = e^{(1+2C_1)T\epsilon^2}\int_{\mathcal{A}_{\varepsilon}} \EE_f\left[ (1 - \phi_T)\right]\Pi(\btheta,\sigma^2)d\btheta d\sigma^2\\
  \leq e^{(1 + 2C_1)T\epsilon^2}\left(\int_{\mathcal{A}_{\varepsilon, T}} \EE_f\left[ (1 - \phi_T)\right]\Pi(\btheta,\sigma^2)d\btheta d\sigma^2 + \int_{(\Theta_{\wtl s_0}\times \mathbb{R}_+)\setminus \mathcal{F}_T(C_2)}\Pi(\btheta,\sigma^2)d\btheta d\sigma^2\right)\\
  \leq e^{(1+2C_1)T\epsilon^2}\left(\sup_{f\in\mathcal{F}_T(C_2); d(f_0, f) > M_1 \epsilon^2}\EE_f\left[ (1 - \phi_T)\right]\int_{\mathcal{A}_{\varepsilon,T}}\Pi(\btheta,\sigma^2)d\btheta d\sigma^2 + e^{-C_2 T\epsilon^2}\left(2 + \frac{b_1}{a_1 - 1}\right)\right)\\
  \leq e^{(1 + 2C_1)T\epsilon^2} e^{- M_1 T\epsilon^2} + e^{-T\epsilon^2(C_2 - 1 - 2C_1)}\left(2 + \frac{b_1}{a_1 - 1}\right)\label{eq:11}
\end{multline}
by using \eqref{proof:test:1} and result (B17) of Lemma B.1.5. By putting together \eqref{eq:12} and \eqref{eq:9}-\eqref{eq:11} and by Lemma B.1.4 we obtain:
\begin{multline}
  \EE_0 \Pi\left(\left.(\btheta,\sigma^2)\in \mathbb{R}^{\wtl g}\times \mathbb{R}_+; \, d(f_0,f)> M_1\epsilon^2\right|y,\mathbf{Z}\right)\\
  \leq e^{- M_1 T\epsilon^2/2} + e^{- (M_1 - 1 - 2C_1) T\epsilon^2} + e^{-T\epsilon^2(C_2 - 1 - 2C_1)}\left(2 + \frac{b_1}{a_1 - 1}\right) + \frac{2}{C_1^2 T \epsilon^2} + e^{-T\epsilon^2(uM_2 - (2 + C_1 + u))}
\end{multline}
\noindent which converges to zero for the values of $M_1, M_2$ and $C_2$ specified above. Finally, since all the constants in the right hand side of \eqref{eq:13:main} do not depend on $(\btheta_0,\sigma_0^2)$ we conclude that
\begin{multline}\label{eq:13:main}
  \sup_{\btheta_0\in\overline\Theta_0,\sigma_0^2\in[\underline{\sigma}^2, \overline{\sigma}^2]}\EE_0 \Pi\left(\left.(\btheta,\sigma^2)\in \mathbb{R}^{\wtl g}\times \mathbb{R}_+; \, d(f_0,f)> M_1\epsilon^2\right|y,\mathbf{Z}\right) \leq e^{- M_1 T\epsilon^2/2} + e^{- (M_1 - 1 - 2C_1) T\epsilon^2}\\
  + e^{-T\epsilon^2(C_2 - 1 - 2C_1)}\left(2 + \frac{b_1}{a_1 - 1}\right) + \frac{2}{C_1^2 T \epsilon^2} + e^{-T\epsilon^2(uM_2 - (2 + C_1 + u))}.
\end{multline}

\indent We now develop the second part of the proof. Denote $\mathcal{A}_2 := \{(\btheta,\sigma^2)\in \mathbb{R}^{\wtl g}\times \mathbb{R}_+; \frac{1}{4}\log\left(\frac{\sigma^2 + \sigma_0^2}{2\sigma_0^2}\right) \leq M_1\epsilon^2\}$. Remark that under Assumption \ref{Ass:1} \textit{(i)} this event is feasible. First, since
\begin{multline}
  d(f_0, f) := -\frac{1}{T}\log \int \sqrt{f_0 f} = -\frac{1}{T}\sum_{t=1}^T \log\int \sqrt{f_{0,t} f_t} \\
  = - \frac{1}{4}\log(\sigma^2\sigma_0^2) + \frac{1}{2} \log\left(\frac{\sigma^2 + \sigma_0^2}{2}\right) + \frac{1}{4(\sigma^2 + \sigma_0^2)}\frac{1}{T}\sum_{t=1}^T [\mathbf{Z}_t'(\btheta - \btheta_0)]^2\\
  = \frac{1}{4}\log\left(\frac{\sigma^2 + \sigma_0^2}{2\sigma_0^2}\right) + \frac{1}{4} \log\left(\frac{\sigma^2 + \sigma_0^2}{2\sigma^2}\right) + \frac{1}{4(\sigma^2 + \sigma_0^2)}\frac{1}{T}\sum_{t=1}^T [\mathbf{Z}_t'(\btheta - \btheta_0)]^2,\label{eq:20:main}
\end{multline}
where $f_t = \mathcal{N}(Z_t\btheta,\sigma^2 )$ and $f_{0,t} = \mathcal{N}(Z_t'\btheta_0,\sigma_0^2)$, then \eqref{eq:13:main} implies that uniformly on $(\btheta_0,\sigma_0^2)\in\overline\Theta_0\times[\underline{\sigma}^2, \overline{\sigma}^2]$:
\begin{align}
  & \EE_0 \Pi\left(\left.\mathcal{A}_2\right|y,\mathbf{Z}\right) \rightarrow 1,\label{eq:14:1}\\
  & \EE_0 \Pi\left(\left.(\btheta,\sigma^2)\in \mathbb{R}^{\wtl g}\times \mathbb{R}_+;  \frac{1}{4(\sigma^2 + \sigma_0^2)}\frac{1}{T}\sum_{t=1}^T [\mathbf{Z}_t'(\btheta - \btheta_0)]^2 \leq M_1\epsilon^2\right|y,\mathbf{Z},\mathcal{A}_2\right) \rightarrow 1.\label{eq:14:2}
\end{align}
Remark that on $\mathcal{A}_2$: $\frac{\sigma^2 + \sigma_0^2}{2\sigma_0^2} \leq e^{4M_1\epsilon^2}$. Then on $\mathcal{A}_2$:
\begin{multline*}
  \|\mathbf{Z}(\btheta - \btheta_0)\|_2^2 = \sum_{t=1}^T [Z_t(\btheta - \btheta_0)]^2 = \frac{1}{4\sigma_0^2\exp{\{4M_1\epsilon^2\}}}\sum_{t=1}^T [\mathbf{Z}_t(\btheta - \btheta_0)]^2 4\sigma_0^2\exp{\{4M_1\epsilon^2\}} \\
  \leq \frac{1}{2(\sigma^2 + \sigma_0^2)}\sum_{t=1}^T [\mathbf{Z}_t(\btheta - \btheta_0)]^2 4\sigma_0^2\exp{\{4M_1\epsilon^2\}}.
\end{multline*}
It follows from this and the law of total probability that
\begin{multline}
  \EE_0 \Pi\left(\left.\btheta\in \mathbb{R}^{\wtl g}; \|\mathbf{Z}(\btheta - \btheta_0)\|_2^2 \leq M T\epsilon^2\right|y,\mathbf{Z}\right)\\
  \geq \EE_0 \Pi\left(\left.\{(\btheta,\sigma^2)\in \mathbb{R}^{\wtl g}\times \mathbb{R}_+; \|\mathbf{Z}(\btheta - \btheta_0)\|_2^2 \leq M T\epsilon^2\} \cap \mathcal{A}_2 \right|y,\mathbf{Z}\right)\\
  \geq \EE_0 \Pi\left(\left.(\btheta,\sigma^2)\in \mathbb{R}^{\wtl g}\times \mathbb{R}_+; \frac{1}{2(\sigma^2 + \sigma_0^2)}\|\mathbf{Z}(\btheta - \btheta_0)\|_2^2 4\sigma_0^2 e^{4M_1\epsilon^2} \leq M T\epsilon^2 \right|y,\mathbf{Z}, \mathcal{A}_2\right)\Pi(\left.\mathcal{A}_2\right|y,\mathbf{Z})\\
  \geq \EE_0 \Pi\left(\left.(\btheta,\sigma^2)\in \mathbb{R}^{\wtl g}\times \mathbb{R}_+; \frac{1}{4(\sigma^2 + \sigma_0^2)}\|\mathbf{Z}(\btheta - \btheta_0)\|_2^2 \leq M_1 T\epsilon^2 \right|y,\mathbf{Z}, \mathcal{A}_2\right)\Pi(\left.\mathcal{A}_2\right|y,\mathbf{Z})\label{eq:19}
\end{multline}
\noindent for every $M \geq 2M_1 4\overline{\sigma}^2e^{4 M_1 \epsilon^2}$. Finally, \eqref{eq:19} converges to one by \eqref{eq:14:1} and \eqref{eq:14:2} uniformly over $(\btheta_0,\sigma_0^2)\in\overline\Theta_0\times[\underline{\sigma}^2, \overline{\sigma}^2]$.

\subsubsection{Proof of Theorem \ref{thm2:contraction:rate:theta}}
By definition of $\Theta_{\wtl s_0}$, for every $\btheta \in \Theta_{\wtl s_0}$ it holds that $s_{(\btheta - \btheta_0)} \leq s_{\btheta} + s_{\btheta_0} \leq M_2 \wtl s_0 + s_0$. Therefore, $\forall \btheta \in \Theta_{\wtl s_0}$
\begin{displaymath}
  \|\mathbf{Z}(\btheta - \btheta_0)\|_2^2 = \frac{\|\mathbf{Z}(\btheta - \btheta_0)\|_2^2}{\|\mathbf{Z}\|_o^2 \|\btheta - \btheta_0\|_2^2} \|\mathbf{Z}\|_o^2 \|\btheta - \btheta_0\|_2^2 \geq \wtl\phi(M_2 \wtl s_0 + s_0)\|\mathbf{Z}\|_o^2 \|\btheta - \btheta_0\|_2^2.
\end{displaymath}
By using this inequality to get the inequality in the fourth line below, we obtain:
\begin{multline*}
  \EE_0 \Pi\left(\left.\btheta\in \mathbb{R}^{\wtl g}; \|\btheta - \btheta_0\|_2^2 \geq \frac{M T\epsilon^2}{\wtl\phi(s_0 + M_2\wtl s_0)\|\mathbf{Z}\|_o^2}\right|y,\mathbf{Z}\right)\\
  = \EE_0 \Pi\left(\left.\btheta\in \Theta_{\wtl s_0}; \|\btheta - \btheta_0\|_2^2 \geq \frac{M T\epsilon^2}{\wtl\phi(s_0 + M_2\wtl s_0)\|\mathbf{Z}\|_o^2}\right|y,\mathbf{Z}\right)\\ \hfill
  + \EE_0 \Pi\left(\left.\btheta\in \Theta_{\wtl s_0}^c; \|\btheta - \btheta_0\|_2^2 \geq \frac{M T\epsilon^2}{\wtl\phi(s_0 + M_2\wtl s_0)\|\mathbf{Z}\|_o^2}\right|y,\mathbf{Z}\right)\\
  \leq \EE_0 \Pi\left(\left.\btheta\in \mathbb{R}^{\wtl g}; \frac{\|\mathbf{Z}(\btheta - \btheta_0)\|_2^2}{\wtl\phi(M_2 \wtl s_0 + s_0)\|\mathbf{Z}\|_o^2 } \geq \frac{M T\epsilon^2}{\wtl\phi(s_0 + M_2\wtl s_0)\|\mathbf{Z}\|_o^2}\right|y,\mathbf{Z}\right) + \EE_0 \Pi\left(\left.\Theta_{\wtl s_0}^c\right|y,\mathbf{Z}\right).
\end{multline*}
\noindent The two terms on the right hand side converge to zero uniformly over $(\btheta_0,\sigma_0^2)\in\overline\Theta_0\times[\underline{\sigma}^2, \overline{\sigma}^2]$ due to Theorem \ref{thm:1} and Lemma B.1.4 in the Supplementary Appendix.

\subsubsection{Proof of Theorem \ref{thm4:out-of-sample1}}
First, remark that $\|\mathbf{Z}\|_0^2 \|\btheta - \btheta\|_2^2 = \max_{1 \leq j \leq G} \|\mathbf{Z}_j\|_{op}^2 \sum_{j=1}^G \|(\btheta - \btheta_0)_j\|_2^2 \geq \sum_{j=1}^G \|\mathbf{Z}_j\|_{op}^2 \|(\btheta - \btheta_0)_j\|_2^2 \geq \sum_{j=1}^G \|\mathbf{Z}_j' (\btheta - \btheta_0)_j\|_2^2$. On the set $\Theta_{\wtl s_0}$ the following upper bound holds:
\begin{multline*}
  |\mathbf{Z}_{\tau - h}'(\btheta - \btheta_0)|^2 = \wtl \phi(s_0 + M_2 \wtl s_0) \frac{|\mathbf{Z}_{\tau - h}'(\btheta - \btheta_0)|^2}{\wtl \phi(s_0 + M_2 \wtl s_0)}
  \leq \frac{\|\mathbf{Z}(\btheta - \btheta_0)\|_{2}^2}{T\|\mathbf{Z}/\sqrt{T}\|_o^2 \|\btheta - \btheta_0\|_2^2} \frac{|\mathbf{Z}_{\tau - h}'(\btheta - \btheta_0)|^2}{\wtl \phi(s_0 + M_2 \wtl s_0)}\\ \hfill
  \leq \frac{\|\mathbf{Z}(\btheta - \btheta_0)\|_{2}^2}{\wtl \phi(s_0 + M_2 \wtl s_0)}\frac{|\mathbf{Z}_{\tau - h}'(\btheta - \btheta_0)|^2}{T\sum_{j=1}^G \|\mathbf{Z}_j' (\btheta - \btheta_0)_j/\sqrt{T}\|_2^2}\\
  \leq \frac{\|\mathbf{Z}(\btheta - \btheta_0)\|_{2}^2}{T\wtl \phi(s_0 + M_2 \wtl s_0)} \sup_{\{\btheta; \,0 \leq s_{\btheta} \leq M_2 \wtl s_0 + s_0\}}\frac{|\mathbf{Z}_{\tau - h}'\btheta|^2}{\sum_{j=1}^G \|\mathbf{Z}_j' \btheta_j/\sqrt{T}\|_2^2} =: \frac{\|\mathbf{Z}(\btheta - \btheta_0)\|_{2}^2}{T\wtl \phi(s_0 + M_2 \wtl s_0)} \eta_0
\end{multline*}
since on $\Theta_{\wtl s_0}$, $s_{(\btheta - \btheta_0)} \leq s_{\btheta} + s_{\btheta_0} \leq M_2 \wtl s_0 + s_0$. We can then use this result to get the second inequality below:
\begin{multline*}
  \EE_0 \Pi\left(\left.\btheta\in \mathbb{R}^{\wtl g}; |\mathbf{Z}_{\tau - h}'(\btheta - \btheta_0)|^2 \geq \eta_0 \frac{M\epsilon^2}{\wtl\phi(s_0 + M_2\wtl s_0)}\right|y,\mathbf{Z}\right)\\
  \leq \EE_0 \Pi\left(\left.\btheta\in \Theta_{\wtl s_0}; |\mathbf{Z}_{\tau - h}'(\btheta - \btheta_0)|^2 \geq \eta_0 \frac{M\epsilon^2}{\wtl\phi(s_0 + M_2\wtl s_0)}\right|y,\mathbf{Z}\right)\\ \hfill
  + \EE_0 \Pi\left(\left.\btheta\in \Theta_{\wtl s_0}^c; |\mathbf{Z}_{\tau - h}'(\btheta - \btheta_0)|^2 \geq \eta_0 \frac{M\epsilon^2}{\wtl\phi(s_0 + M_2\wtl s_0)}\right|y,\mathbf{Z}\right)\\
  \leq \EE_0 \Pi\left(\left.\btheta\in \mathbb{R}^{\wtl g}; \frac{\|\mathbf{Z}(\btheta - \btheta_0)\|_{2}^2}{T\wtl \phi(s_0 + M_2 \wtl s_0)} \eta_0 \geq \eta_0 \frac{M\epsilon^2}{\wtl\phi(s_0 + M_2\wtl s_0)}\right|y,\mathbf{Z}\right)
   + \EE_0 \Pi\left(\left.\Theta_{\wtl s_0}^c\right|y,\mathbf{Z}\right).
\end{multline*}
\noindent The last two terms converge to zero uniformly over $(\btheta_0,\sigma_0^2)\in\overline\Theta_0\times[\underline{\sigma}^2, \overline{\sigma}^2]$ due to Theorem \ref{thm:1} and Lemma B.1.4 in the Supplementary Appendix.

\subsubsection{Proof of Theorem \ref{thm3:out-of-sample1}}
By using the definition of $f_{\mathbf{Z}_{\tau-h},y,\mathbf{Z}}(y_{\tau})$, $f(\cdot|\mathbf{Z}_{\tau-h},y,\mathbf{Z})$, $f_{\mathbf{Z}_{\tau-h},\btheta_0,\sigma_0^2}(\cdot)$ and the Total Variation distance, we have:
\begin{multline*}
  \|f_{\mathbf{Z}_{\tau-h},y,\mathbf{Z}} - f_{\mathbf{Z}_{\tau-h},\btheta_0,\sigma_0^2}\|_{TV} = \bigintsss \left|f(y_{\tau}|\mathbf{Z}_{\tau-h},y,\mathbf{Z}) - f_0(y_{\tau}|\mathbf{Z}_{\tau-h},\btheta_0,\sigma_0^2)\right|d y_{\tau}\\
  = \bigintsss \left|\int \left(f_0(y_{\tau}|\btheta,\sigma^2,\mathbf{Z}_{\tau-h}) - f_0(y_{\tau}|\mathbf{Z}_{\tau-h},\btheta_0,\sigma_0^2)\right)\Pi(\btheta,\sigma^2|y,\mathbf{Z})d\btheta d\sigma^2\right|d y_{\tau}\\
  \leq \bigintsss \int \left|f_0(y_{\tau}|\btheta,\sigma^2,\mathbf{Z}_{\tau-h}) - f_0(y_{\tau}|\mathbf{Z}_{\tau-h},\btheta_0,\sigma_0^2)\right|\Pi(\btheta,\sigma^2|y,\mathbf{Z})d\btheta d\sigma^2 d y_{\tau}\\
  = \bigintsss \left\|f_{\mathbf{Z}_{\tau-h},\btheta,\sigma^2} - f_{\mathbf{Z}_{\tau-h},\btheta_0,\sigma_0^2}\right\|_{TV}\Pi(\btheta,\sigma^2|y,\mathbf{Z})d\btheta d\sigma^2,
\end{multline*}
\noindent where we have used the Fubini's Theorem to get the equality in the last line. Let $\mathcal{A}$ be a subset of the parameter space that will be defined below and recall the notation $\Theta_{\wtl s_0} := \{\btheta\in\Theta; \, s_{\btheta} < M_2 \wtl s_0\}$, where $\wtl  s_0 := \max\{s_0, \log(T)/\log(G)\}$. Hence, by using the upper bound of the Total variation distance: $\|f_{\mathbf{Z}_{\tau-h},\btheta,\sigma^2} - f_{\mathbf{Z}_{\tau-h},\btheta_0,\sigma_0^2}\|_{TV} = 2 -2 \int \min\{f_{\mathbf{Z}_{\tau-h},y,\mathbf{Z}}(y_{\tau}), f_{\mathbf{Z}_{\tau-h},\btheta_0,\sigma_0^2}(y_{\tau})\}dy_{\tau} \leq 2$ we obtain:
\begin{multline}\label{eq:22:supplement}
   \EE_{\mathbf{Z}_{\tau-h}}\EE_0\|f_{\mathbf{Z}_{\tau-h},y,\mathbf{Z}} - f_{\mathbf{Z}_{\tau-h},\btheta_0,\sigma_0^2}\|_{TV}\\
  \leq  \EE_{\mathbf{Z}_{\tau-h}}\EE_0\bigintsss_{\mathcal{A}} \left\|f_{\mathbf{Z}_{\tau-h},\btheta,\sigma^2} - f_{\mathbf{Z}_{\tau-h},\btheta_0,\sigma_0^2}\right\|_{TV} \Pi(\btheta,\sigma^2|y,\mathbf{Z})d\btheta d\sigma^2 + 2\EE_0\bigintsss_{\mathcal{A}^c} \Pi(\btheta,\sigma^2|y,\mathbf{Z})d\btheta d\sigma^2\\
  \leq  \EE_{\mathbf{Z}_{\tau-h}}\EE_0\bigintsss_{\mathcal{A}\cap\Theta_{\wtl s_0}} \left\|f_{\mathbf{Z}_{\tau-h},\btheta,\sigma^2} - f_{\mathbf{Z}_{\tau-h},\btheta_0,\sigma_0^2}\right\|_{TV} \Pi(\btheta,\sigma^2|y,\mathbf{Z})d\btheta d\sigma^2\\ \hfill
  + 2\EE_0\bigintsss_{\Theta_{\wtl s_0}^c}\Pi(\btheta,\sigma^2|y,\mathbf{Z})d\btheta d\sigma^2 + 2\EE_0\Pi(\mathcal{A}^c|y,\mathbf{Z}).
\end{multline}
By using Lemma B.1(i) and Lemma B.5 (i) and (vi) in \cite{GhosalVVVart2017} we get the inequalities
\begin{multline}\label{eq:23:supplement}
  \frac{1}{8}\left\|f_{\mathbf{Z}_{\tau-h},\btheta,\sigma^2} - f_{\mathbf{Z}_{\tau-h},\btheta_0,\sigma_0^2}\right\|_{TV} ^2 \leq \frac{1}{2}d_H^2(f_{\mathbf{Z}_{\tau-h},\btheta,\sigma^2} , f_{\mathbf{Z}_{\tau-h},\btheta_0,\sigma_0^2})\\
  \leq d(f_{\mathbf{Z}_{\tau-h},\btheta,\sigma^2} , f_{\mathbf{Z}_{\tau-h},\btheta_0,\sigma_0^2}) := -\log\int \sqrt{f_{\mathbf{Z}_{\tau-h},\btheta,\sigma^2} f_{\mathbf{Z}_{\tau-h},\btheta_0,\sigma_0^2}},
\end{multline}
\noindent where $d_H^2$ denotes the squared Hellinger distance and $d$ is the Renyi divergence of order $1/2$. Moreover, on $\Theta_{\wtl s_0}$ and by assuming that $\EE_{\mathbf{Z}_{\tau - h}}[\mathbf{Z}_{\tau - h,j}\mathbf{Z}_{\tau - h,k}'] = 0$ for every $j \neq k$, the following holds
\begin{multline*}
  \EE_{\mathbf{Z}_{\tau-h}}|\mathbf{Z}_{\tau - h}'(\btheta - \btheta_0)|^2 = \wtl \phi(s_0 + M_2 \wtl s_0) \frac{\EE_{\mathbf{Z}_{\tau-h}}|\mathbf{Z}_{\tau - h}'(\btheta - \btheta_0)|^2}{\wtl \phi(s_0 + M_2 \wtl s_0)}\\
  \leq \frac{\|\mathbf{Z}(\btheta - \btheta_0)\|_{2}^2}{T\|\mathbf{Z}/\sqrt{T}\|_o^2 \|\btheta - \btheta_0\|_2^2} \left\|\EE_{\mathbf{Z}_{\tau - h}}[\mathbf{Z}_{\tau - h}\mathbf{Z}_{\tau - h}']\right\|_{o} \frac{\|\btheta - \btheta_0 \|_2^2}{\wtl \phi(s_0 + M_2 \wtl s_0)}
  = \frac{1}{T}\sum_{t=1}^T [\mathbf{Z}_t'(\btheta - \btheta_0)]^2\eta
\end{multline*}
since $s_{\btheta - \btheta_0} \leq s_{\btheta} + s_{\btheta_0} \leq M_2 \wtl s_0 + s_0$, $\forall\btheta\in\Theta_{\wtl s_0}$ and where we have used the notation $\eta:=\frac{\left\|\EE_{\mathbf{Z}_{\tau - h}}[\mathbf{Z}_{\tau - h}\mathbf{Z}_{\tau - h}']\right\|_{o}}{\|\mathbf{Z}/\sqrt{T}\|_o^2\wtl \phi(s_0 + M_2 \wtl s_0)}$. By using this result, we have that on $\Theta_{\wtl s_0}$:
\begin{multline}\label{eq:24:supplement}
  -\EE_{\mathbf{Z}_{\tau-h}}\log\int \sqrt{f_{\mathbf{Z}_{\tau-h},\btheta,\sigma^2} f_{\mathbf{Z}_{\tau-h},\btheta_0,\sigma_0^2}}\\ \hfill
  = \frac{1}{4}\log\left(\frac{\sigma_0^2 + \sigma^2}{2\sigma_0^2}\right) + \frac{1}{4}\log\left(\frac{\sigma_0^2 + \sigma^2}{2\sigma^2}\right) + \frac{1}{4(\sigma^2 + \sigma_0^2)}\EE_{Z_{\tau-h}}[\mathbf{Z}_{\tau - h}'(\btheta - \btheta_0)]^2\\
  \leq \frac{1}{4}\log\left(\frac{\sigma_0^2 + \sigma^2}{2\sigma_0^2}\right)\wtl \eta + \frac{1}{4}\log\left(\frac{\sigma_0^2 + \sigma^2}{2\sigma^2}\right)\wtl \eta + \frac{1}{4(\sigma^2 + \sigma_0^2)}\frac{1}{T}\sum_{t=1}^T [\mathbf{Z}_t'(\btheta - \btheta_0)]^2\wtl \eta = \wtl \eta d(f_0,f)
\end{multline}
\noindent where $\wtl \eta := \max\{\eta, 1\}$ and $d(f_0,f)$ is given in \eqref{eq:20:main}.

Define $\mathcal{A} := \{(\btheta,\sigma^2)\in \mathbb{R}^{\wtl g}\times \mathbb{R}_+; d(f_0,f)\leq  M_1\epsilon^2\}$. Hence, from \eqref{eq:23:supplement}, \eqref{eq:24:supplement} and the Fubini's theorem, we obtain:
\begin{multline}\label{eq:21:supplement}
  \EE_{\mathbf{Z}_{\tau-h}} \EE_0\bigintsss_{\mathcal{A}\cap\Theta_{\wtl s_0}} \left\|f_{\mathbf{Z}_{\tau-h},\btheta,\sigma^2} - f_{\mathbf{Z}_{\tau-h},\btheta_0,\sigma_0^2}\right\|_{TV} \Pi(\btheta,\sigma^2|y,\mathbf{Z})d\btheta d\sigma^2\\
  \leq \EE_0\bigintsss_{\mathcal{A}\cap\Theta_{\wtl s_0}} 2\sqrt{2} \sqrt{\wtl \eta d(f_0,f)} \Pi(\btheta,\sigma^2|y,\mathbf{Z})d\btheta d\sigma^2 \leq 2\sqrt{2} \sqrt{\wtl \eta M_1\epsilon^2} \EE_0\Pi(\mathcal{A}|y,\mathbf{Z}).
\end{multline}
By putting together \eqref{eq:22:supplement} and \eqref{eq:21:supplement} we obtain:
\begin{multline*}
  \EE_{\mathbf{Z}_{\tau-h}} \EE_0\|f_{\mathbf{Z}_{\tau-h},y,\mathbf{Z}} - f_{\mathbf{Z}_{\tau-h},\btheta_0,\sigma_0^2}\|_{TV}\\
  \leq 2\sqrt{2} \sqrt{\wtl \eta M_1\epsilon^2} \EE_0\Pi(\mathcal{A}|y,\mathbf{Z}) + 2\EE_0\Pi(\Theta_{\wtl s_0}^c|y,\mathbf{Z}) + 2\EE_0\Pi(\mathcal{A}^c|y,\mathbf{Z})
\end{multline*}
which converges to zero uniformly over $(\btheta_0,\sigma_0^2)\in\overline{\Theta}_0\times [\underline{\sigma}^2, \overline{\sigma}^2]$ under the condition of the theorem and because $\EE_0\Pi(\mathcal{A}^c|y,\mathbf{Z})\rightarrow 0$ by \eqref{eq:13:main} and $\EE_0\Pi(\Theta_{\wtl s_0}^c| y, \mathbf{Z})\rightarrow 0$ by Lemma B.1.4 (uniformly over $(\btheta_0,\sigma_0^2)\in\overline{\Theta}_0\times [\underline{\sigma}^2, \overline{\sigma}^2]$).

\subsection{Stabilization algorithm}\label{sec:stab_alg_appendix}
The stabilization algorithm used in the paper is a slightly modified version of the algorithm proposed in \citet{Andrieu2005} and discussed in \citet{Atchade2011}, which uses re-projections on randomly varying compact sets. Recall that the updating (approximate EM) algorithm described in Section \ref{sec:EB} is:
\begin{displaymath}
\boldsymbol\omega^{(s+1)}=\boldsymbol\omega^{(s)}+a^{(s)} H(\boldsymbol\omega^{(s)},\boldsymbol\phi^{(s+1)})
\end{displaymath}
where we use the transformation $\boldsymbol\omega=0.5\log(\boldsymbol\lambda)$. Let $\{a^{(s)}, s \ge 0\}$ and $\{e^{(s)}, s \ge 0\}$ be two monotone non-increasing sequences of positive numbers. Here we choose $a^{(s)}=1/s^{q}$, with $q=0.8$, and $e^{(s)}=\overline{e}+(1-\overline{e})(1-\varsigma_{s}^{-\alpha_e})$, with $\overline{e}=3$ and $\alpha_e=0.1$. Let $\{\mathsf{K}^{(s)}, s \ge 0\}$ be a monotone increasing sequence of compact subsets of $\boldsymbol\Omega$ such that $\bigcup_{s \ge 0}\mathsf{K}^{(s)}=\boldsymbol\Omega$. Here we set compact subsets of the form $\mathsf{K}^{(s)}=[\max(-\kappa_{s}-1,-c),\kappa_{s}+1]$, where $c>0$. To avoid unstable outcomes due to extremely small numbers in $\boldsymbol\lambda$, we set $c=5$. Let $\widetilde{\boldsymbol\Omega}\times\widetilde{\boldsymbol\Phi}\subset\mathsf{K}^{(s)}\times\boldsymbol\Phi$ and $\Pi : \boldsymbol\Omega\times\boldsymbol\Phi\rightarrow\widetilde{\boldsymbol\Omega}\times\widetilde{\boldsymbol\Phi}$ be a re-projection function, such as $\widetilde{\boldsymbol\Omega}\times\widetilde{\boldsymbol\Phi}=(\widetilde{\boldsymbol\phi},\widetilde{\boldsymbol\omega})$ for an arbitrary point $(\widetilde{\boldsymbol\omega},\widetilde{\boldsymbol\phi})\in\mathsf{K}^{(s)}\times\widetilde{\boldsymbol\Phi}$. Let $\varphi$ be a function such that $\varphi(w)=1-w$, for all $w \ge 0$.

\begin{algorithm}
\caption{Stochatistic approximation with truncation on random boundaries}\label{stab_alg}
\begin{algorithmic}
\State \strut Set $\kappa_{0}=0$, $\nu_{0}=0$, $\varsigma_{0}=0$, $\boldsymbol\omega^{(0)}\in \boldsymbol\Omega$, and $\boldsymbol\phi^{(0)}\in \boldsymbol\Phi^{(0)}$.
\State \strut For $s\ge 1$, compute:
\State \strut \textit{(a)} $\overline{\boldsymbol\phi}\sim P_{\boldsymbol\omega^{(s-1)}}(\boldsymbol\phi^{(s-1)},\cdot)$
\State \strut \textit{(b)} $\overline{\boldsymbol\omega}=\boldsymbol\omega^{(s-1)}+a^{(\varsigma_{s-1}+1)} H(\boldsymbol\omega^{(s-1)},\overline{\boldsymbol\phi})$,
\State where $P_{\boldsymbol\omega}$ is the Markov kernel.
\If {$\overline{\boldsymbol\omega}\in \mathsf{K}^{(\kappa_{s-1})}$ and $\vert\overline{\boldsymbol\omega}-\boldsymbol\omega^{(s-1)}\vert\le e^{(\varsigma_{s-1})}$}
\State $(\boldsymbol\omega^{(s)},\boldsymbol\phi^{(s)})=(\overline{\boldsymbol\omega},\overline{\boldsymbol\phi})$
\State $\kappa_{s}=\kappa_{s-1}$, $\nu_{s}=\nu_{s-1}+1$, $\varsigma_{s}=\varsigma_{s-1}+1$
\Else
\State $(\boldsymbol\omega^{(s)},\boldsymbol\phi^{(s)})=(\widetilde{\boldsymbol\omega},\widetilde{\boldsymbol\phi})\in \widetilde{\boldsymbol\Omega}\times \widetilde{\boldsymbol\Phi}$
\State $\kappa_{s}=\kappa_{s-1}+1$, $\nu_{s}=0$, $\varsigma_{s}=\varsigma_{s-1}+\varphi(\nu_{s-1})$
\EndIf
\State \textbf{end}
\end{algorithmic}
\end{algorithm}

With this algorithm, $\kappa_{s}$ is the index of the active truncation set (also equal to the number of restarts before $s$), $\nu_{s}$ is the number of iterations since the last restart, and $\varsigma_{s}$ is the current index in the step-size sequence. We set $\varphi(w)=1$ for all $w\in\mathbb{N}$, such that $\varsigma_{s}=s$. Hence, if $\overline{\boldsymbol\omega}\notin \mathsf{K}^{(\kappa_{s-1})}$ or $\vert\overline{\boldsymbol\omega}-\boldsymbol\omega^{(s-1)}\vert > e^{(\varsigma_{s-1})}$, we re-initialize the algorithm starting from $(\widetilde{\boldsymbol\omega},\widetilde{\boldsymbol\phi})$, which are obtained by drawing from:
\begin{align*}
\widetilde{\boldsymbol\omega} &\sim \textrm{Uniform}\left(\min(\boldsymbol\omega^{(s-1)},\mathsf{K}^{(s-1)}_{u}), \max(\boldsymbol\omega^{(s-1)},\mathsf{K}^{(s-1)}_{u})\right) \qquad \text{if~} \overline{\boldsymbol\omega}\ge \mathsf{K}^{(s-1)}_{u} \\
\widetilde{\boldsymbol\omega} &\sim \textrm{Uniform}\left(\min(\boldsymbol\omega^{(s-1)},\mathsf{K}^{(s-1)}_{l}), \max(\boldsymbol\omega^{(s-1)},\mathsf{K}^{(s-1)}_{l})\right) \qquad \text{if~} \overline{\boldsymbol\omega}< \mathsf{K}^{(s-1)}_{l}
\end{align*}
where $\mathsf{K}^{(s-1)}_{u}=\kappa_{s-1}+1$ and $\mathsf{K}^{(s-1)}_{l}=\max(-\kappa_{s-1}-1,-c)$, and parameters $\boldsymbol\phi\vert\widetilde{\boldsymbol\omega}$ are drawn from the prior distributions described in Sections \ref{sec:BMIDAS-AGL} and \ref{sec:BMIDAS-AGL-SS}. We then iterate until the acceptance conditions stated in the algorithm are met. Finally, we set the new compact subsets to $\mathsf{K}^{(\kappa_{s-1}+1)}$ and the new sequence of step-size.

\end{document}